\documentclass[10pt,a4paper]{article}
\usepackage[margin=1.2in]{geometry}
\usepackage{microtype}
\usepackage{setspace}
\usepackage{multirow}
\usepackage{amsmath,amssymb}        % matemática
\usepackage{wrapfig}
\usepackage{csvsimple}
\usepackage{booktabs,colortbl}
\usepackage{soul}
\usepackage{geometry}

\usepackage{enumitem} % no preâmbulo

\usepackage{xstring}                 % comparação de strings robusta

\usepackage[most]{tcolorbox}
\usepackage{xcolor} % garante \definecolor

\definecolor{authorRed}{RGB}{180,0,0}
\definecolor{authorBlue}{RGB}{0,90,160}
\definecolor{authorGray}{RGB}{245,245,245}

\tcbset{
  authornote/.style 2 args={
    breakable,
    enhanced,
    colback=authorGray,
    colframe=#1,
    coltitle=#1,
    colbacktitle=white,
    fonttitle=\bfseries,
    title={#2},
    attach boxed title to top left={yshift=-2mm, xshift=2mm},
    boxed title style={
      sharp corners,
      boxrule=0.8pt,
      interior style={fill=white},
      before skip=2pt, after skip=2pt
    },
    left=6mm, right=6mm, top=3mm, bottom=3mm,
    boxrule=0.8pt,
    arc=2pt
  }
}

  {\begin{tcolorbox}[authornote={authorBlue}{#1}]}
  {\end{tcolorbox}}

  {\begin{tcolorbox}[authornote={authorRed}{#1}]}
  {\end{tcolorbox}}
\usepackage[toc,page,header]{appendix}
\usepackage{minitoc}  % IF you are using ICML 
\doparttoc % Tell to minitoc to generate a toc for the parts
\faketableofcontents % Run a fake tableofcontents command for the partocs

\DeclareUnicodeCharacter{FEFF}{} % (protege contra BOM de arquivos Excel/Sheets)

\usepackage{graphicx}
\usepackage{float}
\usepackage{subfig}
\usepackage{booktabs}
\usepackage{multirow}
\usepackage{siunitx}
\usepackage{pgfplotstable}

\PassOptionsToPackage{hyphens}{url}

\usepackage[table]{xcolor}
\usepackage{hyperref}
\hypersetup{colorlinks=true,linkcolor=blue,citecolor=blue,urlcolor=blue}

\makeatletter
\expandafter\def\expandafter\UrlBreaks\expandafter{\UrlBreaks\do\/\do\-}
\makeatother
\usepackage[labelfont=bf,labelsep=period]{caption}
\renewcommand{\figurename}{Fig.} % "Fig." instead of "Figure"

\usepackage[super,sort&compress]{natbib}
\usepackage{multibib} % se quer biblio separada para SI
\newcites{SI}{Supplementary References} % cria \citeSI, \bibliographySI, etc.

\newcommand{\beginsupplement}{%
  \clearpage
  \part*{Supplementary Information}
  \captionsetup{labelfont=bf,labelsep=period}
  \renewcommand{\figurename}{Supplementary Fig.}%
  \renewcommand{\tablename}{Supplementary Table}%
  \pagenumbering{roman}
  \setcounter{page}{1}%
  \setcounter{figure}{0}%
  \setcounter{table}{0}%
  \makeatletter
  \setcounter{NAT@ctr}{0}%
  \makeatother
}

\title{\textbf{CODE-II: A large-scale dataset for artificial\\ intelligence in ECG analysis}}

\author{\small
Petrus E.~O.~G.~B.~Abreu\thanks{Corresponding authors: \texttt{petrusabreu@ufmg.br};\texttt{antonio.ribeiro@ebserh.gov.br};
\texttt{antonio.horta.ribeiro@it.uu.se}}~\,\thanks{Universidade Federal de Minas Gerais (UFMG), Brazil}~,
Gabriela M.~M.~Paix\~{a}o\footnotemark[2]~,
Jiawei Li\thanks{Uppsala University, Sweden}~,
Paulo R.~Gomes\footnotemark[2]~,
Peter W.~Macfarlane\thanks{University of Glasgow, Scotland}~,\\\small
Ana C.~S.~Oliveira\footnotemark[2]~,
Vinícius T. Carvalho\footnotemark[2]~,
Thomas B.~Sch\"{o}n\footnotemark[3]~,
Antonio Luiz P.~Ribeiro~\footnotemark[1] \footnotemark[2] \thanks{Telehealth Center, Hospital das Cl\'{i}nicas, UFMG, Brazil}~,
Ant\^onio H.~Ribeiro\footnotemark[1]~\,\footnotemark[3]
}

\date{}
\begin{document}
\maketitle
\part{}

% ========================== Abstract ========================
\begin{abstract}
Data-driven methods for electrocardiogram (ECG) interpretation are rapidly progressing. Large datasets have enabled advances in artificial intelligence (AI) based ECG analysis, yet limitations in annotation quality, size, and scope remain major challenges. Here we present CODE-II, a large-scale real-world dataset of 2,735,269 12-lead ECGs from 2,093,807 adult patients collected by the Telehealth Network of Minas Gerais (TNMG), Brazil. Each exam was annotated using standardized diagnostic criteria and reviewed by cardiologists. A defining feature of CODE-II is a set of 66 clinically meaningful diagnostic classes, developed with cardiologist input and routinely used in telehealth practice. We additionally provide an open available subset: CODE-II-open, a public subset of 15,000 patients, and the CODE-II-test, a non-overlapping set of 8,475 exams reviewed by multiple cardiologists for blinded evaluation. A neural network pre-trained on CODE-II achieved superior transfer performance on external benchmarks (PTB-XL and CPSC 2018) and outperformed alternatives trained on larger datasets.
\end{abstract}

%====================================================================================
% Introduction - SECTION
%====================================================================================
\section{Introduction}

Cardiovascular diseases are the leading cause of death worldwide, causing about 17.9 million deaths annually \cite{Mensah2023-nn}. The electrocardiogram (ECG) is a simple, non-invasive, low-cost, and widely available tool for screening and monitoring cardiovascular conditions. Despite its broad accessibility, manual interpretation requires highly skilled and well-trained cardiologists, which can be a limiting factor in many healthcare settings. Automated ECG interpretation has a history spanning more than six decades, with pioneering computer-based analyses introduced in the 1960s. While widely used in practice, modest performance of rule based algorithms limits its clinical usage as tool and relegates them to an ancillary role \cite{Rautaharju2016-wr}. In recent years, however, the convergence of large-scale data and artificial intelligence (AI) has transformed the field. Deep neural networks now achieve cardiologist-level performance in clinically relevant diagnostic tasks, including the detection of six abnormalities \cite{Ribeiro2020_nc}, and have further enabled novel applications beyond traditional interpretation, such as Chagas disease screening \cite{Jidling2023-sq}, atrial fibrillation risk prediction \cite{Attia2019-hy}, and AI-derived ECG age as a predictor of mortality risk \cite{Lima2021-mk}. These advances underscore the transformative potential of AI-based ECG analysis, while also highlighting the critical need for large, well-annotated datasets with clinically meaningful labels to support further progress.

There is an intrinsic connection between telehealth and artificial intelligence. Telehealth generates the large volumes of data required to train AI algorithms and its fully digital format provides an ideal setting for integrating AI-based tools to enhance clinical workflows. The Telehealth Network of Minas Gerais (TNMG) \cite{Oliveira2025-bd} is a consolidated public tele-electrocardiography (tele-ECG) service in operation since 2005, currently providing more than 8,000 ECG reports daily to over 1,400 counties across 14 states in Brazil. A total of more than 11 million digital ECGs have been performed\footnote{\url{https://telessaude.hc.ufmg.br}}. From this service, the CODE dataset (CODE-I) \cite{Ribeiro2020_nc} was established, comprising ECGs and annotations provided by doctors in the TNMG from 2010 to 2017. This dataset has been linked to mortality data, the Clinical Outcomes in Digital Electrocardiography (CODE) study \cite{Ribeiro2019-kp}, with many publications regarding the prognostic value of the ECG, using both standard ECG classification and AI-based methods \cite{Paixao2019-lm,Paixao2020-ym,Paixao2021-yh,Lima2021-mk,Paixao2022-ai,Sau2024-as}.

Computerized ECG analysis has been data-driven since its inception in the 1960s, although the availability of large, high-quality datasets emerged only later. Landmark resources shaped the field, from early cohorts such as the MIT-BIH Arrhythmia Database \cite{Moody2001-gg} to more recent large-scale datasets like the PTB-XL \cite{Wagner2020-ht} and the Harvard-Emory \cite{Koscova2025-im}. The CODE dataset \cite{Ribeiro2020_nc} shaped the field in its own way. It comprises more than 2 million ECGs from over 1 million patients. Published in 2020, it is now available for more than 50 research groups. A subset of it, the CODE-15\%, was made completely open and was downloaded more than 59 thousand times by the time of the writing of this paper. The models pre-trained on it (also made openly available\footnote{\url{https://github.com/antonior92/automatic-ecg-diagnosis}}) had a huge impact on research and have been downloaded thousands of times.

One of the main limitations of the original CODE dataset was the restricted number of annotated abnormalities. We focused on six representative classes, which were extracted from free-text medical reports and underwent several rounds of revision to ensure consistency. However, the effort required to curate these six classes made it infeasible, at the time, to extend the process to all abnormalities routinely considered by the telehealth center. Thanks to major improvements in the center's internal operations, such as standardizing report formats and resolving annotation inconsistencies \cite{Oliveira2025-bd}, CODE-II now offers high-quality labels for comprehensive abnormality classes.

The heterogeneity of ECG reports has been an issue since the inception of epidemiological studies in the 1930s. The lack of standardization in ECG classes has impaired comparisons between different populations worldwide. In this setting, ECG classification systems were developed, such as the Minnesota Code \cite{Prineas2016-vy} and NovaCode \cite{Rautaharju1992-ft}, but their use was primarily limited to research, with limitations in clinical practice. The last ECG statement issued by the American Heart Association (AHA) and the American College of Cardiology (ACC) and endorsed by the Heart Rhythm Society and the International Society for Computerized Electrocardiology was released in 2007 to examine the relation of the resting ECG to its technology, increase understanding of how the modern ECG is derived and recorded, and promote standards that will improve the accuracy and usefulness of the ECG in practice \cite{Kligfield2007-fj}. Since then, to the best of our knowledge, no other comprehensive international guideline on ECG reporting has been published.

%Since then, no other international guideline in the ECG report has been published.

With over 40 cardiologists involved in the TNMG tele-ECG service, standardizing the ECG report was essential to ensure consistent, high-quality, and reliable interpretations. This standardization had to balance two priorities: alignment with internationally accepted criteria and adaptation to Brazil's regional and epidemiological context, for instance, accounting for the high prevalence of Chagas disease in certain states. To this end, the ECG CODE classes were defined based on the 2007 guidelines from the AHA and the ACC \cite{Kligfield2007-fj}, and the Brazilian guideline for reporting ECG \cite{Pastore2016-tn,Samesima2022-oj}. Moreover, AI-based classification within the tele-ECG workflow may help prioritize urgent cases with life-threatening abnormalities, improving the efficiency of the reporting process. The CODE-II dataset, collected between January 2019 and December 2022, fully benefits from this standardized approach, enabling high-quality annotation across a wide range of abnormalities.

Our contributions are: (1) we curate and describe a large-scale ECG database: the CODE-II dataset, (2) we describe the classification system used in the telehealth service, (3) we develop a deep neural network-based classification model using this dataset, showcasing its utility for AI applications, (4) we offer a subset of this dataset, the CODE-II-open, as an open public benchmark for AI-based ECG classification models, with annotations derived from the 66 expert-defined CODE diagnostic classes, and (5) we provide a high-quality test set, the CODE-II-test, containing 66 ECG classes and annotated by multiple cardiologists, to support the evaluation of AI algorithms. The proposed CODE classes are clinically meaningful and well-suited for AI-based ECG analysis. This resource also facilitates comparisons with other electronic cohorts and encourages international collaboration in ECG research. The high-quality data will serve as a valuable foundation for developing robust AI models for automated ECG interpretation.

%====================================================================================
% RESULTS - SECTION
%====================================================================================
\section{Results}

%------------------------------------------------------------------------------------
% CODE-II cohort - SUBSECTION
%------------------------------------------------------------------------------------
\subsection{CODE-II dataset}

CODE-II is a large-scale, real-world dataset of 12-lead ECG exams collected and annotated by the TNMG from January 2019 to December 2022, originally comprising over 3 million exams. After applying the filtering and quality-control procedures described in the Methods section, a curated set of 2,735,269 exams from 2,093,807 unique patients was retained. This refined cohort constitutes the CODE-II dataset.

\begin{figure}[ht]
	\centering{
		\includegraphics[scale=1]{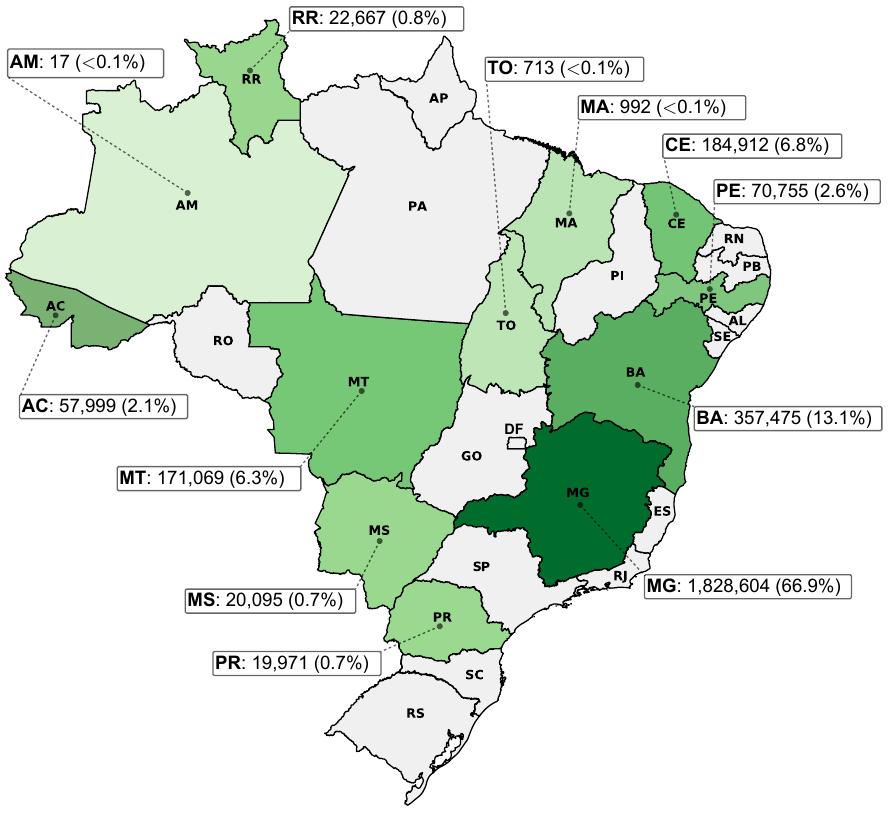} %\vspace{-4mm}
		\caption{Map of Brazil illustrating the absolute and relative numbers of ECGs for each state served by TNMG in the CODE-II dataset.}
		\label{Figure1}
	}
\end{figure}

During the CODE-II collection period, TNMG’s tele-ECG service supported 12 Brazilian states, with exam volumes across states shown in Fig.\,\ref{Figure1}. Based on the indications provided by the healthcare professionals who submitted the exams to the TNMG analysis center, approximately 73.2\% of the exams were related to elective cases, 25.3\% to urgent cases, and 1.5\% to preferential cases. These exams were collected mainly in primary health care centers, followed by hospitals, emergency departments, and ambulances.

Among the 2,093,807 patients, 59.1\% (1,237,632) were female and 40.9\% (856,175) were male. As some patients underwent multiple exams---1,663,122 had a single ECG and 430,685 had two or more ECGs (Supplementary Fig.\,4a)---only the first exam per patient was considered in analyses involving clinical or demographic characterization of the population. The overall mean age was 53.6 years, with a mean of 53.0 years for female patients and 54.4 years for male patients.

Self-reported clinical comorbidities of patients, categorized by age (18--59 years and 60 years or older at the time of the exam) and sex, are summarized in Supplementary Table\,1. Hypertension was the most prevalent comorbidity, affecting nearly half of the cohort (47.7\%), followed by diabetes mellitus (14.0\%), smoking (8.2\%), dyslipidemia (4.8\%), previous myocardial infarction (1.8\%), Chagas disease (1.1\%), and chronic obstructive pulmonary disease (0.7\%). Patients could report more than one comorbidity, and medication usage was also considered to infer the presence of hypertension, dyslipidemia, and diabetes, as detailed in the Supplementary Materials.

In addition to these comorbidities, clinical indications for ECG exams were recorded using checkbox fields, allowing multiple symptoms to be reported per patient. The most frequent reasons were routine examinations and chest pain, followed by preoperative risk assessment, palpitations, dyspnea, and ``other''. The full distribution of clinical indications is presented in Supplementary Fig.\,3. Notably, routine exams and chest pain were the leading causes across all demographic groups, with younger female patients (aged 18--59) showing the highest proportions across all categories. A more granular characterization and further descriptive details of the cohort, including extended tables and figures on patient- and exam-level attributes, are provided in the Supplementary Materials.

%------------------------------------------------------------------------------------
% The CODE diagnostic classes and its distribution in the population - SUBSECTION
%------------------------------------------------------------------------------------
\subsection{The CODE diagnostic classes and its distribution in the population}

Each ECG in the CODE-II dataset was annotated by a certified cardiologist with one or more diagnostic labels from a standardized set of 66 classes, known as the CODE diagnostic classes, of which 65 are not mutually exclusive and 1 (Normal case) is assigned exclusively. These classes were developed to ensure consistency and clinical relevance across all ECG reports, enabling the analysis of cardiovascular findings on a large scale. As described in the Methods section, these labels encompass a wide spectrum of normal and abnormal findings. They can be grouped into 10 clinically meaningful categories: Pacemaker, Normal, Technical issues, Sinus Rhythm, Arrhythmia, Atrioventricular Conduction Disorders, Chamber Hypertrophy, Intraventricular Conduction Disturbances, Ischemia/Infarction, and Miscellaneous Conditions.

\begin{figure}[htb] 
	\centering{
		\includegraphics[scale=0.9]{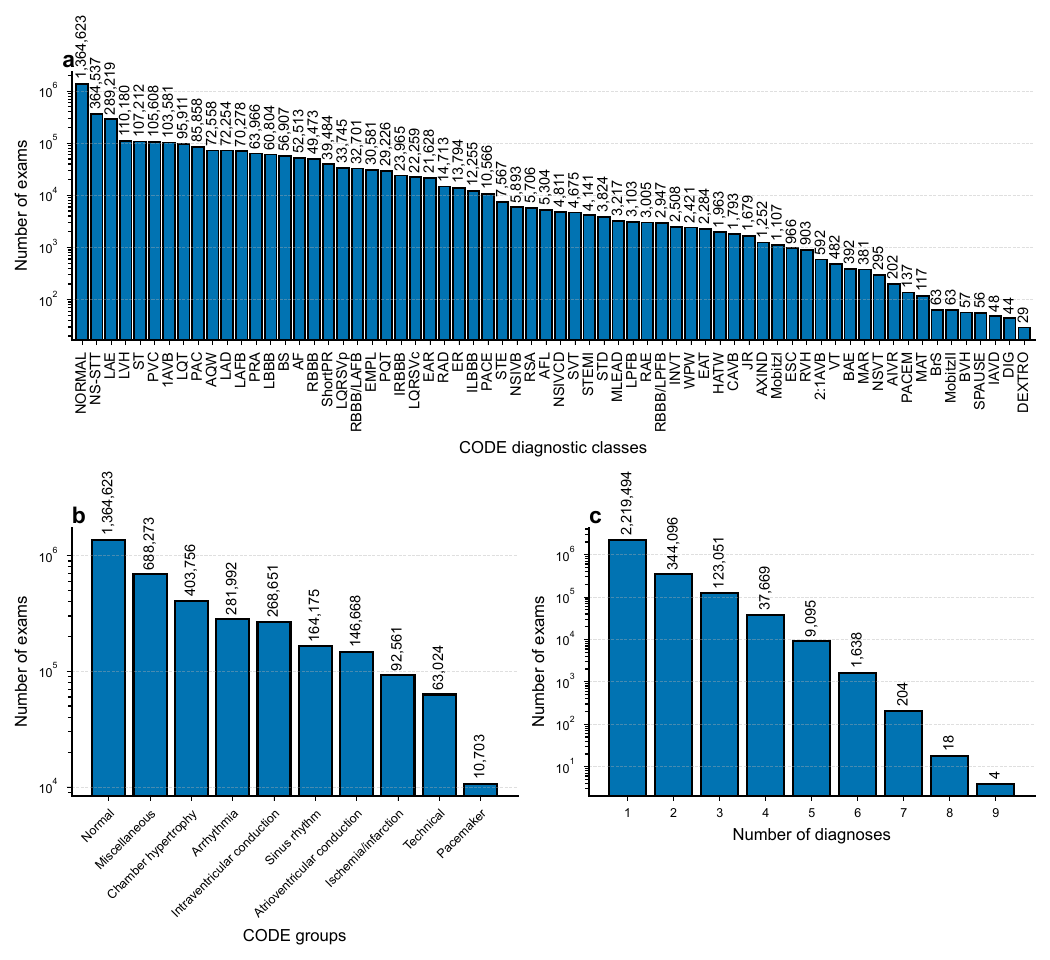} \vspace{-8mm}
		\caption{Summary of diagnostic class frequencies and group combinations in CODE-II. (a) Number of ECG exams associated with each diagnostic class in the CODE-II dataset. Diagnostic classes are not mutually exclusive; multiple diagnoses may be assigned to a single exam, except for Normal ECGs, which are exclusive. (b) Number of exams per each CODE diagnostic group. Exams may belong to more than one group. (c) Distribution of the number of diagnostic classes assigned per exam. \textit{All panels use a logarithmic scale on the y-axis}.}
		\label{Figure2}
	}
\end{figure}

Figure\,\ref{Figure2}a shows the distribution of all 66 CODE diagnostic classes according to their frequency in the dataset. For visualization purposes, each class is referenced by a shorthand, hereafter referred to as the CODE label; the corresponding formal diagnostic statements appear in Fig.\,\ref{Figure8}, and the label--statement mapping is given in Supplementary Table\,2. The most common findings included normal ECG (NORMAL: 1,364,623 exams, 49.9\%), nonspecific ST-T abnormality (NS-STT: 364,537 exams, 13.3\%), left atrial enlargement (LAE: 289,219 exams, 10.6\%), left ventricular hypertrophy (LVH: 110,180 exams, 4.0\%), and sinus tachycardia (ST: 107,212 exams, 3.9\%). These findings illustrate the diverse clinical spectrum captured by the dataset, ranging from routine screenings to structural and electrical abnormalities. Less frequent but clinically relevant conditions, such as atrial fibrillation (AF: 52,513 exams, 1.9\%) and complete bundle branch blocks---left bundle branch block (LBBB: 60,804 exams, 2.2\%) and right bundle branch block (RBBB: 49,473 exams, 1.8\%)---were also well represented. The complete distribution of all 66 classes, together with their associated labels and the training-validation split that will be used in subsequent analysis, is provided in Supplementary Table\,2.

The distribution of exams across the 10 CODE diagnostic groups was also analyzed. As illustrated in Fig.\,\ref{Figure2}b, normal was the most prevalent category (1,364,623 exams, 49.9\%), followed by miscellaneous conditions (688,273 exams, 25.2\%), chamber hypertrophy (403,756 exams, 14.8\%), arrhythmia (281,992 exams, 10.3\%), and intraventricular conduction disturbances (268,651 exams, 9.8\%). Since normal is the only CODE group composed of a single diagnosis that is mutually exclusive with the remaining 65 diagnostic classes, we further examined the number of exams exclusively assigned to each group. Supplementary Fig.\,6 presents this distribution, the 10 most frequent combinations of non-exclusive group assignments (among 251 distinct patterns observed), and the distribution of the number of CODE groups assigned per exam. This group-level analysis highlights not only the clinical diversity of the dataset but also its potential for developing and evaluating diagnostic models across a wide range of cardiovascular patterns, from benign findings to complex arrhythmogenic and structural conditions.

Additionally, we evaluated the number of diagnostic classes assigned per ECG. As illustrated in Fig.\,\ref{Figure2}c, most exams were annotated with only one CODE class, with decreasing frequencies for exams containing multiple labels. This distribution reflects the fact that, while many ECGs show isolated findings, a substantial subset presents with two or more co-occurring abnormalities. This highlights the need for AI models capable of handling multilabel classification tasks.

To minimize bias due to repeated exams from the same individuals, we performed an additional analysis using only the first ECG per patient (Supplementary Table\,3). The relative frequencies remained consistent.

%------------------------------------------------------------------------------------
% CODE-II Test: a curated test set for model evaluation - SUBSECTION
%------------------------------------------------------------------------------------
\subsection{CODE-II-test: a curated test set for model evaluation}

The CODE-II-test dataset consists of 8,475 ECGs collected between 2018 and 2025 from unique patients not present in the CODE-II dataset. Each exam was annotated by multiple certified cardiologists using the 66 CODE diagnostic classes. Final labels were assigned following predefined agreement and majority-based rules: 6,309 exams (74.4\%) achieved complete concordance among reviewers, while 2,166 exams (25.6\%) were resolved using the majority rule. To ensure balanced representation, additional care was taken to include cases covering rare diagnostic classes. This curated dataset was specifically designed to provide a reliable benchmark for evaluating the performance of AI models using high-quality, expert-reviewed ECGs.

Among the 8,475 patients included in the CODE-II-test, 57.5\% (4,871) were female and 42.5\% (3,604) were male. The overall mean age was 55.5 years, with similar averages for female (55.3 years) and male (55.7 years) patients. Patients were stratified into two age groups: 18--59 years, comprising 58.6\% of the cohort (2,898 female and 2,068 male), and 60 years or older, comprising 41.4\% (1,973 female and 1,536 male). No patients under 18 years of age were included.

The distribution of the 66 CODE diagnostic classes and their aggregation into 10 diagnostic groups highlight the broad spectrum of electrocardiographic findings represented in the dataset. Normal ECGs were the most frequent (3,398 exams, 40.1\%), followed by abnormalities such as nonspecific ST-T changes (969 exams, 11.4\%) and left atrial enlargement (912 exams, 10.7\%), alongside a wide range of other clinically relevant abnormalities. A complete description and detailed characterization of this cohort, including extended tables and figures on patient- and exam-level attributes, are provided in the Supplementary Materials.

%------------------------------------------------------------------------------------
% Baseline AI model for CODE diagnostic classification - SUBSECTION
%------------------------------------------------------------------------------------
\subsection{Baseline AI model for CODE diagnostic classification}

We developed a deep neural network to classify ECG exams according to the 66 CODE diagnostic classes (Fig.\,\ref{Figure8}). It is built upon a convolutional residual architecture previously employed by the TNMG research group to identify clinically relevant abnormalities in rhythm and morphology. Adapted from the ResNet model originally introduced for image classification \cite{He2016-dx,He2016-sh}, the architecture was modified to process one-dimensional ECG signals and has demonstrated expert-level performance in earlier studies \cite{Ribeiro2020_nc}.

The model was trained on the large-scale CODE-II dataset and evaluated on the independent, curated CODE-II-test set. The datasets originate from the same telehealth network (TNMG), but are non-overlapping, containing exams collected from different patients to allow for an unbiased assessment. In addition to evaluations based on the CODE diagnostic classes, we also explored the utility of this model’s encoder when applied to external ECG datasets annotated with alternative diagnostic schemes.

%------------------------------------------------------------------------------------
% Performance of the model on the CODE-II Test - SUBSECTION
%------------------------------------------------------------------------------------
\subsection{Performance of the model on the CODE-II-test}

\begin{figure}[H]
	\centering{
		\includegraphics[scale=1]{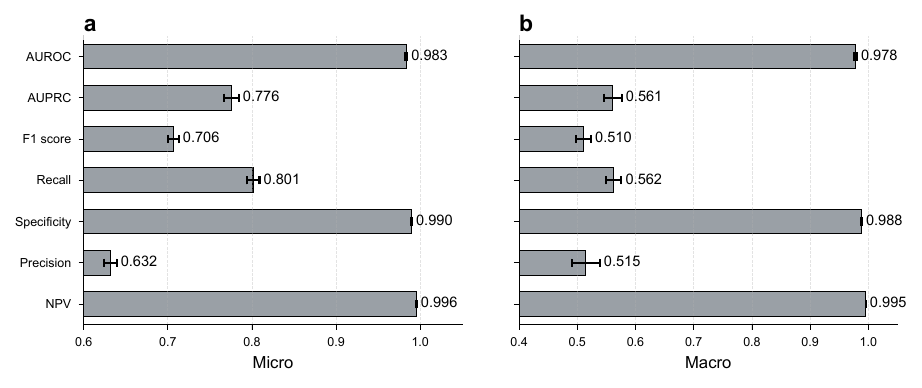} \vspace{-6mm}
		\caption{Global performance of the model on the CODE-II-test dataset. Panels show (a) micro- and (b) macro-averaged results for AUROC, AUPRC, F1 score, Recall, Specificity, Precision, and NPV. Threshold-dependent metrics were computed after applying class-specific thresholds selected to maximize the F1-score. Bars represent the mean metric values, with numerical values shown to the right of each bar, and horizontal lines denoting 95\% confidence intervals estimated from 1,000 bootstrap resamples.}
		\label{Figure3}
	}
\end{figure}

For threshold-independent metrics, we report the Area Under the Receiver Operating Characteristic Curve (AUROC) and the Area Under the Precision–Recall Curve (AUPRC, also referred to as Average Precision, AP). The model achieved a micro-averaged AUROC (micro-AUROC) of 0.983 and AUPRC of 0.776, and a macro-averaged AUROC (macro-AUROC) of 0.978 and AUPRC of 0.561, indicating robust discriminative ability, with consistently strong performance across common classes and greater variability among rare categories. When applying class-specific thresholds that maximize the F1-score, the model reached a micro-F1 of 0.706, precision of 0.632, recall (sensitivity) of 0.801, specificity of 0.990, and negative predictive value (NPV) of 0.996, reflecting solid global performance with a tendency toward higher sensitivity at the cost of lower precision. These thresholds were fixed from the validation set and applied directly to the test set, ensuring a fair evaluation while making threshold-dependent metrics sensitive to shifts in prevalence and calibration between cohorts. Macro-averaged scores were lower, with a macro-F1 of 0.510, precision of 0.515, recall of 0.562, specificity of 0.988, and NPV of 0.995, highlighting the greater difficulty of detecting rare conditions when all classes are equally weighted. These metrics are summarized in Fig.\,\ref{Figure3}, where horizontal lines represent 95\% confidence intervals computed from 1,000 bootstrap resamples. Overall, the model performs robustly across the diagnostic spectrum, although results for rare classes remain more variable and should be interpreted with caution due to class imbalance.

\begin{figure}[H] %htpb
	\centering{
		\includegraphics[scale=0.95]{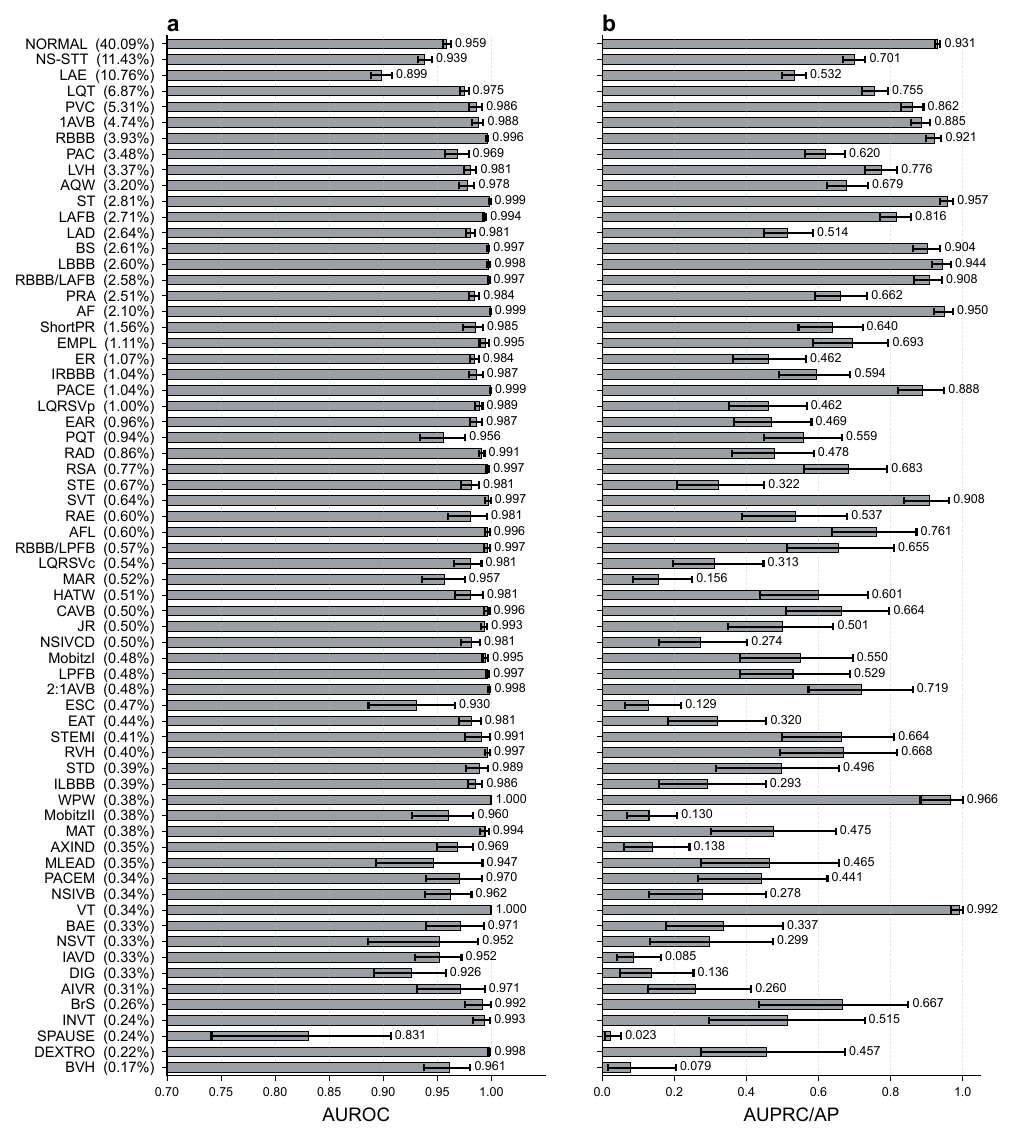} \vspace{-8mm}
		\caption{Per-class threshold-independent performance of the model on the CODE-II-test. Panels show (a) AUROC and (b) AUPRC (average precision) with 95\% confidence intervals computed from 1,000 bootstrap resamples. Bars indicate the mean metric value for each diagnostic class, with the numerical value shown to the right of each bar. The horizontal lines represent the corresponding confidence intervals. Diagnostic classes are ordered by their prevalence in the test set (shown in parentheses), where prevalence divided by 100 corresponds to the expected AUPRC of a random classifier.}
		\label{Figure4}
	}
\end{figure}

Per-class analysis revealed strong discriminative ability across both common and rare diagnoses (Fig.\,\ref{Figure4}). For example, the ECG within normal limits for age and sex (NORMAL class), representing approximately 40.1\% of the test set, achieved an AUPRC of 0.931—more than twice its prevalence (0.401)—underscoring the model's reliability in identifying exams without abnormalities. The model also performed well for moderately prevalent but clinically important abnormalities: left bundle branch block (LBBB, 2.6\%) reached an AUPRC of 0.944, and atrial fibrillation (AF, 2.6\%) achieved an AUPRC of 0.950. Even in rare but critical conditions such as ST-elevation myocardial infarction (STEMI, 0.4\%), the model obtained an AUPRC of 0.647, substantially above prevalence, demonstrating its ability to prioritize time-sensitive cases despite limited representation. By contrast, performance was more modest in extremely rare classes; for instance, isorhythmic atrioventricular dissociation (IAVD, 0.3\%) achieved an AUPRC of only 0.074, illustrating ongoing challenges in detecting sparsely represented conditions. The horizontal lines in Fig.\,\ref{Figure4} represent 95\% confidence intervals computed from 1,000 bootstrap resamples. Full per-class results, including AUPRC, AUROC, prevalence, and threshold-dependent metrics, are provided in Supplementary Table\,9.

We compared the  F1-max thresholds with an alternative threshold-selection rule based on Youden’s J statistic. Under Youden’s J, micro-averaged precision decreased to 0.228 and recall increased to 0.946, yielding a lower F1 of 0.368 (macro precision 0.182, recall 0.945, F1 0.274), which are reported with those obtained with F1-max thresholds in Supplementary Table 10 (micro/macro). At the class level, we compare just 3 of 66 diagnostic classes—multifocal atrial tachycardia (MAT), isorhythmic atrioventricular dissociation (IAVD), and digitalis effect (DIG)—for which F1-max thresholds selected on validation produced degenerate values on Test (precision, recall, and F1 equal to zero) and show that Youden’s J yielded non-zero values. In these cases, the Youden-selected thresholds were several orders of magnitude smaller than the validation F1-max thresholds, enabling positive predictions and non-zero precision/recall/F1 (Supplementary Table 11). Taken together, these comparisons show that thresholds chosen on validation can materially alter threshold-dependent performance estimates when applied to a different test distribution. Full per-class metrics under the primary protocol are provided in Supplementary Table 9; comparative results for threshold choice appear in Supplementary Tables 10 and 11.

Regarding cardiologist performance on CODE-II-test, results are uniformly high under a fair-scope protocol aligned with the agreement/majority labeling rules used to define the ground-truth labels (Supplementary Table 8). Among high-volume reviewers ($\geq 100$ exams; n = 20 cardiologists), the mean micro-F1 was 0.966 (mean micro-recall 0.978; mean micro-precision 0.956), and the mean macro-F1 was 0.944, with macro precision and recall likewise high. Performance varies with workload and breadth of classes encountered: the highest-volume reviewer (8,368 exams; 65 classes) achieved micro-F1 0.869 (precision 0.821; recall 0.924) and macro-F1 0.844 (precision 0.805; recall 0.915), lower than the high-volume mean and reflecting more realistic conditions under greater workload and broader diagnostic coverage. The model, in turn, maintains strong ranking ability (AUROC/AUPRC), while threshold-dependent aggregate metrics are sensitive to thresholds chosen on validation and applied unchanged to the Test set with different class prevalence and score calibration—under F1-max thresholds, micro-F1 0.706 (precision 0.632; recall 0.801) and macro-F1 0.512 (precision 0.517; recall 0.562) (Supplementary Tables 9--11). A formal head-to-head comparison is not entirely fair since the annotators have been used to generate the ground-truth labels.

%------------------------------------------------------------------------------------
% Scaling laws - SUBSECTION
%------------------------------------------------------------------------------------
\subsection{Scaling laws in CODE-II: how training dataset size influences model performance}

\begin{figure}[ht]
	\centering{
		\includegraphics[scale=0.9]{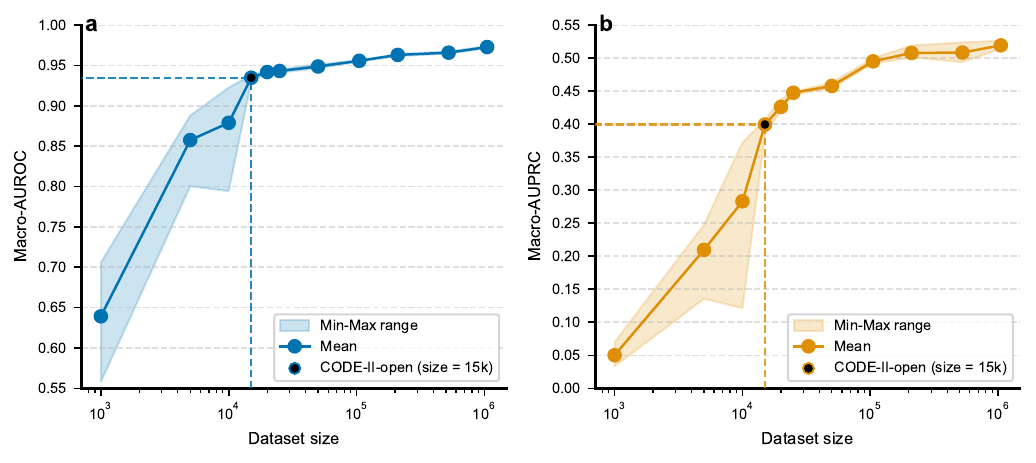} \vspace{-4mm}
		\caption{Scaling law analysis of model performance. (a) Macro-AUROC and (b) macro-AUPRC as a function of dataset size on the fixed CODE-II-test. The curves represent the mean performance across three independent runs, with shaded areas indicating the range between the minimum and maximum values.}
		\label{Figure5}
	}
\end{figure}

The effect of dataset size on model performance, a behavior consistent with previously observed scaling laws in deep learning \cite{Kaplan2020scalinglaws}, is shown in Fig.\,\ref{Figure5} for both macro-AUROC and macro-AUPRC/AP on the fixed CODE-II-test. Dataset sizes ranged from small subsets of 1,000 patients up to 1,050,000 patients, corresponding to approximately 50\% of the patient-level dataset, with the CODE-II-open also included in the analysis. As expected, larger datasets for model development consistently led to improved performance, although gains diminished as dataset size increased. Each curve represents the mean performance across three independent runs for a given dataset size, while the shaded area indicates the range between the minimum and maximum values observed across these runs.

As shown in Fig.\,\ref{Figure5}, model performance (macro-AUROC and macro-AUPRC) improved sharply with increasing dataset size up to around 15,000 patients, after which the rate of improvement became progressively smaller. This transition coincides with the CODE-II-open, confirming it as the inflection point of the scaling curve. Variability across runs was higher at the smallest dataset sizes and narrowed as the dataset grew, supporting the robustness of the observed trend beyond the inflection region. These findings provide empirical evidence for the importance of dataset scale in ECG classification and quantify the performance gap between publicly available and full-scale training resources. Importantly, they also demonstrate that most of the attainable performance can be achieved with substantially fewer data than the total available. Together with the analysis on CODE-II-open, these results highlight the practical trade-off between dataset size and marginal performance gains, offering guidance for future model development and data-sharing initiatives in ECG research.

%------------------------------------------------------------------------------------
% Comparative evaluation on external ECG benchmarks - SUBSECTION
%------------------------------------------------------------------------------------
\subsection{Comparative evaluation on external ECG benchmarks}

To further assess the generalizability of the representations learned by our model, we fine-tuned the model pre-trained on CODE-II on two external ECG classification benchmark datasets: PTB-XL \cite{Wagner2020-ht}, comprising 21,837 clinical 12-lead ECG recordings, and CPSC 2018 \cite{Liu2018-lx}, containing 9,831 12-lead recordings. For PTB-XL, experiments were conducted under two training settings: (i) full-data training, using 100\% of the training data, and (ii) few-shot learning, using only 5\% and 10\% of the training data, in order to evaluate the ability of the pre-trained model to adapt and generalize from limited labeled data. This design simulates real-world scenarios with limited labeled data and enables a comprehensive assessment of model transferability.

We benchmark our model against a set of supervised models trained from scratch and publicly available pre-trained ECG models. Full model names, citations, and implementation details are provided in the Methods section. To evaluate the impact of pre-training, we also compare each of the pre-trained model with a version trained from scratch on the external dataset using randomly initialized weights.

\begin{figure}[htpb] %htpb
	\centering{
		\includegraphics[scale=0.85]{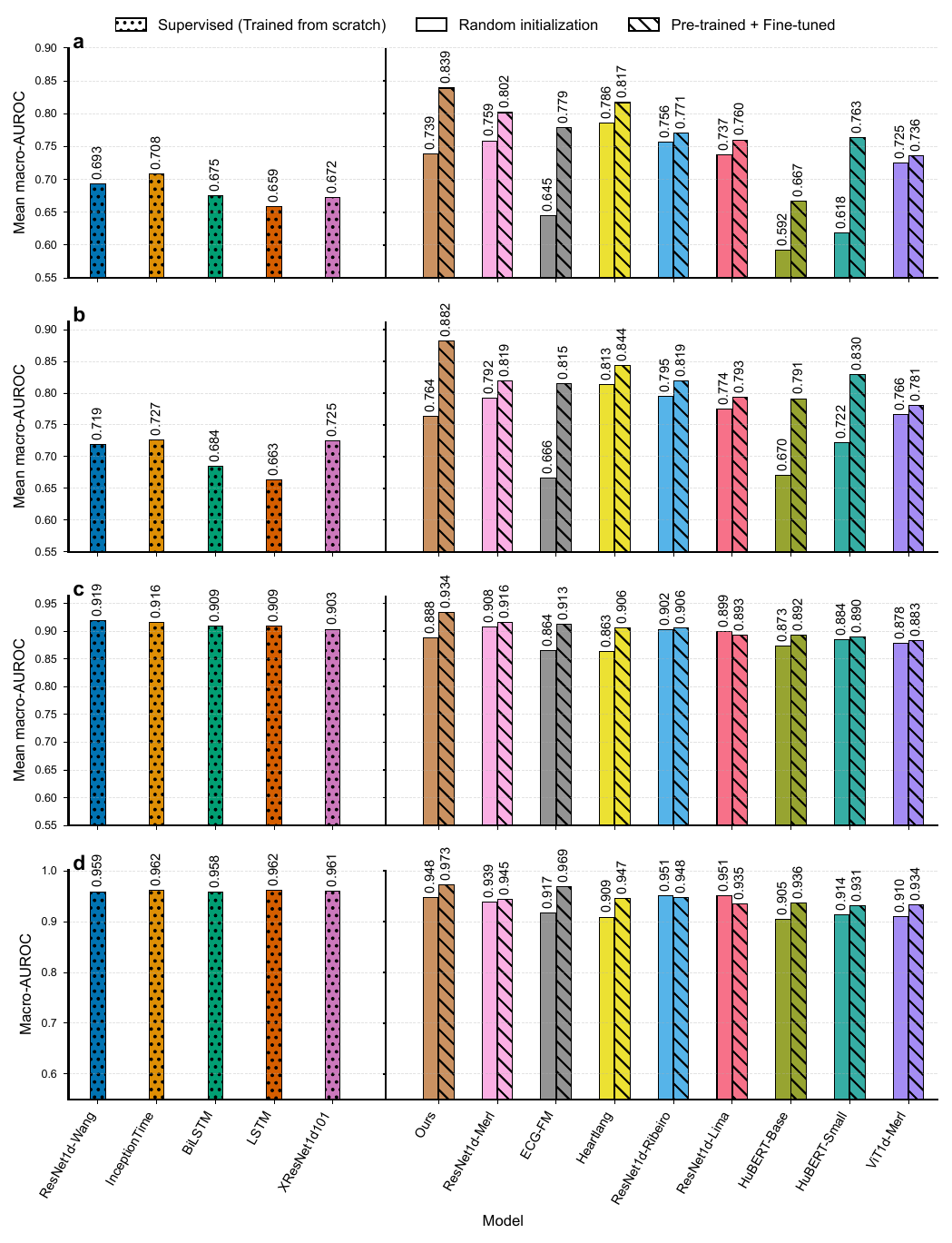} \vspace{-4mm}
		\caption{Generalization performance of supervised and pre-trained models across external ECG benchmarks. Each panel summarizes model performance across training regimes ranging from few-shot to full-data settings: (a–c) PTB-XL mean macro-AUROC, obtained by averaging each model's macro-AUROC across the five diagnostic taxonomies (diagnostic, subclass, superclass, format, and rhythm) when trained with 5\%, 10\%, and 100\% of the data; and (d) CPSC 2018 macro-AUROC. Colors denote distinct model architectures, while bar patterning indicates the training regime: dotted (supervised models trained from scratch), diagonal-hatched (pre-trained models fine-tuned on the target dataset), and solid (same architecture trained from scratch, with random initialization, used to assess the effect of pre-training). Across most architectures, fine-tuned pre-trained models achieve higher performance than their randomly initialized counterparts.}
		\label{Figure6}
	}
\end{figure}

We report in Supplementary Data\,6 the average macro-AUROC across three independent runs for all evaluated models, with the performance range (maximum-minimum) indicated in parentheses. The number of model parameters is also reported in Supplementary Data\,6. As shown in Supplementary Table\,12, our model consistently achieved the highest average macro-AUROC across all 5 PTB-XL diagnostic categories and the CPSC 2018 dataset (average macro-AUROC of 0.9405) under full-data training, demonstrating strong robustness as reflected by an average range of 0.003. It outperformed both the supervised and pre-trained baselines. Under the more challenging few-shot learning setting, our model ranked first or second in 8 out of 10 scenarios. Notably, although supervised models perform competitively under full-data training, their performance declined significantly under few-shot conditions, highlighting the superior transferability of pre-trained models. Overall, our model was the only model that achieved an average macro-AUROC above 0.9 across all datasets (0.9071), surpassing the second-best model, Heartlang, by 3.27\%. Compared to the randomly initialized counterpart, our pre-trained model showed a substantial improvement of 8.68\%, thereby demonstrating the effectiveness of pre-training on the CODE-II dataset, see Supplementary Data\,6. Importantly, CODE-II is only 30.01\% the size of HuBERT's pre-training datasets (9.1 million ECGs), and our model achieved superior performance with only 7.85\% of ECG-FM, 18.27\% of Heartlang, and 23.62\% of the parameters of HuBERT-Small. Figure\,\ref{Figure6} summarizes model performance from the few-shot to full-data regimes, reporting the mean macro-AUROC across all 5 PTB-XL diagnostic categories for 5\%, 10\%, and 100\% of the training data (Figs.\,\ref{Figure6}a–c) and the results for the CPSC 2018 dataset (Fig.\,\ref{Figure6}d).

% \begin{todo}[Figure being prepared]
% This figure will include four subfigures (a–d) summarizing model performance from few-shot to full-data regimes across all categories. Panels (a–c) report the mean macro–AUROC on PTB-XL, averaged over the five category schemes (diagnostic, subclass, superclass, form, rhythm), when training with 5\%, 10\%, and 100\% of the PTB-XL training set, respectively. Panel (d) shows macro–AUROC for all models on the CPSC 2018 dataset.
% \end{todo}

%------------------------------------------------------------------------------------
% CODE-II Mini: benchmark dataset and a brief model comparison - SUBSECTION
%------------------------------------------------------------------------------------
\subsection{CODE-II-open: benchmark dataset and a brief model comparison}

The CODE-II-open dataset, which will be publicly released as part of this work, comprises 15,000 unique patients and constitutes a carefully curated subset of the CODE-II dataset. Its main characteristics are similar to those of the full dataset, with a detailed description provided in the Supplementary Material. Despite being smaller than the full CODE-II dataset, CODE-II-open is a substantial public benchmark, comparable to existing open datasets, including PTB-XL (21,837 ECGs from 18,885 patients, Germany) and CPSC 2018 (9,831 ECGs from 9,458 patients, China). Crucially, CODE-II-open inherits the rigorous data-curation procedures and the clinically meaningful CODE diagnostic classes established by certified cardiologists at TNMG, which have been applied and tested in routine telecardiology practice for over a decade.

% Using the same evaluation pipeline and external models described in the subsection \textit{Comparative evaluation on external ECG benchmarks}, we conducted a brief analysis---less extensive than the external benchmarking---to illustrate the representativeness of the CODE-II-open as a benchmark dataset. Specifically, the models were trained and validated on CODE-II-open and evaluated on CODE-II-test.

Using the same evaluation pipeline and external models described in the previous subsection, we conducted a concise analysis to assess the representativeness of the CODE-II-open dataset as a benchmark. Supplementary Data\,7 reports macro-AUROC and macro-AUPRC for three independent runs of all evaluated models, along with the performance range (maximum--minimum) in parentheses. The mean values across runs are summarized in Supplementary Table~15 and illustrated in Fig.~\ref{Figure7}. Models were trained either from scratch or fine-tuned from pre-trained versions, and the best-performing model based on validation macro-AUROC on CODE-II-open was evaluated on CODE-II-test. Our baseline architecture trained from scratch on CODE-II-open achieved a macro-AUROC of 0.951 and a macro-AUPRC of 0.410, whereas the same architecture pre-trained on the full CODE-II dataset reached 0.978 and 0.561, respectively. Despite being trained on only approximately 0.72\% of the available data, the CODE-II-open model matched or outperformed several well-established ECG models under their best-performing regimes, indicating that the dataset is sufficiently rich and diverse to support the development of competitive ECG classifiers. The performance gap observed between our model trained on the full CODE-II dataset and the same architecture trained on CODE-II-open follows the scaling-law trends presented in Fig.\,\ref{Figure5}.

\begin{figure}[ht]
	\centering{
		\includegraphics[scale=0.85]{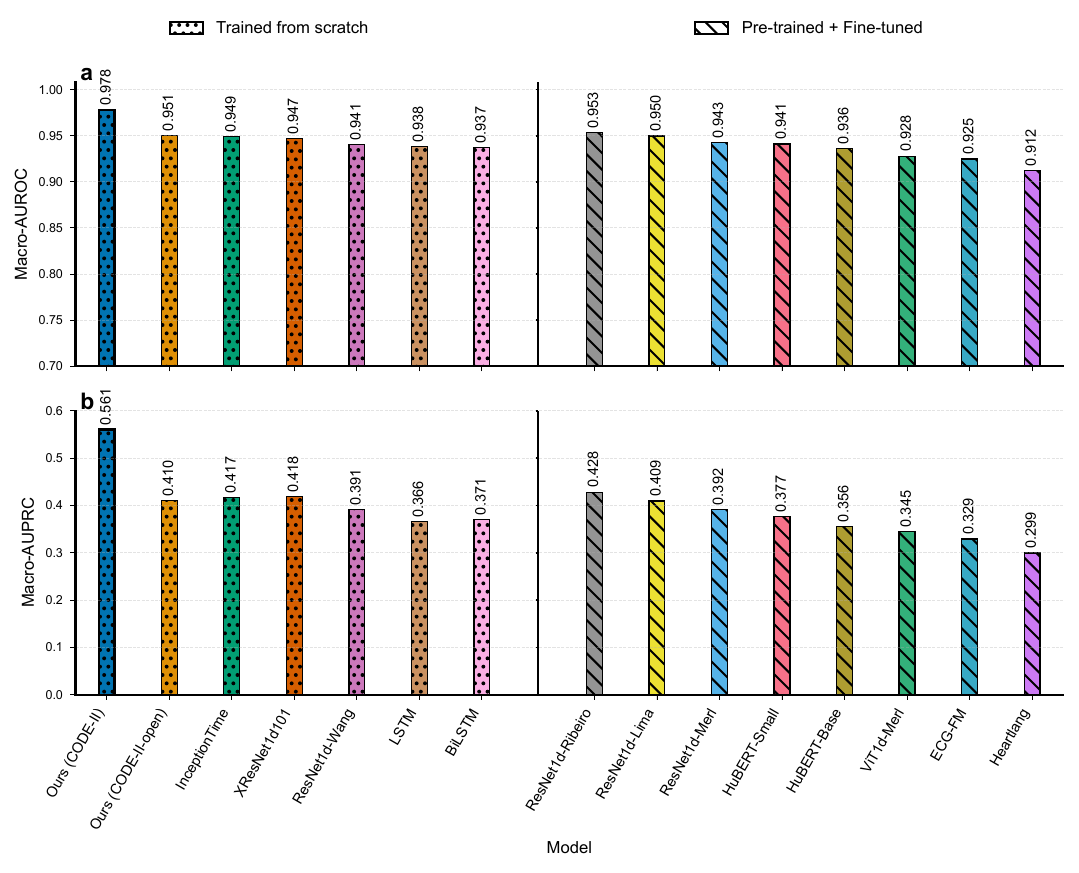} \vspace{-4mm}
		\caption{Models evaluated on CODE-II-open and CODE-II-test. Panels show (a) macro-AUROC and (b) macro-AUPRC for all models trained on the CODE-II-open dataset and evaluated on the CODE-II-test set. Models were either trained from scratch or fine-tuned from pre-trained versions. Colors indicate distinct model architectures, while bar patterning denotes the training regime: dotted for models trained from scratch and diagonal-hatched for pre-trained models fine-tuned on CODE-II-open. For reference, the baseline model trained from scratch on the full CODE-II dataset is also included.}
		\label{Figure7}
	}
\end{figure}

% \begin{todo}[Figure being prepared]
% This figure will summarize the results referenced above from Supplementary Table~15. It will show side-by-side macro-averaged AUROC and AUPRC for (i) our model trained on the full CODE-II cohort and (ii) our model and literature baselines trained on CODE-II-open, all evaluated on CODE-II-test.
% \end{todo}

These results position the CODE-II-open as both a realistic and valuable public benchmark for ECG classification research. Beyond enabling transparent model evaluation, it provides an accessible foundation for future studies on model pre-training, fine-tuning, and cross-dataset generalization.

%====================================================================================
% Discussion SECTION
%====================================================================================
\section{Discussion}

Our work provides a comprehensive resource for advancing AI-based ECG analysis. A key strength is the CODE-II dataset, characterized by high-quality annotations derived from standardized diagnostic criteria and reviewed by cardiologists with extensive expertise in ECG interpretation. The introduction of 66 diagnostic classes—the CODE diagnostic classes—developed in collaboration with clinical experts, captures a broad spectrum of ECG findings while providing a clinically meaningful and interpretable label space that aligns with cardiological reasoning and remains suitable for model development. The open release of the CODE-II-open dataset, a representative subset, and the rigorously curated \mbox{CODE-II-test}, reviewed by multiple cardiologists, provides the research community with robust benchmarks for developing and evaluating automated ECG classification models within this framework. Together, these resources combine diagnostic breadth, clinical heterogeneity, and expert curation to establish a new reference standard for automated ECG analysis. The performance on the CODE-II-test provides direct insights into how such algorithms could be deployed to enhance and streamline large-scale telehealth services.

Building on these foundations, CODE-II represents a substantial advance in scope and design compared with the original CODE dataset (CODE-I) \cite{Ribeiro2020_nc}. CODE-I was a landmark study showing that deep neural networks could outperform clinicians on 12-lead ECG interpretation, but it was restricted to six rhythm and conduction classes such as atrial fibrillation, first-degree atrioventricular block, sinus bradycardia and tachycardia, and bundle branch blocks. CODE-II overcomes these limitations by expanding the label space from 6 to 66 classes, capturing a broader range of ECG abnormalities not represented in CODE-I. It also introduces new subsets: while CODE-I provided the CODE-15\% subset and a test set of 827 ECGs, CODE-II adds two purpose-built resources—the CODE-II-open, a public subset of 15,000 unique patients, and the CODE-II-test, a non-overlapping collection of 8,475 unique patients designed for blind evaluation under standardized criteria. By pre-defining clinically stratified training and validation splits in CODE-II-open, the dataset mitigates sampling bias and enables fairer algorithmic comparisons, analogous to PTB-XL \cite{Wagner2020-ht}. Initial evaluations suggest that models trained on CODE-II achieve high performance across many more diagnostic categories than what was possible with CODE-I, highlighting the advantages of a richer and more informative label space.

A key demonstration of the value of the CODE-II dataset lies in its ability to support models that generalize effectively to external datasets. When fine-tuned on established ECG benchmarks such as PTB-XL \cite{Wagner2020-ht} and CPSC 2018 \cite{Liu2018-lx}, our proposed model pre-trained on CODE-II consistently outperformed both supervised models trained from scratch and other publicly available pre-trained ECG models. This advantage was evident not only under full-data training but also in few-shot settings, where CODE-II pre-training enabled strong performance even when only a small fraction of the available training data was used. Such robustness highlights the transferability of the representations learned from CODE-II, reflecting the breadth and clinical fidelity of its diagnostic classes. Importantly, our results demonstrate that models pre-trained on CODE-II consistently outperform alternatives---even those with substantially more parameters or trained on larger datasets such as HuBERT \cite{Coppola2024-lz}, ECG-FM \cite{McKeen2024-fm}, or Heartlang \cite{Jin2025-ym}. These improvements can be attributed primarily to the quality of the CODE-II data and the breadth of its 66 diagnostic classes rather than to architectural or training innovations, as underscored by the limited gains observed when training comparable models on the six classes of CODE-I. This underscores the combined value of a well-designed model architecture and a high-quality, clinically curated dataset, highlighting that label quality and a clinically meaningful diagnostic system may prove more important than dataset size alone in achieving robust and generalizable AI tools for ECG clinical analysis.

There is a heterogeneity in the meaning of an ECG diagnosis, or diagnostic class as defined in this study, since it may reflect a direct measurement of the ECG intervals (for example, QT prolongation or a first-degree AV block), a disease diagnosed primarily through ECG, like atrial fibrillation or complete AV block, or a disease or condition that requires confirmation by non-electrocardiographic means, like an acute  myocardial infarction, pericarditis or hyperkalemia \cite{Committee_Members2001-ph}. This blending of signal patterns and clinical diagnoses can hinder both interpretability and generalizability, especially in the context of machine learning. The CODE diagnostic classes address this limitation by providing a clearer separation: they are primarily defined based on recognizable electrocardiographic patterns, leaving the recognition of a specific medical condition to the clinical reasoning, taking into account age, sex, clinical symptoms, and comorbidities. This structure preserves the richness of expert interpretation while offering a more consistent and algorithm-friendly label space. Notably, the CODE diagnostic classes have been tested and refined over more than a decade of use in a large-scale telehealth service, ensuring their practicality and robustness in real-world settings.

While machine learning is enabling novel applications of the ECG---such as predicting patient prognosis or detecting conditions traditionally beyond the reach of standard electrocardiography---it is important to emphasize that these advances build upon, rather than diminish, the value of conventional ECG interpretation. Traditional diagnostic frameworks encapsulate nearly a century of accumulated clinical knowledge, with human experts having defined categories of abnormalities such as atrial fibrillation, atrioventricular block, and ischemic changes that remain central to both clinical reasoning and machine learning. Training AI models on these well-established electrocardiographic categories allows algorithms to inherit a wealth of prior knowledge that guides pattern recognition and ensures interpretability. Indeed, even the most recent ECG foundation models rely on classic diagnostic labels curated by experts \cite{Han2024-pq}. For example, a recent \textit{Nature Communications} study defined 60 arrhythmia and ECG finding classes for an ECG model \cite{Lai2023-af}, while the ECGFounder model was trained on 150 categories from the Harvard-Emory ECG database, which reflect established ECG diagnostic standards \cite{Li2024-vg}. These examples highlight that cutting-edge AI initiatives extend rather than replace the traditional ECG interpretive framework. The CODE-II diagnostic classes were developed in this same spirit, embedding accumulated clinical wisdom into a structure that is both meaningful for physicians and suitable for model development using artificial intelligence.

Beyond methodological advances, the implications of CODE-II extend directly to clinical practice and telecardiology. By enabling the development of AI models that combine diagnostic accuracy with operational efficiency, the dataset opens new possibilities for optimizing workflows in large-scale telehealth services. Such models can facilitate scheduling systems that prioritize urgent cases, reduce the likelihood of diagnostic errors, and allow physicians respond more rapidly to straightforward exams---for example, by automatically identifying normal tracings and directing specialist attention to those requiring further evaluation. They can also serve as valuable decision-support tools for non-specialists working in resource-limited settings, providing a reliable second opinion and enabling earlier recognition of critical conditions. However, effectively deploying these tools requires going beyond conventional performance metrics. For instance, when the focus is on classifying abnormalities, a false positive---when the model erroneously labels a normal ECG as abnormal---may generate unnecessary alarms, additional referrals, or redundant testing, potentially overwhelming services with benign cases. Conversely, a false negative---when a true abnormality such as a serious arrhythmia or ischemic change goes undetected---is even more concerning, as it can delay urgent treatment with potentially severe consequences. Recent studies illustrate this trade-off, showing that AI models can substantially reduce false negatives compared to human readers while modestly increasing false positives in ambulatory ECG monitoring \cite{Johnson2025-ev}. To ensure that AI models trained on CODE-II genuinely benefit telehealth workflows, it is essential to evaluate not only global performance metrics but also the clinical and operational impact of errors at both macro and micro levels. In designing and evaluating AI for telecardiology, sensitivity (recall) and precision (positive predictive value) should be balanced according to clinically motivated thresholding strategies tailored to specific clinical applications. For example, we previously reported preliminary results for normal-ECG detection \cite{Abreu2025-ff}, highlighting that a high-precision threshold is particularly relevant for minimizing missed abnormalities and avoiding delays in clinical intervention. Conversely, a high-recall threshold is more appropriate for urgent abnormalities such as ST-elevation patterns, where capturing all potential cases---even at the expense of precision---is essential to avoid missing life-threatening events. In future work, we will formalize these operating modes and assess their impact on workflow efficiency, leveraging CODE-II to guide models optimized for real-world deployment.

Despite these advances, certain limitations should be acknowledged. CODE-II is derived from a single national telehealth system, and although it encompasses millions of exams, its population characteristics may not fully capture the diversity of global clinical settings. Moreover, the CODE diagnostic classes were intentionally designed to reflect recognizable electrocardiographic patterns that are well suited for machine learning but do not replace the broader clinical reasoning that integrates symptoms, comorbidities, and complementary examinations. The dataset also reflects the inherent variability of real-world ECG acquisition, including artifacts (e.g., noise from movement or electrode contact) and device heterogeneity, which, while increasing ecological validity, can introduce additional challenges for AI model development. Recognizing these limitations does not diminish the value of CODE-II; rather, it highlights key directions for future research.

As demonstrated in our external evaluations on datasets from Germany (PTB-XL) and China (CPSC 2018), models trained on CODE-II generalized effectively across distinct populations, underscoring its potential to capture clinically transferable representations. Future multinational collaborations will be essential to extend such validation across diverse healthcare systems and ensure global generalizability. Another promising avenue is the integration of ECG with multimodal clinical data, extending the utility of CODE-II beyond diagnostic classification toward prognostic modeling and risk stratification---similar to prior efforts in CODE-I, where a subset of records was linked to mortality outcomes \cite{Ribeiro2020_nc}. Further investigation into threshold selection and calibration---including probability scaling and prevalence-aware adjustment---is an important next step, given the observed shifts in class prevalence and score calibration between the validation and test sets, which make threshold-dependent metrics particularly sensitive.

Comprehensive dataset descriptions---including patient demographics, comorbidities, symptoms, and ECG interval measurements stratified by sex and age---are provided in the Supplementary Materials to facilitate broader scientific use. A brief analysis of these electrocardiographic measurements revealed physiologically consistent sex- and age-related differences, in line with prior population-based studies and known physiological patterns \cite{Mason2007-uy,Rijnbeek2014-gx,Palhares2017-eb,Pinto2017-ba}. Such consistency underscores the clinical validity of the CODE-II dataset and supports its potential for secondary research beyond diagnostic modeling, including studies on electrophysiological variability, population health, and disease risk profiling. Taken together, the rigorously curated CODE-II dataset, its 66-class diagnostic system refined through more than a decade of telehealth practice, and the publicly available CODE-II-open and CODE-II-test subsets establish a foundation for global collaboration and continuous progress in AI-driven cardiology. In the era of emerging \textit{ECG foundation models} trained on increasingly large datasets, CODE-II offers a curated, clinically meaningful, and expertly annotated benchmark that can strengthen this growing research area by providing robust, interpretable diagnostic labels grounded in real-world practice.

\section{Methods}

%------------------------------------------------------------------------------------
% CODE-II SUBSECTION
%------------------------------------------------------------------------------------
\subsection{Tele-electrocardiogram system}

The TNMG tele-ECG system was developed by the network’s in-house team. Instead of transmitting images of the ECG tracings, the system captures electrocardiographic signals directly from digital electrocardiographs of different models, enabling efficient management and allowing further processing of the acquired ECG data \cite{Oliveira2025-bd}.

A primary care professional performs the ECGs remotely using digital electrocardiograms. This professional, typically a nursing technician, applies a standardized clinical questionnaire that includes the type of exam (urgent, elective or preferential), date of birth, sex (recorded as male or female, corresponding to biological sex), clinical indication, symptoms, current medications, and self-reported comorbidities: smoking, hypertension, diabetes, dyslipidemia, Chagas disease, previous myocardial infarction, and chronic obstructive pulmonary disease. Electrocardiography machines record standard 12-lead ECG signals, typically sampled at rates of 300 Hz, 500 Hz, 600 Hz, or 1000 Hz. Each exam may consist of multiple tracings, in general 2 to 4, each lasting 7 to 12 seconds, captured successively.

An in-house software platform was developed and is integrated with electrocardiography machines to capture ECG tracings, upload them with the patient's clinical history, and transmit them via the Internet to the Telehealth Center at the Hospital das Clínicas of UFMG. All received ECGs are converted to a custom format that meets the current needs of healthcare operations and academic research. These exams are preprocessed for automatic measurement extraction and then organized into a priority queue based on the user-suggested priority, time of receipt, and ECG data. The web platform for ECG reports displays the ECG tracings, clinical data, and automatic measurements of the ECG waves and intervals provided by the University of Glasgow ECG system and integrated into our software. The platform has several tools for magnifying, filtering, and correcting the automatic measurements. Urgent cases are indicated, with abnormal measurements displayed in red, and borderline values highlighted in orange. All ECGs are analyzed by a certified cardiologist, who selects one or more ECG CODE classes. A decision support system was developed to allow the selection of only those diagnosis classes compatible with the original or corrected measurements \cite{Gomes2021-jj}. When the cardiologist completes the report, a PDF file becomes available for download at the original remote point, including the identification, clinical data, ECG tracings, and the full report. A monthly audit by senior cardiologists of a sample of randomly selected ECGs is conducted to ensure the quality and consistency of the cardiological expertise.

%------------------------------------------------------------------------------------
% CODE diagnostic classes -  SUBSECTION
%------------------------------------------------------------------------------------
\subsection{CODE diagnostic classes}

The CODE diagnostic classes were introduced in 2018 to replace free-text ECG reports. The main goal was to standardize ECG diagnosis among the cardiologist team and establish a homogeneous report for primary care physicians.  The criteria used for ECG diagnosis were based on the standards of the American Heart Association, American College of Cardiology, the Heart Rhythm Society and the International Society for Computerized Electrocardiology for ECG interpretation \cite{Kligfield2007-fj} and the Brazilian guidelines for reporting ECG \cite{Pastore2016-tn,Samesima2022-oj}. A total of 66 ECG classes were defined and grouped into 10 categories: Pacemaker, Normal, Technical, Sinus Rhythm, Arrhythmia, Atrioventricular Conduction, Chamber Hypertrophy, Intraventricular Conduction, Ischemia/Infarction, and Miscellaneous (see Fig.\ref{Figure8}). The distribution and prevalence of exams for each of the 66 CODE classes are shown in Supplementary Table 2. A comprehensive description and representative examples of all CODE classes are available at: \url{https://code.telessaude.hc.ufmg.br/}.

\begin{figure}[htp] %p
	\centering{
		\includegraphics[scale=0.98]{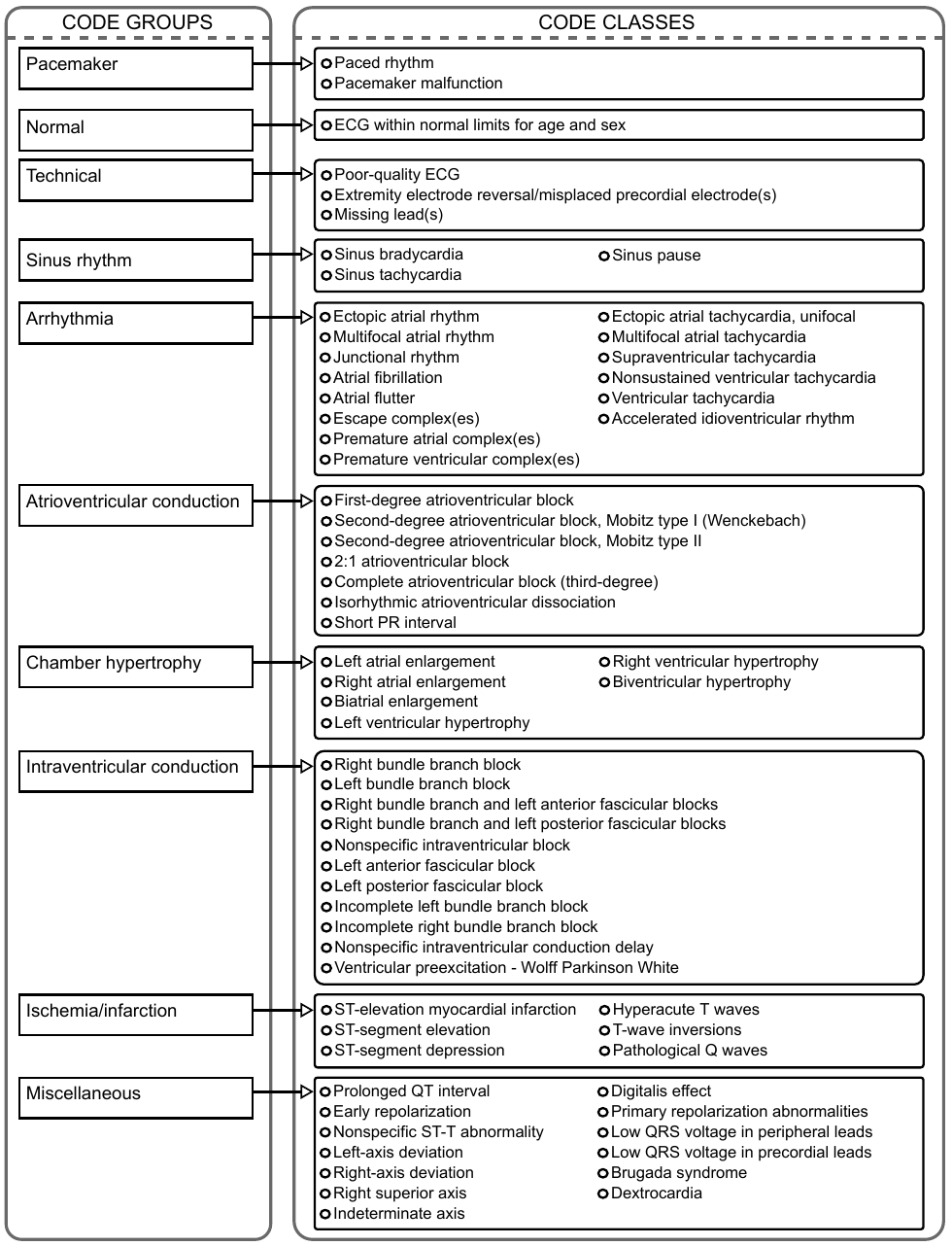} %\vspace{-4mm}
		\caption{Schematic representation of the 10 groups and their corresponding 66 ECG CODE classes.}
		\label{Figure8}
	}
\end{figure}

This standardized labeling framework ensured consistent annotation of ECG exams at scale, supporting both routine clinical workflows and the development of AI models. All diagnostic labels used in this study were derived from this coding system, ensuring alignment with clinical standards and reproducibility of findings.

%------------------------------------------------------------------------------------
% Inclusion criteria and organization of CODE-II dataset -  SUBSECTION
%------------------------------------------------------------------------------------
\subsection{Inclusion criteria and organization of CODE-II dataset}

All ECGs performed between January 2019 and December 2022 were included for evaluation. The dataset underwent two main preprocessing stages: one to assess the consistency of key exam and report information, and another to ensure the technical integrity of the ECG tracings.

\begin{figure}
\begin{centering}
    \includegraphics[height=0.45\textwidth]{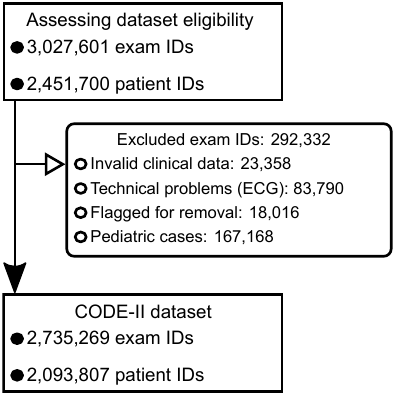} %\vspace{-4mm}
    \includegraphics[height=0.45\textwidth]{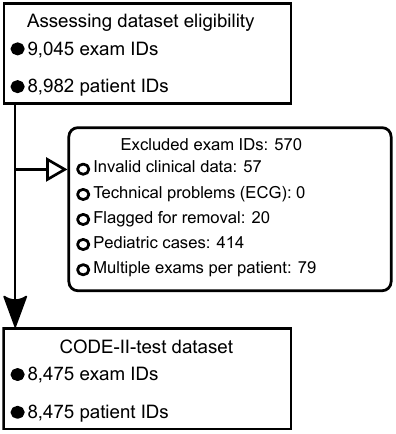} %\vspace{-4mm}
		\caption{Flowchart of the CODE-II dataset (\textbf{left}) and CODE-II-test dataset (\textbf{Right}) .}
		\label{Figure9}
\end{centering}
		
\end{figure}

In the first preprocessing stage, we applied exclusion criteria based on the exam date, the patient’s date of birth, and sex. Exams with inconsistent information were removed by discarding the corresponding exam identifiers (IDs), which were deemed invalid clinical data. In the second stage, we removed exam IDs for which ECG tracings were missing or had insufficient duration in their essential leads (I, II, V1--V6) for analysis. These cases were deemed technical problems. Additionally, exams previously flagged for removal due to being linked to external research projects or because their medical reports consisted solely of medical observations or requests were removed. Exams from patients under 18 years old were also removed and deemed pediatric cases (Fig.\,\ref{Figure9}).

Notably, prior to applying the preprocessing stages described above, all records underwent an internal curation process conducted by a TNMG data specialist. This process aimed to identify and link records belonging to the same individual and assign a unique patient ID, even when discrepancies in patient identifiers were present. This step helped correct inconsistencies and prevent unnecessary data loss. The specialist had access to personally identifiable information on patients from both the TNMG system and the Brazilian Public Health System (SUS), which enabled the application of multiple verification steps to accurately detect, link, and correct records. Importantly, these personal identifiers are not included in the final dataset due to privacy regulations and are accessible only to authorized personnel involved in internal data validation.

The resulting dataset, named CODE-II, is organized into three components. (i) Clinical/exam metadata: exam ID and date; patient identifiers (ID, date of birth, sex); exam location; clinical history; current medications; comorbidities; and the clinical indication for the ECG. Patient age was computed as the difference between exam date and date of birth, converted to years by dividing by 365.25 to account for leap years. (ii) Reporting and labeling data: exam and report IDs (linking to the metadata), report upload dates, cardiologist IDs, report type (original or revised), diagnostic labels and their CODE class IDs, and ECG-derived cardiac measurements. (iii) Raw signal data: 12-lead ECG waveforms accompanied by technical metadata, including the exam identifier; a sequential index for each tracing within the exam (starting at 1 and preserving the acquisition order); the sampling rate; and the signal resolution. Each tracing corresponds to a consecutive recording from the same clinical session. The total number of tracings in an exam can be inferred from the highest sequential index observed for that exam.

%------------------------------------------------------------------------------------
% The CODE-II Test -  SUBSECTION
%------------------------------------------------------------------------------------
\subsection{The CODE-II-test}

To enable the development and benchmarking of new AI applications, we curated a high-quality test dataset, named CODE-II-test, from audited ECG cases within TNMG. The dataset comprises 8,475 12-lead ECGs from unique patients and was specifically designed as an independent benchmark for evaluating AI-based diagnostic models. Special attention was taken to include exams spanning even the rarer diagnostic classes, thereby covering the full spectrum of the 66 CODE classes---normal findings, technical issues that preclude analysis, and diverse abnormalities. To prevent information leakage, \mbox{CODE-II-test} is non-overlapping with the CODE-II dataset at the patient level (i.e., no patient appears in both), avoiding any transfer of patient-specific features when evaluating models trained on CODE-II. The same preprocessing pipeline used for CODE-II (Fig.\,\ref{Figure9}a) was applied, with an additional exclusion criterion retaining only the first ECG per patient (Fig.\,\ref{Figure9}b).

A total of 46 certified cardiologists independently participated in the annotation process, each contributing to a variable number of exams depending on availability and audit allocation. Diagnostic classes were assigned using the standardized 66 CODE diagnostic classes. Final labels were determined using two predefined criteria: (i) \textit{agreement}, defined as complete concordance among all reviewers across all diagnoses assigned to an exam; or (ii) \textit{majority rule}, defined as at least two reviewers agreeing on at least one diagnosis for the exam.

The CODE-II-test dataset was curated to include only essential metadata: exam ID, exam date, patient ID, date of birth, sex, age, specialist IDs, and the final assigned diagnoses. This design ensures both high-quality annotations and strict patient-level separation from the training and validation sets, enabling its use as an independent benchmark for AI evaluation.

%------------------------------------------------------------------------------------
% Architecture and training of the AI model for CODE classification -  SUBSECTION
%------------------------------------------------------------------------------------
\subsection{Architecture and training of the AI model for CODE classification}

\begin{figure}[ht]
	\centering{
		\includegraphics[scale=1]{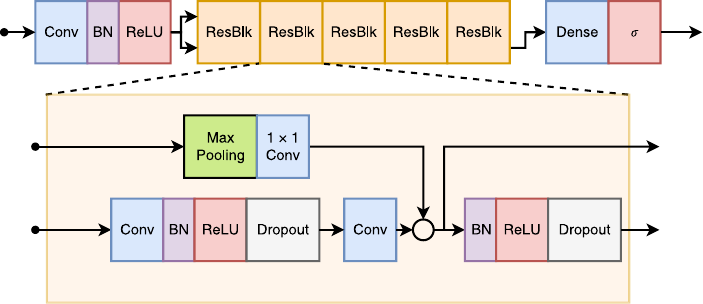} %\vspace{-4mm}
		\caption{Deep neural network model architecture.}
		\label{Figure10}
	}
\end{figure}

To showcase the potential of the dataset for AI application and also for establishing a baseline for future developments, we train and test a baseline model. The architecture of the baseline model, illustrated in Fig.\,\ref{Figure10}, is a deep convolutional neural network based on a residual architecture adapted from ResNet \cite{He2016-dx,He2016-sh,Ribeiro2020_nc}. The model receives as input 8-lead ECG signals of 4096 samples and begins with an initial convolutional layer that applies 64 filters while preserving the number of samples. This is followed by batch normalization and ReLU activation, and then includes 5 residual blocks. All convolutional layers in the model, including the initial layer and those within the residual blocks, use a kernel size (filter length) of 17. Each residual block consists of 2 convolutional layers, followed by batch normalization, ReLU activations, and dropout (with a rate of 0.5). Subsampling by a factor of 4 is applied in each block, reducing the number of samples to [4096, 1024, 256, 64, 16] across the network. The number of filters increases across the blocks according to the configuration [64, 128, 196, 256, 320]. Max pooling and 1x1 convolutions are included in the skip connections to ensure that the number of samples and filters matches those from the signals in the main branch. The output from the final residual block is flattened and passed to a fully connected layer, which produces independent probabilities for each diagnostic class through a sigmoid activation function, in line with the multilabel nature of this classification task.

A preprocessing pipeline was applied to ensure data quality and input consistency. Since each ECG exam may contain multiple tracings recorded during the same clinical session, each lasting between 7 and 12 seconds and associated with the same set of diagnostic classes (CODE labels), we retained up to 4 tracings per exam to provide representative and balanced inputs (see Supplementary Fig. 5). All signals were resampled to 400 Hz and either zero-padded or trimmed to 4096 samples per lead. From the standard 12-lead ECG, we selected the 8 essential leads (I, II, and V1--V6), preserving core diagnostic information while reducing input dimensionality. The resulting 8-lead signals for each tracing were then used as inputs to the model.

The model was trained using a binary cross-entropy loss function with equal weight assigned to each class. Optimization was performed using the Adam algorithm with an initial learning rate of 0.001 and no weight decay. A learning rate scheduler reduced the rate by a factor of 0.1 when the validation loss plateaued for 5 consecutive epochs, down to a minimum learning rate of $1\times10^{-7}$. The training was conducted for up to 70 epochs or until the learning rate dropped below this minimum predefined value, and the final model was selected based on the lowest validation loss observed during this process.

Training and validation were conducted using the CODE-II dataset, comprising a total of 2,735,269 ECG exams. We ensured that the training and validation sets had no overlapping patients. For each diagnosis class, we defined a target of approximately 20\% of associated exams for the validation set. To meet this target while preserving patient exclusivity, we employed a complexity-aware splitting heuristic that alternates patients between training and validation sets in order to balance diagnostic complexity. This strategy minimizes the likelihood of conflicts arising from exams associated with multiple diagnoses. Without such constraints, validation performance could be inflated due to memorization of patient-specific features, resulting in less robust hyperparameter tuning and unreliable early stopping decisions. The dataset was split into 2,226,443 exams (81.4\%) for training and 508,826 exams (18.6\%) for validation. The per-diagnosis distributions achieved by this splitting strategy are presented in Supplementary Table 2.

Given that each ECG exam may include up to 4 tracings, it is important to clarify how they are handled throughout model development. During training and validation, each tracing, comprising 8 leads and associated with the same diagnostic classes, is treated as an independent input, effectively serving as a form of data augmentation. In the testing phase, the model independently evaluates each tracing and generates a probability for each diagnostic class. For each exam, the final probabilities for each class are computed by averaging the class-specific outputs across all available tracings. These aggregated probabilities (i.e., the model’s output), which range from 0 to 1, are used to determine which diagnostic classes are considered present in the exam. A class is classified as present when its aggregated probability exceeds a predefined threshold. In this work, the threshold for each class was selected from its precision-recall curve on the CODE-II validation set, which is part of the model development process. The final performance was then assessed on the independent CODE-II-test dataset. The threshold selection strategies, along with the evaluation metrics, are detailed in the following subsection.

%------------------------------------------------------------------------------------
% Evaluation metrics and threshold selection -  SUBSECTION
%------------------------------------------------------------------------------------
\subsection{Evaluation metrics and threshold selection}

To evaluate model performance on the CODE-II-test, we employed metrics commonly used in multilabel classification tasks. Threshold-independent metrics included the Area Under the Receiver Operating Characteristic Curve (AUROC) and the Area Under the Precision–Recall Curve (AUPRC). Both were computed per diagnostic class as well as in micro- and macro-averaged forms. AUROC quantifies the model’s ability to discriminate between positive and negative instances across all decision thresholds and is less affected by class imbalance. In contrast, AUPRC emphasizes the balance between precision and recall (sensitivity), making it particularly informative for rare diagnostic classes. To contextualize AUPRC values, we report the prevalence of each diagnostic class in the test set, which serves as a theoretical reference point for interpretation, as it represents the AUPRC expected from a random classifier. For example, an AUPRC of 0.60 is considered strong for a class with 5\% prevalence, since the ratio between the AUPRC and the class prevalence ($0.60/0.05$) indicates that the model performs 12 times better than random. In contrast, for a class with 50\% prevalence, the same score corresponds to a much smaller improvement ($0.60/0.50 = 1.2$), suggesting only a modest gain over chance.

Threshold-dependent metrics comprised precision, recall, specificity, F1-score, and negative predictive value (NPV). These were calculated per diagnostic class and summarized globally using two complementary schemes: micro-averaging, which aggregates true positives, false positives, and false negatives across all classes; and macro-averaging, which averages metrics computed independently for each class.

For threshold selection, we adopted a class-specific strategy that maximizes the F1-score, a widely used approach that identifies operating points balancing precision and recall within each diagnostic class. This serves as a standardized baseline, particularly suitable when both false positives and false negatives carry clinical relevance. Class-specific thresholds were chosen on the validation set and then fixed and applied to the CODE-II-test set without further tuning. While this preserves an honest evaluation, it also makes threshold-dependent metrics sensitive to shifts in class prevalence and score calibration between validation and test. To illustrate the effect of threshold choice without expanding the scope of the study, we report a limited comparison using Youden’s J statistic---which selects the threshold that best separates positives from negatives by balancing recall and specificity---applied to micro/macro aggregates and a small subset of classes. These results are exploratory and reported for illustration; the primary evaluation protocol is based on class-specific F1-max thresholds selected on validation and applied unchanged to the CODE-II-test.

% Although not the focus of the present study, clinically motivated thresholding strategies represent important directions for tailoring AI systems to specific diagnostic priorities. For example, a high-precision threshold is particularly relevant for identifying normal conditions, where false positives (i.e., incorrectly labeling an abnormal ECG as normal) could result in missed diagnoses and delayed clinical intervention. In this scenario, a possible strategy would enforce a minimum precision of 95\%, a method we have previously reported in preliminary results for detecting normal ECGs \cite{Abreu2025-ff}. Conversely, a high-recall threshold is more appropriate for urgent abnormalities, such as ST-elevation patterns, where capturing all potential cases---even at the expense of precision---is essential to avoid missing life-threatening events. In future work, we will further investigate both strategies to develop clinically meaningful operating modes.

%------------------------------------------------------------------------------------
% Evaluation protocol on external ECG benchmarks -  SUBSECTION
%------------------------------------------------------------------------------------
\subsection{Evaluation protocol on external ECG benchmarks}

To assess the generalizability of the representations learned by our model, we conducted a series of experiments using our model pre-trained on CODE-II. The goal was to evaluate whether this model, originally developed for multilabel classification with CODE diagnostic classes, could yield transferable and discriminative features when applied to external public ECG datasets with different populations and diagnostic labels.

We selected two publicly available ECG datasets for this evaluation. The first, the Physikalisch-Technische Bundesanstalt ECG dataset (PTB-XL), comprises 21,837 clinical 12-lead ECGs from 18,885 patients collected in Germany \cite{Wagner2020-ht}. Diagnostic annotations are provided at multiple levels of granularity, including 44 distinct diagnostic classes, which can be aggregated into broader superclass and subclass categories. Additionally, this dataset includes 19 form statements describing ECG waveform morphology and 12 rhythm statements related to cardiac rhythm. To further evaluate model transferability, we constructed more challenging training scenarios by using only 5\% and 10\% subsets of the PTB-XL training set. These subsets maintain a label distribution similar to that of the full dataset, allowing for a controlled assessment of performance under limited data availability. The second dataset, the China Physiological Signal Challenge 2018 (CPSC 2018), consists of 9,831 ECG recordings from 9,458 patients in China, annotated into 9 diagnostic classes, covering common arrhythmias and conduction disorders \cite{Liu2018-lx}. Both datasets were selected due to their expert-reviewed labels, diagnostic diversity, and frequent use in benchmarking ECG classification methods. Importantly, they differ from CODE-II not only in labeling schemes but also in patient populations and acquisition environments, making them well-suited for evaluating the robustness and transferability of the learned representations. Moreover, the ability of the model to generalize across these heterogeneous datasets indirectly reflects the diagnostic breadth and quality of the CODE-II dataset used for training.

We compared our trained model with two baseline settings: (i) supervised models trained from scratch, and (ii) large pre-trained ECG models fine-tuned on our target dataset. The supervised models are both effective and lightweight, with parameter counts ranging from 0.45 million to 2.35 million. These include LSTM \cite{Hochreiter1997-tu}, BiLSTM \cite{Graves2005-hr}, XResNet1d101 \cite{He2019-jc}, ResNet1d-Wang \cite{Wang2017-mo}, and InceptionTime \cite{Ismail_Fawaz2020-oo}. The pre-trained ECG models, including ResNet1d-Lima \cite{Lima2021-mk}, ResNet1d-Ribeiro \cite{Ribeiro2020_nc}, ResNet1d-Merl and ViT1d-Merl \cite{Liu2024-me}, ECG-FM \cite{McKeen2024-fm}, Heartlang \cite{Jin2025-ym}, and the HuBERT-Small and HuBERT-Base models \cite{Coppola2024-lz}, are publicly accessible and encompass both Transformer-based and ResNet-based architectures, with parameter counts ranging from 3.87 million to 92.83 million. These models have demonstrated strong generalization capabilities and were pre-trained on both public and access-on-demand ECG datasets. For example, HuBERT was pre-trained on 9.1 million ECG recordings from datasets such as CODE-I \cite{Ribeiro2020_nc}, MIMIC-IV ECG \cite{Gow2023-lc}, and the Chapman-Shaoxing ECG dataset \cite{Zheng2020-qq}, which is nearly 4 times the size of CODE-II. ResNet1d-Merl and ViT1d-Merl were both trained using the MERL framework, which applies a multimodal learning strategy to approximately 0.8 million ECG-report pairs from the MIMIC-IV ECG dataset. ResNet1d-Merl employs a 1D ResNet18 as the ECG encoder, while ViT1d-Merl substitutes this component with a Vision Transformer-based encoder.

We established a standardized evaluation pipeline to ensure fair and reproducible comparisons. Following the PTB-XL setup \cite{Wagner2020-ht}, we adopted the same 8:1:1 train/validation/test split and preprocessing pipeline. Models were trained for up to 100 epochs with early stopping based on validation macro-AUROC, using a patience of 5. The model with the highest validation AUROC was selected for final evaluation. We used binary cross-entropy loss and optimized with Rectified Adam. For supervised baselines, we adopted implementations and hyperparameters from \cite{Strodthoff2021-nd}, fixing the learning rate to 0.001 for consistency. For pre-trained models, we appled full-model fine-tuning and replaced the original projection head with a linear classification layer to enable downstream evaluation. To ensure compatibility with the respective pre-training settings, ECG signals were resampled to match the frequency and lead configurations used in each model: 100 Hz for Heartlang, 500 Hz for ResNet1d-Merl, ViT1d-Merl, HuBERT-Small, HuBERT-Base and ECG-FM, and 400 Hz for ResNet1d-Lima, ResNet1d-Ribeiro, and ours. Given the sensitivity of large models to learning rate, we performed a grid search over [0.001, 0.0001, 0.00001, 0.000001] on the validation set to identify the optimal value for each model. We adopted macro-AUROC as the primary evaluation metric on the test set, as it is threshold-independent and widely used, thereby avoiding bias introduced by fixed threshold selection. Detailed descriptions and configurations of the baselines are provided in Supplementary Tables 13 and 14.

%------------------------------------------------------------------------------------
% CODE-II Mini -  SUBSECTION
%------------------------------------------------------------------------------------
\subsection{CODE-II-open}

The CODE-II-open dataset is a curated subset of the full CODE-II cohort, developed as part of this study to serve as a public benchmark for training, validation, and reproducibility of deep learning models for ECG classification based on the 66 expert-defined CODE diagnostic classes. While the CODE-II dataset was used for model development and internal evaluation, the CODE-II-open was specifically created to support reproducible experimentation and external use.

CODE-II-open comprises 15,000 12-lead ECG exams, each corresponding to the first exam of a unique patient, and collectively accounting for approximately 0.72\% of all patients included in the full CODE-II dataset. These exams were selected from the same sets used to train and validate our baseline model for the 66 CODE diagnostic classes, ensuring consistency with the original study design. To support reproducible experimentation and reflect the proportions used in model development, the dataset was partitioned into 12,000 exams for training and 3,000 for validation. The selection process was guided by the diagnostic classes, with efforts made to approximate an 80/20 class-wise split between the training and validation sets, while also improving overall class distribution balance within the CODE-II-open dataset. The per-diagnosis distributions achieved using this strategy are provided in Supplementary Table 5.

Each exam in the CODE-II-open dataset is accompanied by two main sets of information. The first includes metadata related to the ECG acquisition, such as the exam ID, exam date, patient ID, date of birth, sex, reported comorbidities, and clinical indication for the ECG. Patient age was recalculated as the difference between exam and birth dates, expressed in years by dividing the result in days by 365.25 to account for leap years. The second set comprises diagnostic and reporting data, including the report upload date, type of report (original or revised), electrocardiographic measurements, and the set of diagnostic classes assigned to the exam. These diagnoses correspond to the 66 CODE diagnostic classes, including their respective class IDs and labels.

For each exam, up to four raw 12-lead ECG tracings are provided, corresponding to recordings captured during the same clinical session. These tracings last between 7 and 12 seconds and were sampled at 300 Hz, 500 Hz, 600 Hz, or 1000 Hz, depending on the acquisition device. Tracings were selected using a preprocessing pipeline that excluded corrupted or structurally inconsistent signals to ensure a minimum quality standard, with up to four recordings retained per exam. Although the signals are shared in their raw form, users can replicate the preprocessing steps used in this study by accessing the publicly available scripts at \url{https://github.com/antonior92/ecg-preprocessing}, which support a variety of preprocessing configurations. In our pipeline, we adopted a specific setup: resampling to 400 Hz, adjusting signal length to 4096 samples, applying high-pass filtering to remove baseline drift, and using a 60 Hz notch filter to suppress powerline interference. Additionally, only the eight essential leads (I, II, V1--V6) were retained for modeling purposes.
%A comprehensive characterization of the CODE-II Mini dataset, including patient demographics, comorbidities, and diagnostic labels, is presented in the Supplementary Material.

%------------------------------------------------------------------------------------
% Scaling laws experiments -  SUBSECTION
%------------------------------------------------------------------------------------
\subsection{Scaling law experiments}

We investigated the effect of dataset size used for model development on model performance through a series of scaling law experiments. In this analysis, we restricted the CODE-II dataset to the first ECG recorded for each patient, resulting in 2,093,807 unique patients and corresponding exams. From this patient-level dataset, we generated multiple training–validation splits of increasing size, including subsets ranging from 1,000 to 50,000 patients, the publicly released CODE-II-open (15,000 patients), and approximately 5\%, 10\%, 25\%, and 50\% of the full patient-level dataset. All splits maintained an approximately 80/20 train–validation ratio by applying the splitting heuristic proposed in this study, which aims to preserve this proportion within each diagnosis class and ensure patient exclusivity across training and validation sets. CODE-II-open was designed with 15,000 patients to align with the inflection point of the scaling curve, where performance transitions from steep to incremental gains. To better capture this transition and evaluate post-inflection behavior, experiments were extended up to approximately 50\% of the dataset. All evaluations were performed on the fixed CODE-II-test set to enable fair and consistent comparisons.

For all scaling law experiments, the primary experimental variable was the dataset size used for model development, while the model architecture and training hyperparameters were kept fixed. For each dataset size, three independent runs were performed using different random seeds (1, 2, and 3) to assess variability arising from model weight initialization and the random ordering of training examples into mini-batches at the start of each epoch. Model performance was evaluated on the fixed CODE-II-test set using macro-AUROC and macro-AUPRC. For each dataset size, we report the mean, minimum, and maximum values of these metrics across the three runs.

%====================================================================================
% Data availability - SECTION
%====================================================================================
\section{Data availability}

The CODE-II-open dataset, a publicly available subset of the full CODE-II database comprising 15,000 single-patient ECG exams annotated with the 66 expert-defined CODE diagnostic classes, is available at (\textcolor{orange}{LINK: that will be available upon publication}). The CODE-II-open and the CODE-II-test (multi-cardiologist annotations) dataset will be relaset for non-commercial use (and with a copyleft open-source license). Access to the full CODE-II dataset (used for model training and validation)  is restricted and requests should be forwarded to the Telehealth Network of Minas Gerais. Supplementary Information provides extended descriptions of the datasets, along with additional figures and tables detailing patient- and exam-level characteristics. The source data underlying the figures and tables in the main text are provided as a Source Data file, except for those derived from restricted datasets, which are available upon request under the same access conditions described above.

%====================================================================================
% Code availability - SECTION
%====================================================================================
\section{Code availability}
Code for ECG signal preprocessing is openly available at \url{https://github.com/antonior92/ecg-preprocessing}. The neural network architecture used in this study is based on the model originally proposed for the CODE-I dataset and available at \url{https://github.com/antonior92/automatic-ecg-diagnosis}. In addition, we evaluated several publicly available large pre-trained ECG models, including ResNet1d-Lima \url{https://github.com/antonior92/ecg-age-prediction}, ResNet1d-Ribeiro \url{https://github.com/antonior92/automatic-ecg-diagnosis}, ResNet1d-MERL and ViT1d-MERL \url{https://github.com/cheliu-computation/MERL-ICML2024}, ECG-FM \url{https://github.com/bowang-lab/ecg-fm}, HeartLang \url{https://github.com/PKUDigitalHealth/HeartLang}, and HuBERT-Small and HuBERT-Base \url{https://github.com/Edoar-do/HuBERT-ECG}. Corresponding references are provided throughout the manuscript. The full model weights trained on the complete CODE-II dataset are licensed to UFMG, which is responsible for their stewardship and controlled distribution.
%====================================================================================
% REFERENCES - SECTION
%====================================================================================
\begingroup
\small
\bibliographystyle{naturemag}
\bibliography{references}
\endgroup

%====================================================================================
% Acknowledgments - SECTION
%====================================================================================
\section{Acknowledgments}

This research was partially supported by the Brazilian agencies National Council for Scientific and Technological Development (CNPq), Coordination for the Improvement of Higher Education Personnel (CAPES), and Minas Gerais State Foundation for Research Support (FAPEMIG), as well as by the collaborative project Telecardiology and Artificial Intelligence, which is also partially funded through the MCTI/CNPq call no. 14/2023 for International Scientific, Technological, and Innovation Research Projects. PEOGBA is a CNPq scholarship holder (Brazil). AHR is partially supported by the eSSENCE strategic collaborative research programme. ALPR is supported in part by CNPq, FAPEMIG, CIIA-S (Innovation Center on Artificial Intelligence for Health), and IATS-CARE (Institute for Health Assessment and Translation for Chronic and Neglected Diseases of High Relevance). TBS was partially supported by \emph{Kjell och M{\"a}rta Beijer Foundation}. \textcolor{orange}{[We kindly invite the remaining authors to include their respective acknowledgement details].} The funders had no role in the design of the study, data collection, analysis, or interpretation, manuscript preparation, or the decision to submit the article for publication.

%====================================================================================
% Author contributions - SECTION
%====================================================================================
\section{Author contributions}

\begin{itemize}\small
    \item  \textbf{Petrus E. O. G. B. Abreu:} Conceptualization, Data Curation, Methodology, Formal Analysis, Validation, Visualization, Writing – Original Draft
    \item \textbf{Gabriela M. M. Paix\~ao:} Conceptualization, Data Curation, Methodology, Validation, Writing – Original Draft
    \item \textbf{Jiawei Li:} Validation, Methodology, Formal Analysis, Visualization, Writing – Review \& Editing
    \item \textbf{Paulo R. Gomes:} Conceptualization, Data Curation, Methodology, Writing – Review \& Editing
    \item \textbf{Peter W. Macfarlane:} Validation, Writing – Review \& Editing
    \item \textbf{Ana C.S. Oliveira:} Data Curation, Writing – Review \& Editing
    \item \textbf{Vin\'icius T. Carvalho:} Data Curation, Writing – Review \& Editing
    \item \textbf{Thomas B. Sch\"on:} Supervision, Methodology, Writing – Review \& Editing
    \item \textbf{Antonio Luiz P. Ribeiro:} Conceptualization, Data Curation, Validation, Methodology, Supervision, Funding Acquisition,  Writing – Original Draft
    \item \textbf{Ant\^onio H. Ribeiro:} Conceptualization, Methodology, Supervision, Formal Analysis, Funding Acquisition, Writing – Original Draft
\end{itemize}

%====================================================================================
% Competing interests - SECTION
%====================================================================================
\section{Competing interests}
AHR holds indirect equity options and serves as the technical advisor for Einthoven Tecnologia LTDA.

%====================================================================================
% Ethics statement - SECTION
%====================================================================================
\section{Ethics statement}

This study was conducted in accordance with all relevant ethical regulations and was approved by the Research Ethics Committee of the Universidade Federal de Minas Gerais (CAAE: 85892325.1.0000.5149). Analyses were performed using two datasets from the Telehealth Network of Minas Gerais (TNMG): the CODE-II dataset, used for model development, and the CODE-II Test dataset, used for independent evaluation. Both datasets consist of anonymized 12-lead ECG recordings and associated metadata collected during routine clinical care. As this was a secondary analysis of anonymized data, the Research Ethics Committee waived the requirement for individual informed consent. No identifiable images of human participants are included in this study. All researchers involved in this work signed confidentiality and data use agreements prior to accessing the data.

%====================================================================================
% Additional Information - SECTION
%====================================================================================
\section{Additional Information} 
\textbf{Supplementary information} is available for this paper at (\textcolor{orange}{LINK: that will be available upon publication}). \\

%====================================================================================
%====================================================================================
%====================================================================================
%====================================================================================
% Reset page and counters
%\setcounter{equation}{0}
%\renewcommand{\theequation}{S.\arabic{equation}}%
%\setcounter{figure}{0}
%\renewcommand{\thefigure}{S.\arabic{figure}}%
%\setcounter{table}{0}
%\renewcommand{\thetable}{S.\arabic{table}}%
%\setcounter{section}{0}
%\newpage
%\appendix
%\pagenumbering{roman} 
%\part{}
%~ 
%\vspace{10pt}
%\parttoc % Insert the document TOC
%\setcounter{page}{0 }
%\newpage
%\section{Problem formulation}
%\newpage
%\section{Blabla}

%====================================================================================
%                            SUPPLEMENTARY MATERIAL
%====================================================================================
% ======= Início do Suplemento dentro do mesmo .tex =======
\beginsupplement

\appendix
%====================================================================================
% CODE-II Dataset: Extended Description - SECTION
%====================================================================================
\section{CODE-II Dataset: Extended Description}

This section provides supplementary information supporting the description presented in the main manuscript about the curated CODE-II dataset, composed of 2,735,269 ECG exams from 2,093,807 unique patients collected and annotated by TNMG between January 2019 and December 2022. The figures and tables below expand the dataset characterization by offering a more granular view of patient-level attributes and exam characteristics, as well as presenting additional information available in the dataset but not detailed in the main text.

%------------------------------------------------------------------------------------
% General Characteristics - SUBSECTION
%------------------------------------------------------------------------------------
\subsection{General Characteristics}

Supplementary Fig.\,\ref{Figure:SupplementaryFigure1} presents the distribution of the 2,735,269 ECG exams across the four-year data collection period, providing temporal context. The annual volume remained relatively stable, with a slight decline in 2020, likely reflecting the impact of the COVID-19 pandemic on routine healthcare services.

We also stratified the 2,093,807 patients by age group and sex, considering only their first ECG exam. Among patients aged 18–59 years (62.2\% of all patients), females accounted for 796,170 (61.1\%) and males for 506,382 (38,9\%). In the older group ($\geq 60$ years; 37.8\% of patients), there were 441,462 females (55.8\%) and 349,793 males (44.2\%). Overall, based on first exams, 59.1\% of patients were female (1,237,632) and 40.9\% were male (856,175). The full age distribuition by sex is show in Supplementary Fig.\,\ref{Figure:SupplementaryFigure2}.

Self-reported clinical comorbidities of the CODE-II dataset, stratified by age (18–59 years and $\geq 60$ years) and sex, are detailed in Supplementary Table\,\ref{Table:SupplementaryTable1}. As reported in the main text, hypertension was the most prevalent comorbidity, affecting 47.7\% of patients, followed by diabetes mellitus (14.0\%), smoking (8.2\%), dyslipidemia (4.8\%), previous myocardial infarction (1.8\%), Chagas disease (1.1\%), and chronic obstructive pulmonary disease (0.7\%). Patients could report more than one comorbidity. Comorbidities and medication usage were recorded using checkboxes marked only when present, meaning that unmarked fields may indicate either missing information or the true absence of the condition or medication use. Medication data were also used to infer the presence of certain conditions: patients reporting the use of calcium channel blockers, diuretics, and/or angiotensin-converting enzyme inhibitors were regarded as having hypertension; those using statins as having dyslipidemia; and those taking insulin or oral hypoglycemic agents as having diabetes.

% \begin{table}[htp] %h!
% 	\centering % Centraliza a tabela
% 	\caption{Self-reported clinical comorbidities of patients in the CODE-II dataset, stratified by age group (18–59 years and $\geq60$ years) and sex, based on the first ECG exam. For hypertesion, dyslipidemia, and diabetes, comorbidity status also inferred from reported medication usage.}
% 	\label{Table:exams_with_mult_reports_LAUDOS_files_7_6} % Label para referência cruzada
% 	\resizebox{\textwidth}{!}{%
% 		\csvreader[separator=tab,
% 		tabular=|c|c|c|c|c|c|c|c|c|c|c|,
% 		table head=\hline id\_laudo & id\_exame & laudo\_inclusao & especialista & retificado & retificacao & tipo & retificacao\_id & diag\_ids & diag & classe\_pred \\ \hline,
% 		late after line=\\ \hline
% 		]{tables/SupplementaryTable1.tsv}{}% Caminho para o CSV
% 		{\csvcoli & \csvcolii & \csvcoliii & \csvcoliv & \csvcolv & \csvcolvi & \csvcolvii & \csvcolviii & \csvcolix & \csvcolx & \csvcolxi} % Valores das colunas
% 	}
% \end{table}

Clinical indications for ECG exams were recorded using checkbox fields, allowing multiple symptoms to be reported per patient. The most frequently indicated reasons were routine examinations and chest pain, followed by preoperative risk assessment, palpitations, dyspnea, and “other”. The “other” category corresponded to a free-text field in which healthcare professionals could specify additional symptoms not covered by predefined options; this field was completed in 6.4\% of exams. The full distribution of clinical indications, stratified by age and sex, is shown in Supplementary Fig.\,\ref{Figure:SupplementaryFigure3}.

In addition, we investigated how many ECG exams each patient underwent over the study period. As illustrated in Supplementary Fig.\,\ref{Figure:SupplementaryFigure4}a, a majority (1,663,122 patients, 79.4\%) had a single exam, although a significant proportion underwent multiple recordings, with some patients having more than ten exams. Supplementary Fig.\,\ref{Figure:SupplementaryFigure4}b shows the number of signal records per exam. These represent multiple consecutive tracings (lasting between 7-12 seconds each) recorded during a single clinical session. All tracings were available to TNMG physicians for diagnostic interpretation, and each exam was annotated with one or more labels from the 66 CODE diagnostic classes.

To illustrate the final number of tracings retained in the CODE-II dataset after applying the pipeline that selects up to four tracings per exam to ensure representative and balanced inputs, Supplementary Fig.\,\ref{Figure:SupplementaryFigure5} presents a summary of this distribution.

The 66 CODE diagnostic classes used by physicians to annotate ECGs in the CODE-II dataset were introduced in the main manuscript, together with an overview of their clinical relevance and frequency distribution. Here, we provide a detailed summary of these diagnostic classes, including their descriptions, an associated label (defined solely for ease of reference in figures), their overall prevalence in the dataset, and the class-wise distribution across the training and validation sets (approximately 80/20 split), based on the patient-exclusive split strategy adopted during model development. These details are presented in Supplementary Table\,\ref{Table:SupplementaryTable2}.

We also present a group-level analysis of diagnostic label assignments. The 66 classes are grouped into 10 clinically meaningful diagnostic groups. Supplementary Fig.\,\ref{Figure:SupplementaryFigure6} provides an analysis of: (a) the number of exams exclusively assigned to a single diagnostic group, (b) the 10 most frequent non-exclusive group combinations observed across all exams, and (c) the distribution of the number of CODE groups assigned per exam. This figure complements the main text by quantifying the combinatorial diversity of diagnostic group assignments.

To further characterize the diagnostic landscape of the dataset, we analyzed the distribution of diagnostic groups at the patient level. Supplementary Table\,\ref{Table:SupplementaryTable3} presents this analysis using only the first ECG exam per patient, stratified by age and sex. Group assignments are not mutually exclusive—except for the Normal group, which is exclusive by definition---allowing patients to appear in multiple categories. This table complements the population-level description in the main manuscript and provides additional insight into the prevalence of diagnostic groups across demographic subgroups.

We also analyzed electrocardiographic interval measurements extracted from the 12-lead ECG signals in the CODE-II dataset. These measurements revealed clear sex- and age-related differences, consistent with known physiological patterns and prior population-based studies \citeSI{Palhares2017-eb,Pinto2017-ba}. Female patients consistently exhibited higher resting heart rates and longer corrected QT intervals (QTc), while male patients showed longer depolarization and conduction intervals, including P-wave duration, PR interval, and QRS duration. For instance, median heart rates among women were 73 bpm (18-59 years) and 72 bpm (60+ years), compared to 68 bpm in both age groups for men. Median QTc intervals ranged from 420 to 427 ms in women and from 410 to 420 ms in men. In contrast, median PR intervals and QRS durations were longer in men across both age groups (e.g., PR: 152–158 ms vs 146–152 ms; QRS: 96–98 ms vs 90–92 ms in men vs women). These findings are consistent with prior Brazilian population studies, such as the ELSA-Brasil study and a large primary care cohort \citeSI{Palhares2017-eb,Pinto2017-ba}, and large international studies have likewise emphasized the impact of age and sex on ECG norms \citeSI{Mason2007-uy,Rijnbeek2014-gx}. Additionally, we observed an age-related leftward shift of the frontal QRS axis, particularly among women, in line with known aging effects on cardiac electrophysiology. Although these electrocardiographic measures are not the central focus of this study, their internal consistency and physiological plausibility further support the representativeness of the CODE-II dataset. Readers interested in more in-depth analyses of ECG interval characteristics are referred to comprehensive population-based studies \citeSI{Mason2007-uy,Rijnbeek2014-gx}. Supplementary Fig.\,\ref{Figure:SupplementaryFigure7} displays the full distribution of ECG intervals stratified by age group and sex. Notably, the proportion of missing values for these measurements is very low in the CODE-II dataset, with fewer than 0.16\% missing in any subgroup when analyzed by sex.

%====================================================================================
% CODE-II Dataset: Extended Description - SECTION
%====================================================================================
\section{CODE-II-open Dataset Characterization}

This supplementary section complements the overview provided in the main manuscript, offering a detailed characterization of the CODE-II-open dataset. We describe the demographic and clinical attributes of the 15,000 patients included, along with their associated diagnostic annotations. Additional information is provided regarding comorbidity prevalence, clinical indications for the ECG exams, and electrocardiographic measurements, all stratified by age and sex whenever relevant. Figures and tables included herein support reproducibility and transparency for studies leveraging the CODE-II-open dataset as a public benchmark.

%------------------------------------------------------------------------------------
% General Characteristics - SUBSECTION
%------------------------------------------------------------------------------------
\subsection{General Characteristics}

To provide a comprehensive overview of the CODE-II-open dataset, we analyzed key characteristics of its 15,000 unique patients and their corresponding ECG exams. These exams were selected from the full CODE-II dataset and span the period between January 2019 and December 2022, thus preserving the original dataset’s temporal coverage, as illustrated in Supplementary Fig.\,\ref{Figure:SupplementaryFigure8}. While the full dataset includes exams from 12 Brazilian states, the CODE-II-open subset covers exams from 11 states; see Supplementary Fig.\,\ref{Figure:SupplementaryFigure9}. This geographic distribution reflects the dataset’s targeted selection strategy, which prioritized diagnostic class diversity and patient uniqueness, without explicitly enforcing territorial representation.

Among the 15,000 patients included in the CODE-II-open dataset, 52.9\% (7,937) were female and 47.1\% (7,063) were male. The overall mean age was 55.2 years, with similar distributions for female and males (55.1 and 55.4 years, respectively), as illustrated in Supplementary Fig.\,\ref{Figure:SupplementaryFigure10}. Patients were categorized into two age groups: 18–59 years (57.8\% of the cohort), with 4,727 (54.5\%) females and 3,943 (45.5\%) males; and 60 years or older (42.2\% of the cohort), with 3,210 (50.71\%) females and 3,120 (49.29\%) males. No patients under 18 years of age were included.

The CODE-II-open dataset contains detailed information on patient comorbidities, recorded through checkbox fields marked only when a condition was present. These conditions were either self-reported or inferred from medication use, while unmarked fields indicated either the absence of the condition or missing information. Hypertension was the most frequently reported comorbidity, affecting approximately 47.5\% of patients, followed by diabetes mellitus (13.3\%), smoking (8.9\%), and dyslipidemia (4.8\%). Less common conditions included previous myocardial infarction (2.3\%), Chagas disease (2.1\%), and chronic obstructive pulmonary disease (COPD, 1.0\%). Comorbidity prevalence increased with age, as evidenced by the higher number of reported cases among patients aged 60 or older. For instance, hypertension cases rose by 39\% among females (from 1,676 to 2,326) and by 72\% among males (from 1,151 to 1,974). Similarly, diabetes cases increased by 80\% in females (from 429 to 771) and by 106\% in males (from 260 to 535) when comparing patients aged 18–59 to those aged 60 or older. Regarding exam priority, 69.7\% of the ECGs were recorded as elective, 29.3\% as urgent, and 1.0\% as preferential. Supplementary Table\,\ref{Table:SupplementaryTable4} details comorbidity distributions across age groups and sex.

As previously mentioned, the CODE-II-open dataset includes only the first ECG for each of the 15,000 patients. Each exam may contain multiple consecutive tracings, recorded during a single clinical session and lasting between 7-12 seconds (Supplementary Fig.\,\ref{Figure:SupplementaryFigure12}). In accordance with the preprocessing procedure outlined in the main manuscript, we retained up to four tracings per exam (Supplementary Fig.\,\ref{Figure:SupplementaryFigure13}), selected to ensure representative and balanced inputs, given that all tracings share the same diagnostic labels. This standardized subset of tracings constitutes the final version of the dataset made available for model development and evaluation.

Each ECG exam in the CODE-II-open dataset may be associated with one or more diagnostic labels from the 66 expert-defined CODE classes (Supplementary Fig.\,\ref{Figure:SupplementaryFigure14}), which were introduced by TNMG cardiologists to replace free-text ECG reports and promote diagnostic standardization. A comprehensive summary of these diagnostic classes—including their descriptions, overall prevalence in the dataset, and class-wise distribution across training and validation sets—is presented in Supplementary Table\,\ref{Table:SupplementaryTable5}. Importantly, the training and validation subsets in CODE-II-open were derived exclusively from the corresponding partitions of the full CODE-II dataset (approximately 80/20 split), which were used to develop the baseline model described in the main manuscript. This approach ensures consistency with the original study design and full compatibility between datasets.

The 66 diagnostic classes can be grouped into 10 clinically meaningful CODE categories, for which we also analyzed exam distributions within the CODE-II-open dataset. The most frequent diagnostic groups were Normal (44.1\%), Miscellaneous (21.1\%), and Arrhythmia (18.7\%), followed by Intraventricular Conduction Disorders, Chamber Hypertrophy, and Atrioventricular Conduction Disorders (Supplementary Fig.\,\ref{Figure:SupplementaryFigure15}a). Most exams (85.8\%) were associated with one or two diagnostic classes, while only a small fraction (1.7\%) included five or more (Supplementary Fig.\,\ref{Figure:SupplementaryFigure15}b). To provide a more detailed group-level view of diagnostic label assignment, we further examined the number of exams exclusively assigned to each group. Notably, the Normal group is the only mutually exclusive CODE category, composed of a single diagnostic class. Supplementary Fig.\,\ref{Figure:SupplementaryFigure16} presents the number of exams exclusively assigned to each group, the ten most frequent combinations of non-exclusive group assignments (out of 147 observed patterns), and the distribution of the number of groups assigned per exam.

The CODE-II-open dataset also includes electrocardiographic interval measurements derived from the 12-lead ECG signals. These include heart rate, P-wave duration, PR interval, QRS duration, QRS axis, and corrected QT interval (QTc), with distributions stratified by age group and sex. Although not the primary focus of this dataset, these cardiac measures provide additional evidence of internal consistency. As observed in the full CODE-II dataset, the distributions in CODE-II-open reveal physiologically plausible differences by age and sex. For instance, resting heart rate and QTc intervals were generally higher in females, while males showed longer conduction intervals such as PR and QRS durations. The median QTc ranged from 421–430 ms in women and 412–424 ms in men; QRS durations were longer in males across both age groups. These findings mirror those from the full dataset and are consistent with known population-level trends. The proportion of missing values for these measurements is very low, with fewer than 0.2\% missing in any age-sex subgroup. Supplementary Fig.\,\ref{Figure:SupplementaryFigure17} summarizes the full distribution of cardiac measurements across age and sex groups.

%====================================================================================
% CODE-II Dataset: Extended Description - SECTION
%====================================================================================
\section{CODE-II-test Dataset Overview}

This supplementary section provides an expanded description of the CODE-II-test dataset, developed as a high-quality benchmark for evaluating AI-based ECG classification models. It comprises 8,475 exams from unique patients, collected by TNMG between 2018 and 2025, all distinct from those included in the CODE-II dataset. To ensure data quality and consistency, the same rigorous preprocessing pipeline used for CODE-II was applied. Each exam was reviewed by multiple certified cardiologists and annotated using the standardized set of 66 CODE diagnostic classes. Final diagnoses were derived according to predefined agreement and majority rules, as detailed in the Methods section of the main manuscript. In addition to dataset characterization, we report cardiologist performance and baseline model results on this test set. The supplementary figures and tables provide a comprehensive overview, reinforcing the role of CODE-II-test as a robust and transparent benchmark for model evaluation under expert annotation.

%------------------------------------------------------------------------------------
% General Characteristics - SUBSECTION
%------------------------------------------------------------------------------------
\subsection{General Characteristics}

To provide a comprehensive characterization of the CODE-II-test dataset, we analyzed the demographic and clinical attributes of its 8,475 unique patients. The exams were collected between 2018 and 2025, with their annual distribution shown in Supplementary Fig.\,\ref{Figure:SupplementaryFigure18}, and originated from 15 Brazilian states, as illustrated in Supplementary Fig.\,\ref{Figure:SupplementaryFigure19}. Both the collection period and the geographic coverage are broader than those of the CODE-II dataset, reflecting the design of the CODE-II-test to capture a diverse set of exams with multiple high-quality annotations.

The dataset comprised 57.5\% females (4,871) and 42.5\% males (3,604), with an overall mean age of 55.5 years (55.3 in females and 55.7 in males). No patients under 18 years of age were included. Most patients (58.6\%) were aged 18–59 years (2,898 females and 2,068 males), while 41.4\% were 60 years or older (1,973 females and 1,536 males). These demographic distributions are broadly consistent with those of the CODE-II training and validation cohorts but reflect the distinct curation process of the CODE-II-test, designed as an independent benchmark under expert annotation. The age distribution stratified by sex is presented in Supplementary Fig.\,\ref{Figure:SupplementaryFigure20}.

We next examined clinical comorbidities, which were recorded through checkbox fields, completed only when a condition was present, either self-reported or inferred from medication use. Hypertension (43.9\%), Chagas disease (14.4\%), and diabetes mellitus (11.6\%) were the most frequent conditions, while smoking (7.3\%), dyslipidemia (6.0\%), previous myocardial infarction (2.2\%), and COPD (0.9\%) were less common. Prevalence patterns varied by age: several conditions, such as hypertension and diabetes, were more frequent among older patients, whereas smoking showed higher prevalence in younger groups. For example, hypertension was reported in 1,244 females aged 60+ compared with 1,006 in younger females, while smoking affected 212 younger males compared with 144 older males. Regarding exam priority, most ECGs were recorded as elective (73.5\%), followed by urgent (26.1\%) and preferential (0.4\%). Supplementary Table\,\ref{Table:SupplementaryTable6} provides full comorbidity distributions stratified by age and sex.

The clinical indications for performing ECG exams were also recorded through checkbox fields, with multiple indications allowed per patient. Routine examinations were the most frequent (40.9\%), followed by chest pain (16.8\%), “other” reasons (9.1\%)—a free-text field that allowed healthcare professionals to specify indications not covered by predefined options—palpitations (8.0\%), dyspnea (6.8\%), and preoperative surgical risk assessment (6.7\%). These patterns were generally consistent across age and sex groups, with some variation, such as dyspnea being more common in older patients. The full distribution of indications stratified by sex and age is shown in Supplementary Fig.\,\ref{Figure:SupplementaryFigure21}.

Each patient contributed only one ECG exam, consistent with the role of the CODE-II-test as a fixed, patient-exclusive evaluation set. Each exam may comprise multiple consecutive tracings, recorded during a single clinical session and lasting between 7 and 12 seconds (Supplementary Fig.\,\ref{Figure:SupplementaryFigure22}). In accordance with the preprocessing strategy described in the main manuscript, up to four tracings were retained per exam in the final dataset to ensure representativeness while preserving standardized inputs (Supplementary Fig.\,\ref{Figure:SupplementaryFigure23}). This design guarantees strict patient-level independence from the training and validation datasets, reinforcing its role as a reproducible benchmark for model evaluation.

Diagnostic annotations were based on the 66 standardized CODE classes (Supplementary Fig.\,\ref{Figure:SupplementaryFigure24}). Of these, 65 are not mutually exclusive, while the Normal class is uniquely assigned. This standardized structure ensures consistency and clinical interpretability across annotations. Supplementary Table\,\ref{Table:SupplementaryTable7} summarizes all classes with their labels, descriptions, prevalence, and dataset distribution.

As outlined in the main manuscript, the 66 diagnostic classes can be aggregated into 10 clinically meaningful CODE groups. Within the CODE-II-test dataset, the most frequent groups were Normal, Miscellaneous, and Arrhythmia (Supplementary Fig.\,\ref{Figure:SupplementaryFigure25}a). Most exams were associated with one or two diagnostic classes, whereas only a small proportion involved more complex profiles with multiple labels (Supplementary Fig.\,\ref{Figure:SupplementaryFigure25}b). At the group level, the Normal category remained the only mutually exclusive group, consisting of a single diagnostic class. Supplementary Fig.\,\ref{Figure:SupplementaryFigure26} presents the number of exams exclusively assigned to each group, the ten most common non-exclusive group combinations, and the overall distribution of group counts per exam.

%------------------------------------------------------------------------------------
% General Characteristics - SUBSECTION
%------------------------------------------------------------------------------------
\subsection{Cardiologist Performance Evaluation}

A total of 46 certified cardiologists participated in the annotation of the CODE-II-test dataset, each contributing to a variable number of exams depending on their availability and audit allocation. Participation was markedly heterogeneous: while some specialists reviewed thousands of exams, others contributed to only a few (Supplementary Fig.\,\ref{Figure:SupplementaryFigure27}). Every exam was independently reviewed by at least two cardiologists, with most being assessed by two or three reviewers, and a smaller fraction by four or more (Supplementary Fig.\,\ref{Figure:SupplementaryFigure28}). The diagnostic process relied on the standardized set of 66 CODE classes, which include normal findings, technical issues preventing analysis, and a broad range of abnormalities. Importantly, the Normal class was assigned exclusively, while all other classes could co-occur within the same exam.

Final diagnoses were determined using two predefined decision rules. Complete concordance among all reviewers across all assigned labels defined the agreement rule, which accounted for 6,309 exams (74.4\%). For the remaining 2,166 cases (25.6\%), the majority rule was applied, requiring at least two reviewers to agree on at least one diagnosis. As a result, every exam in the dataset was assigned a definitive diagnostic label set, ensuring comprehensive coverage under expert supervision. Most exams in the CODE-II-test dataset were associated with one or two diagnostic labels, while a smaller subset presented more complex profiles with multiple abnormalities (Supplementary Fig.\,\ref{Figure:SupplementaryFigure25}b). This distribution, described in detail in the General Characteristics section, reflects the variability of clinical findings captured in the dataset.

Cardiologist performance was computed by comparing individual review records with the final ground-truth label set for each exam, as defined above for the CODE-II-test. Under a fair-scope analysis, evaluation was restricted to diagnostic classes that a cardiologist actually encountered; classes outside a cardiologist’s experience—i.e., never assigned by that cardiologist and never present as true labels in the CODE-II-test exams they reviewed—were excluded from both per-class and aggregate computations. We derived per-class confusion counts (TP, FP, FN, TN) and then summarized cardiologist-level performance using micro- and macro-averaged precision, recall (sensitivity), specificity, F1 score, and NPV over the set of classes each cardiologist encountered (Supplementary Table\,\ref{Table:SupplementaryTable8}).

Supplementary Table\,\ref{Table:SupplementaryTable8} summarizes cardiologist-level performance under the fair-scope definition. Because very small workloads can inflate---sometimes perfect---scores, cardiologists with few exams should be interpreted with caution; we therefore highlight those with $\geq100$ reviewed exams (n = 20 cardiologists). Among these cardiologists, the mean micro-F1 score was 0.966 (mean micro-recall 0.978; mean micro-precision 0.956), and the mean macro-F1 score was 0.944, with macro-precision and recall likewise high. Specificity and NPV were near-perfect across cardiologists, which is expected given the exam-level sparsity of labels: most exams carry only one or two diagnostic labels, and multiple labels per exam are relatively uncommon, so for any given class the majority of exam-class pairs are true negatives. The adjudication procedure (agreement by $\geq2$ cardiologists on $\geq1$ diagnosis per exam) further guarantees at least one positive label while leaving most other classes negative, reinforcing high specificity and NPV. This adjudication aligns the final ground-truth labels with cardiologist agreement and therefore helps explain the uniformly strong performance observed among high-volume annotators.

%------------------------------------------------------------------------------------
% General Characteristics - SUBSECTION
%------------------------------------------------------------------------------------
\subsection{Model Performance Evaluation}

This subsection compiles the full evaluation tables for the baseline model on the CODE-II-test set and provides brief context to interpret them. Detailed interpretation appears in the main manuscript. We report both threshold-independent metrics—Area Under the Receiver Operating Characteristic Curve (AUROC) and Area Under the Precision–Recall Curve (AUPRC, also referred to as Average Precision, AP)—and threshold-dependent metrics—precision, recall (sensitivity), specificity, F1-score, and NPV. Unless otherwise noted, threshold-dependent metrics are computed with class-specific thresholds selected on the CODE-II validation split via F1-max and then applied unchanged to the Test set.

At the aggregate level, the model shows strong discriminative ability (micro-AUROC 0.983, micro-AUPRC 0.776; macro-AUROC 0.978, macro-AUPRC 0.552), with consistently high performance in common classes and greater variability across rare classes. With F1-max thresholds, the model attains micro-F1 0.706 (precision 0.632, recall 0.801) and macro-F1 0.512 (precision 0.517, recall 0.562), reflecting the expected gap when all classes—common and rare—are weighted equally.

Per-class results illustrate both common and rare diagnoses. For example, the Normal class (40.1\% of Test) achieves AUPRC 0.931 (well above prevalence 0.401), while clinically important abnormalities such as LBBB (2.6\%) and AF (2.6\%) reach AUPRC 0.944 and 0.950, respectively. Even in rare but time-critical conditions such as STEMI (0.4\%), the model attains AUPRC 0.647, substantially above prevalence (see Supplementary Table\,\ref{Table:SupplementaryTable9}).

Because threshold choice can materially affect threshold-dependent metrics when validation and test distributions differ in class prevalence and score calibration, we also provide a direct aggregate comparison between the primary F1-max thresholds and an alternative rule based on Youden’s J statistic---both selected on validation and applied unchanged to Test. As expected, AUROC/AUPRC remain unchanged, whereas precision/recall/F1 shift notably; these micro/macro comparisons are reported in Supplementary Table\,\ref{Table:SupplementaryTable10}.

At the class level, three rare diagnoses---multifocal atrial tachycardia (MAT), isorhythmic atrioventricular dissociation (IAVD), and digitalis effect (DIG)---yield degenerate metrics under the F1-max rule on the Test set (precision, recall, and F1 equal to zero). For these classes, Youden’s J selects substantially smaller thresholds, permitting positive predictions and non-zero precision/recall/F1. Side-by-side results are reported in Supplementary Table\,\ref{Table:SupplementaryTable11}.

As brief context relative to specialists, the previous subsection (“Cardiologist Performance Evaluation”) reports uniformly high reviewer performance under a fair-scope protocol (see Supplementary Table\,\ref{Table:SupplementaryTable8}), which is expected because the CODE-II-test ground-truth was adjudicated by agreement/majority rules---thereby aligning reference labels with cardiologist concordance and yielding near-perfect specificity/NPV at the exam-class level. By contrast, the model’s threshold-dependent metrics reflect class-specific thresholds selected on the validation split and applied unchanged to Test, which differs in class prevalence and score calibration; this depresses precision/recall/F1 in some settings, as shown by the Youden’s-J sensitivity analysis (Supplementary Table\,\ref{Table:SupplementaryTable10}) and by the three rare classes with degenerate values under F1-max (Supplementary Table\,\ref{Table:SupplementaryTable11}). A formal head-to-head comparison is therefore beyond scope here; in future work we will examine threshold-selection and calibration strategies under distribution shift, alongside clinically motivated operating modes tailored to specific use cases.

%====================================================================================
% CODE-II Dataset: Extended Description - SECTION
%====================================================================================
\section{External Evaluation and Model Benchmarking}

This supplementary section provides additional details on the external evaluation experiments described in the main manuscript. We assessed the generalization capabilities of the model pre-trained on the CODE-II dataset by fine-tuning it on two widely used public ECG classification benchmarks: PTB-XL and CPSC 2018. These datasets differ from CODE-II in terms of patient population, diagnostic labeling schemes, and acquisition settings, thus offering a robust framework for evaluating the transferability of the learned representations. We conducted experiments under both full-data and few-shot training scenarios (using 5\% and 10\% of the training data) and compared our model against supervised models trained from scratch as well as state-of-the-art pre-trained ECG models. The results presented in the following figures and tables demonstrate how the combination of a high-quality, expertly curated dataset and a robust model architecture can yield highly transferable and diagnostically meaningful representations across diverse clinical scenarios.

%------------------------------------------------------------------------------------
% General Characteristics - SUBSECTION
%------------------------------------------------------------------------------------
\subsection{General Characteristics}

Detailed performance results for all evaluated models on PTB-XL and CPSC 2018 under full-data training are summarized in Supplementary Table\,\ref{Table:SupplementaryTable12}, illustrating differences across model families, including end-to-end supervised baselines and pre-trained approaches. The complete set of results---including additional few-shot training experiments (5\% and 10\%) on PTB-XL, improvements over random initialization, and per-run values---is provided in Supplementary Data~6 (external Excel file, \textit{Supplementary\_Data\_6.xlsx}).

Supplementary Table\,\ref{Table:SupplementaryTable13} summarizes the number of parameters, as well as the size and source of the pre-training datasets used for each model included in the benchmarking. These details help contextualize the representational capacity of the models and the scale of data involved in their original training, which may influence downstream performance in external evaluation tasks.

Details about the fine-tuning configurations used for each pre-trained model—including learning rates, input signal properties, and fine-tuning strategies—are provided in Supplementary Table\,\ref{Table:SupplementaryTable14}. These parameters were selected to ensure comparability while accommodating model-specific requirements and published recommendations.

\newpage
%====================================================================================
% CODE-II Dataset: Extended Description - SECTION
%====================================================================================
\section{Benchmarking on CODE-II-open}

This supplementary section provides additional details on the benchmarking experiments performed using the CODE-II-open dataset, as described in the main manuscript. The purpose of this analysis was to illustrate the representativeness of CODE-II-open as a benchmark dataset and to compare the performance of our baseline architecture with reference ECG models under consistent evaluation settings. The same evaluation pipeline adopted for the external benchmarking experiments was used here to ensure comparability across studies.

%------------------------------------------------------------------------------------
% General Characteristics - SUBSECTION
%------------------------------------------------------------------------------------
\subsection{General Characteristics}

Supplementary Table\,\ref{Table:SupplementaryTable15} reports the results of our baseline model and of models from the literature under their best-performing regime on CODE-II-open, either trained from scratch or fine-tuned from pre-trained versions. For reference, the table also includes the performance of our baseline model when trained on the full CODE-II dataset. These results demonstrate that the CODE-II-open dataset is sufficiently rich and diverse to train competitive ECG classifiers and support its role as a public benchmark for future model development.

%------------------------------------------------------------------------------------
% Scaling laws experiments -  SUBSECTION
%------------------------------------------------------------------------------------
\section{List of Supplementary Figures}
\begin{figure}[H]
	\centering{
		\includegraphics[scale=1]{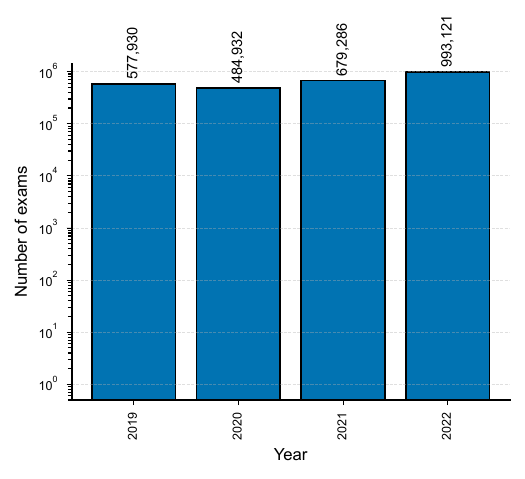} \vspace{-4mm}
		\caption{Number of ECG exams per year in the CODE-II dataset.}
		\label{Figure:SupplementaryFigure1}
	}
\end{figure}

\begin{figure}[H]
	\centering{
		\includegraphics[scale=1]{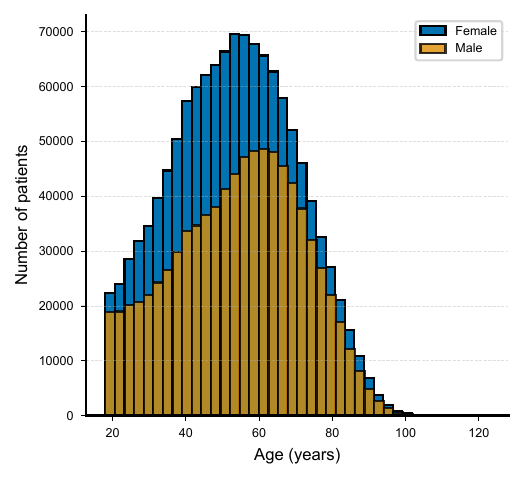} \vspace{-4mm}
		\caption{Age distribution of patients at their first ECG exam in the CODE-II dataset, stratified by sex.}
		\label{Figure:SupplementaryFigure2}
	}
\end{figure}

\begin{figure}[H]
	\centering{
		\includegraphics[scale=0.91]{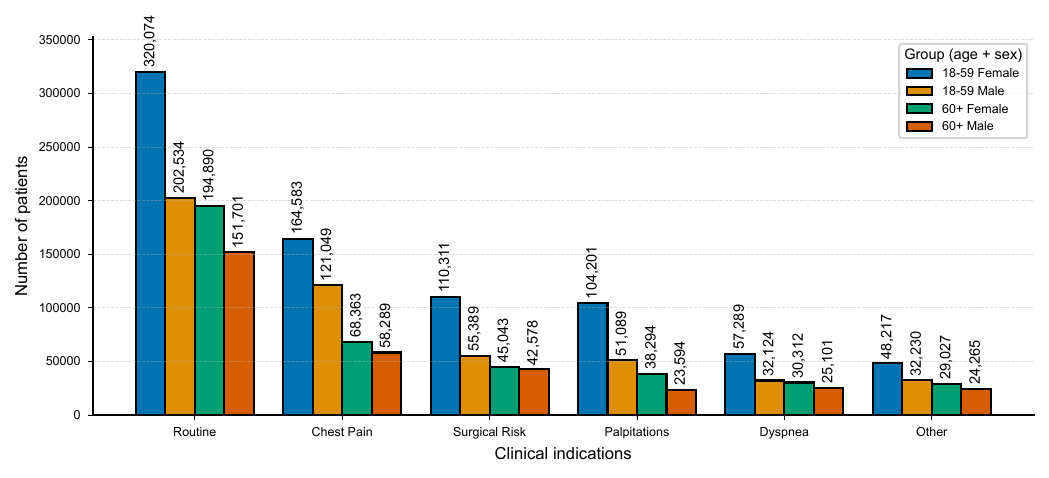} \vspace{-8mm}
		\caption{Clinical indications for exams, stratified by age group and sex, in the CODE-II dataset.}
		\label{Figure:SupplementaryFigure3}
	}
\end{figure}

% Example list with hyperlinks to labels you will define below
\begin{enumerate}[label={}, leftmargin=*]
\item \textbf{Supplementary Fig.\,\ref{Figure:SupplementaryFigure1}:} Number of ECG exams per year in the CODE-II dataset.

\item \textbf{Supplementary Fig.\,\ref{Figure:SupplementaryFigure2}:} Age distribution of patients at their first ECG exam in the CODE-II dataset, stratified by sex.

\item \textbf{Supplementary Fig.\,\ref{Figure:SupplementaryFigure3}:} Clinical indications for exams, stratified by age group and sex, in the CODE-II dataset.

\item \textbf{Supplementary Fig.\,\ref{Figure:SupplementaryFigure4}:} Summary of exam and signal record distributions in the CODE-II dataset. (a) Distribution of the number of ECG exams per patient. (b) Distribution of the number of signal records per exam. Each exam may include multiple consecutive tracings recorded during the same clinical session.

\item \textbf{Supplementary Fig.\,\ref{Figure:SupplementaryFigure5}:} Final distribution of the number of tracings per exam in the CODE-II dataset after applying the pipeline that retains up to four tracings per exam.

\item \textbf{Supplementary Fig.\,\ref{Figure:SupplementaryFigure6}:} Analysis of group-level diagnostic label assignments in the CODE-II dataset. (a) Number of exams exclusively assigned to each of the 10 CODE groups. (b) Ten most frequent non-exclusive group combinations, among 251 distinct combinations observed. Group names were abbreviated (using the first four letters or a combination of initials) and joined by “+” to denote co-occurrence. (c) Distribution of the number of CODE groups assigned per exam, indicating how frequently exams belong to multiple groups.

\item \textbf{Supplementary Fig.\,\ref{Figure:SupplementaryFigure7}:} Median and interquartile range (IQR) of cardiac measurements stratified by age group and sex in the CODE-II dataset. (a) Heart rate (beats per minute); (b) P wave duration (milliseconds), reflecting atrial depolarization; (c) PR interval (ms), reflecting atrioventricular conduction time; (d) QRS duration (ms), representing ventricular depolarization; (e) QRS axis (degrees), showing the mean frontal plane electrical axis; and (f) Corrected QT interval (QTc, ms), adjusted for heart rate.

\item \textbf{Supplementary Fig.\,\ref{Figure:SupplementaryFigure8}:} Number of ECG exams per year in the CODE-II-open dataset.

\item \textbf{Supplementary Fig.\,\ref{Figure:SupplementaryFigure9}:} Geographic distribution of ECG exams in the CODE-II-open dataset across Brazilian states.

\item \textbf{Supplementary Fig.\,\ref{Figure:SupplementaryFigure10}:} Age distribution of patients in the CODE-II-open dataset, stratified by sex.

\item \textbf{Supplementary Fig.\,\ref{Figure:SupplementaryFigure11}:} Distribution of ECG exam indications in the CODE-II-open dataset by age group and sex.

\item \textbf{Supplementary Fig.\,\ref{Figure:SupplementaryFigure12}:} Distribution of the number of signal records per exam in the CODE-II-open dataset. Each exam may include multiple consecutive tracings recorded during the same clinical session.

\item \textbf{Supplementary Fig.\,\ref{Figure:SupplementaryFigure13}:} Final distribution of the number of tracings per exam in the CODE-II-open dataset after applying the pipeline that retains up to four tracings per exam.

\item \textbf{Supplementary Fig.\,\ref{Figure:SupplementaryFigure14}:} Number of exams per diagnostic class in the CODE-II-open dataset. Class assignments are not mutually exclusive: a single exam may be associated with multiple diagnoses, except for Normal ECGs, which are assigned exclusively.

\item \textbf{Supplementary Fig.\,\ref{Figure:SupplementaryFigure15}:} Distribution of CODE-II-open ECG exams by diagnostic group and number of assigned diagnoses. (a) Number of exams per CODE diagnostic group (note that exams may appear in multiple groups), and (b) the number of exams with a given number of assigned CODE diagnoses.

\item \textbf{Supplementary Fig.\,\ref{Figure:SupplementaryFigure16}:} Analysis of group-level diagnostic label assignments in the CODE-II-open dataset. (a) Number of exams exclusively assigned to each of the 10 CODE groups. (b) Ten most frequent non-exclusive group combinations, among 147 distinct combinations observed. Group names were abbreviated (using the first four letters or a combination of initials) and joined by “+” to denote co-occurrence. (c) Distribution of the number of CODE groups assigned per exam, indicating how frequently exams belong to multiple groups.

\item \textbf{Supplementary Fig.\,\ref{Figure:SupplementaryFigure17}:} Median and interquartile range (IQR) of electrocardiographic measurements in the CODE-II-open dataset, stratified by age group and sex. (a) Heart rate (beats per minute); (b) P wave duration (milliseconds), reflecting atrial depolarization; (c) PR interval (ms), reflecting atrioventricular conduction time; (d) QRS duration (ms), representing ventricular depolarization; (e) QRS axis (degrees), showing the mean frontal plane electrical axis; and (f) Corrected QT interval (QTc, ms), adjusted for heart rate.

\item \textbf{Supplementary Fig.\,\ref{Figure:SupplementaryFigure18}:} Annual distribution of ECG exams in the CODE-II-test dataset (2018–2025).

\item \textbf{Supplementary Fig.\,\ref{Figure:SupplementaryFigure19}:} Geographic distribution of ECG exams in the CODE-II-test dataset across Brazilian states.

\item \textbf{Supplementary Fig.\,\ref{Figure:SupplementaryFigure20}:} Age distribution of patients in the CODE-II-test dataset, stratified by sex.

\item \textbf{Supplementary Fig.\,\ref{Figure:SupplementaryFigure21}:} Distribution of ECG exam indications in the CODE-II-test dataset by age group and sex.

\item \textbf{Supplementary Fig.\,\ref{Figure:SupplementaryFigure22}:} Number of signal records per exam in the CODE-II-test dataset.

\item \textbf{Supplementary Fig.\,\ref{Figure:SupplementaryFigure23}:} Final distribution of the number of tracings per exam in the CODE-II-test dataset after applying the pipeline that retains up to four tracings per exam.

\item \textbf{Supplementary Fig.\,\ref{Figure:SupplementaryFigure24}:} Distribution of ECG exams per diagnostic class in the CODE-II-test dataset. Class assignments are not mutually exclusive: a single exam may be associated with multiple diagnoses, except for Normal ECGs, which are assigned exclusively.

\item \textbf{Supplementary Fig.\,\ref{Figure:SupplementaryFigure25}:} Distribution of ECG exams in the CODE-II-test dataset by diagnostic group and by number of assigned diagnoses. (a) Number of exams per CODE diagnostic group (exams may appear in multiple groups). (b) Distribution of exams by number of assigned CODE diagnoses.

\item \textbf{Supplementary Fig.\,\ref{Figure:SupplementaryFigure26}:} Group-level diagnostic label assignments in the CODE-II-test dataset. (a) Number of exams exclusively assigned to each of the 10 CODE groups. (b) Ten most frequent non-exclusive group combinations, among 106 distinct combinations observed. Group names were abbreviated (using the first four letters or a combination of initials) and joined by “+” to denote co-occurrence. (c) Distribution of the number of CODE groups assigned per exam, indicating how frequently exams belong to multiple groups.

\item \textbf{Supplementary Fig.\,\ref{Figure:SupplementaryFigure27}:} Distribution of the number of exams annotated per cardiologist in the CODE-II-test dataset (n = 46 cardiologists).

\item \textbf{Supplementary Fig.\,\ref{Figure:SupplementaryFigure28}:} Distribution of the number of cardiologists per exam in the CODE-II-test dataset.

\end{enumerate}

%------------------------------------------------------------------------------------
% Scaling laws experiments -  SUBSECTION
%------------------------------------------------------------------------------------
\section{List of Supplementary Tables}
\begin{enumerate}[label={}, leftmargin=*]
\item \textbf{Supplementary Table\,\ref{Table:SupplementaryTable1}:} Self-reported clinical comorbidities of patients in the CODE-II dataset, stratified by age group (18–59 years and $\geq60$ years) and sex, based on the first ECG exam. For hypertesion, dyslipidemia, and diabetes, comorbidity status also inferred from reported medication usage.

\item \textbf{Supplementary Table\,\ref{Table:SupplementaryTable2}:} Summary of selected CODE diagnostic classes in the CODE-II dataset, including class descriptions, labels, and the number and percentage of exams in the full dataset ($n = 2{,}735{,}269$), as well as class-wise counts of unique patients and ECG exams in the training and validation sets obtained using the proposed patient-exclusive stratified split. The table reports the 10 most prevalent classes plus 5 representative rare classes for illustration. The complete per-class distribution for all 66 CODE classes is provided in Supplementary Data 1 (external Excel file, Supplementary\_Data\_1.xlsx).

\item \textbf{Supplementary Table\,\ref{Table:SupplementaryTable3}:} Group-level distribution of CODE diagnostic assignments at the patient level, stratified by age (18–59 and $\geq 60$ years) and sex. Columns report counts by stratum, totals, the percentage of patients within each diagnosis group (by sex and overall), and the percentage relative to all patients in CODE-II. Diagnostic groups are not mutually exclusive; patients may belong to multiple groups, except for the Normal group, which is exclusive by definition.

\item \textbf{Supplementary Table\,\ref{Table:SupplementaryTable4}:} Prevalence of self-reported or inferred comorbidities across age and sex groups in the CODE-II-open dataset.

\item \textbf{Supplementary Table\,\ref{Table:SupplementaryTable5}:} Summary of selected CODE diagnostic classes in the CODE-II-open dataset, including class descriptions, labels, and the number and percentage of exams in the dataset ($n = 15{,}000$), as well as class-wise counts of unique patients and ECG exams in the training and validation sets obtained using the proposed patient-exclusive stratified split. The table reports the 10 most prevalent classes plus 5 representative rare classes for illustration. The complete per-class distribution for all 66 CODE classes is provided in Supplementary Data 2 (external Excel file, Supplementary\_Data\_2.xlsx).

\item \textbf{Supplementary Table\,\ref{Table:SupplementaryTable6}:} Prevalence of self-reported or inferred comorbidities across age and sex groups in the CODE-II-test dataset.

\item \textbf{Supplementary Table\,\ref{Table:SupplementaryTable7}:} Summary of selected CODE diagnostic classes in the CODE-II-test dataset, including class descriptions, labels, and the number and percentage of exams in the dataset ($n = 8{,}475$). The table reports the 10 most prevalent classes plus 5 representative rare classes for illustration. The complete per-class distribution for all 66 CODE classes is provided in Supplementary Data 3 (external Excel file, Supplementary\_Data\_3.xlsx).

\item \textbf{Supplementary Table\,\ref{Table:SupplementaryTable8}:} Cardiologist-level performance on the CODE-II-test under the fair-scope definition. For each cardiologist, we report the number of exams reviewed, the number of diagnostic classes encountered, and micro- and macro-averaged precision, recall, specificity, F1 score, and NPV computed only over classes actually seen by that cardiologist. To avoid small-sample inflation, the table lists cardiologists with $\geq 100$ reviewed exams. The complete list, including all cardiologists, is provided in Supplementary Data 4 (external Excel file, Supplementary\_Data\_4.xlsx).

\item \textbf{Supplementary Table\,\ref{Table:SupplementaryTable9}:} Per-class model performance on the CODE-II-test under the F1-max thresholding rule. For each selected diagnostic class, we report Test prevalence (number and percentage), threshold-independent metrics (AUROC, AUPRC), and threshold-dependent metrics (F1-score, {Recall}/Sensitivity, Specificity, Precision, and NPV) computed using class-specific thresholds selected on the CODE-II validation split and applied unchanged to the test set. All metrics are reported with 95\% confidence intervals estimated from 1,000 bootstrap resamples (shown in parentheses). The table lists the 10 most prevalent classes and 5 representative rare classes for illustration. The complete results for all 66 classes are provided in Supplementary Data 5 (external Excel file, Supplementary\_Data\_5.xlsx).

\item \textbf{Supplementary Table\,\ref{Table:SupplementaryTable10}:} Aggregate model performance on the CODE-II-test under two threshold rules: the primary F1-max thresholds and an alternative based on Youden’s J (both selected on the CODE-II validation split and applied unchanged to the Test set). Micro- and macro-averaged metrics are reported---Precision, Recall, F1-score, AUROC, and AUPRC---for each rule.

\item \textbf{Supplementary Table\,\ref{Table:SupplementaryTable11}:} Per-class model performance on the CODE-II-test for the three diagnostic classes that produced degenerate metrics under the F1-max rule (precision, recall, and F1 equal to zero). For each class, we report Test-set prevalence (number and \%), AUROC and AUPRC, and---under two thresholding rules, F1-max and Youden's J, both selected on the CODE-II validation split and applied unchanged to the Test set---the selected threshold and the resulting Precision, Recall, Specificity, F1-score, and NPV.

\item \textbf{Supplementary Table\,\ref{Table:SupplementaryTable12}:} Macro-AUROC results for all supervised and pre-trained models on the five PTB-XL tasks---diagnosis, subclass, superclass, form, and rhythm---and on the full CPSC 2018 dataset. All models were trained using 100\% of the available training data. For each model, we report the mean macro-AUROC over three runs, with the run-to-run variation shown in parentheses (maximum minus minimum). The table also includes the average macro-AUROC across tasks and the total number of model parameters. The complete set of results, including each model under few-shot training (5\% and 10\%) on PTB-XL and the improvement relative to its randomly initialized counterpart (when applicable), is provided in Supplementary Data~6 (external Excel file, \textit{Supplementary\_Data\_6.xlsx}).

\item \textbf{Supplementary Table\,\ref{Table:SupplementaryTable13}:} Model sizes and pre-training datasets for baseline ECG models. The table reports the number of parameters (estimated from full-data training on PTB-XL diagnostic classes), the size of the pre-training set (in ECG records or ECG–report pairs), and the corresponding data sources used for model pre-training.

\item \textbf{Supplementary Table\,\ref{Table:SupplementaryTable14}:} Training and fine-tuning configurations for pre-trained ECG models. For each model, we report the learning rates used under full-data and few-shot scenarios, the input frequency and segment length, and the adopted fine-tuning strategy. When applicable, model-specific exceptions are noted. The same configurations were used for both PTB-XL and CPSC 2018 experiments unless otherwise specified. Input signals shorter than the target input length were symmetrically zero-padded, while longer signals were symmetrically truncated.

\item \textbf{Supplementary Table\,\ref{Table:SupplementaryTable15}:} Performance of the baseline and literature ECG models on the CODE-II-open dataset under their best-performing configuration, either trained from scratch or fine-tuned from pre-trained versions. The same evaluation pipeline used for external benchmarking was applied to ensure comparability across experiments. Metrics are reported as macro–AUROC and macro–average precision (macro–AP). The performance of the baseline model trained on the full CODE-II cohort is shown for reference.

\end{enumerate}

%------------------------------------------------------------------------------------
% Scaling laws experiments -  SUBSECTION
%------------------------------------------------------------------------------------
\section{List of Supplementary Data (external files)}
\begin{enumerate}[label={}, leftmargin=*]
\item \textbf{Supplementary Data 1:} Full distribution of the 66 CODE diagnostic classes in the CODE-II dataset, including class descriptions, labels, and the number and percentage of exams in the full dataset ($n = 2{,}735{,}269$), as well as class-wise counts of unique patients and ECG exams in the training and validation sets obtained using the proposed patient-exclusive stratified split. \textit{Data are provided as a separate downloadable Excel file} (File: \texttt{Supplementary\_Data\_1.xlsx}).

\item \textbf{Supplementary Data 2:} Full distribution of the 66 CODE diagnostic classes in the CODE-II dataset, including class descriptions, labels, and the number and percentage of exams in the dataset ($n = 15{,}000$), as well as class-wise counts of ECG exams in the training and validation sets obtained using the proposed patient-exclusive stratified split. \textit{Data are provided as a separate downloadable Excel file} (File: \texttt{Supplementary\_Data\_2.xlsx}).

\item \textbf{Supplementary Data 3:} Full distribution of the 66 CODE diagnostic classes in the CODE-II dataset, including class descriptions, labels, and the number and percentage of exams in the dataset ($n = 8{,}475$). \textit{Data are provided as a separate downloadable Excel file} (File: \texttt{Supplementary\_Data\_3.xlsx}).

\item \textbf{Supplementary Data 4:} Complete cardiologist-level performance on the CODE-II-test under the fair-scope definition, including all reviewers (n = 46 cardiologists). Columns include cardiologist ID, exams reviewed, diagnostic classes seen, and micro- and macro-averaged precision, recall, specificity, F1 score, and NPV computed over classes encountered by each cardiologist. \textit{Data are provided as a separate downloadable Excel file} (File: \texttt{Supplementary\_Data\_4.xlsx}).

\item \textbf{Supplementary Data 5:} Per-class model performance on the CODE-II-test under the F1-max thresholding rule. For all 66 diagnostic classes, we report Test prevalence (number and percentage), threshold-independent metrics (AUROC, AUPRC), and threshold-dependent metrics (F1-score, {Recall}/Sensitivity, Specificity, Precision, and NPV) computed using class-specific thresholds selected on the CODE-II validation split and applied unchanged to the test set. All metrics are reported with 95\% confidence intervals estimated from 1,000 bootstrap resamples (shown in parentheses). \textit{Data are provided as a separate downloadable Excel file} (File: \texttt{Supplementary\_Data\_5.xlsx}).

\item \textbf{Supplementary Data 6:} Macro-AUROC results for all supervised and pre-trained models on PTB-XL (across the diagnostic, subclass, superclass, form, and rhythm categories) and on CPSC 2018. Both datasets were evaluated under full-data training, while few-shot training (5\% and 10\%) was applied only to PTB-XL. Reported values are means over three runs, with the run-to-run range width (maximum minus minimum) shown in parentheses. For each model, we also report the average macro-AUROC across tasks, the improvement relative to its randomly initialized counterpart (when applicable), and the total number of model parameters. A second worksheet provides per-run results. \textit{Data are provided as a separate downloadable Excel file} (File: \texttt{Supplementary\_Data\_6.xlsx}).

\item \textbf{Supplementary Data 7:} Performance of the baseline and reference ECG models on the CODE-II-open dataset under their best-performing configurations, either trained from scratch or fine-tuned from pre-trained weights. The same evaluation pipeline used for external benchmarking was applied to ensure comparability across experiments. Metrics are reported as macro-AUROC and macro-AUPRC for each of the three independent runs, together with their mean and performance range (maximum–minimum) across runs. \textit{Data are provided as a separate downloadable Excel file} (File: \texttt{Supplementary\_Data\_7.xlsx}).

\end{enumerate}

% ================================
% Full content sections
% ================================
\newpage
\section*{Supplementary Figures}

\begin{figure}[H]
	\centering{
		\includegraphics[scale=0.9]{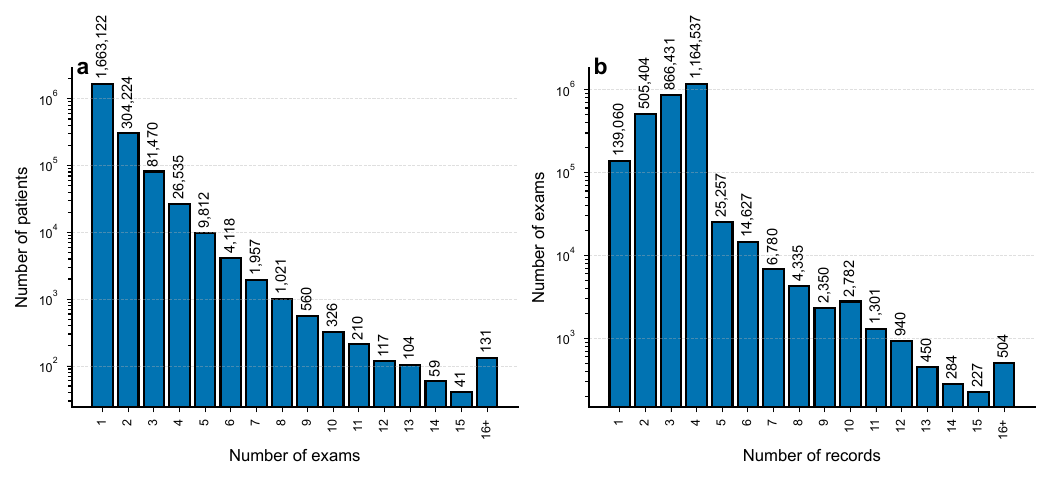} \vspace{-8mm}
		\caption{Summary of exam and signal record distributions in the CODE-II dataset. (a) Distribution of the number of ECG exams per patient. (b) Distribution of the number of signal records per exam. Each exam may include multiple consecutive tracings recorded during the same clinical session.}
		\label{Figure:SupplementaryFigure4}
	}
\end{figure}

\begin{figure}[H]
	\centering{
		\includegraphics[scale=1]{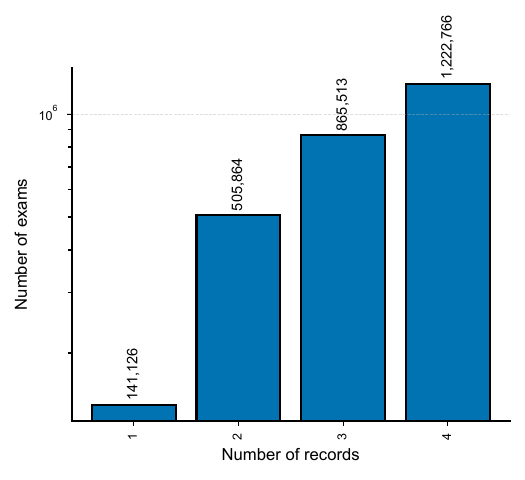} \vspace{-4mm}
		\caption{Final distribution of the number of tracings per exam in the CODE-II dataset after applying the pipeline that retains up to four tracings per exam.}
		\label{Figure:SupplementaryFigure5}
	}
\end{figure}

\begin{figure}[H]
	\centering{
		\includegraphics[scale=0.9]{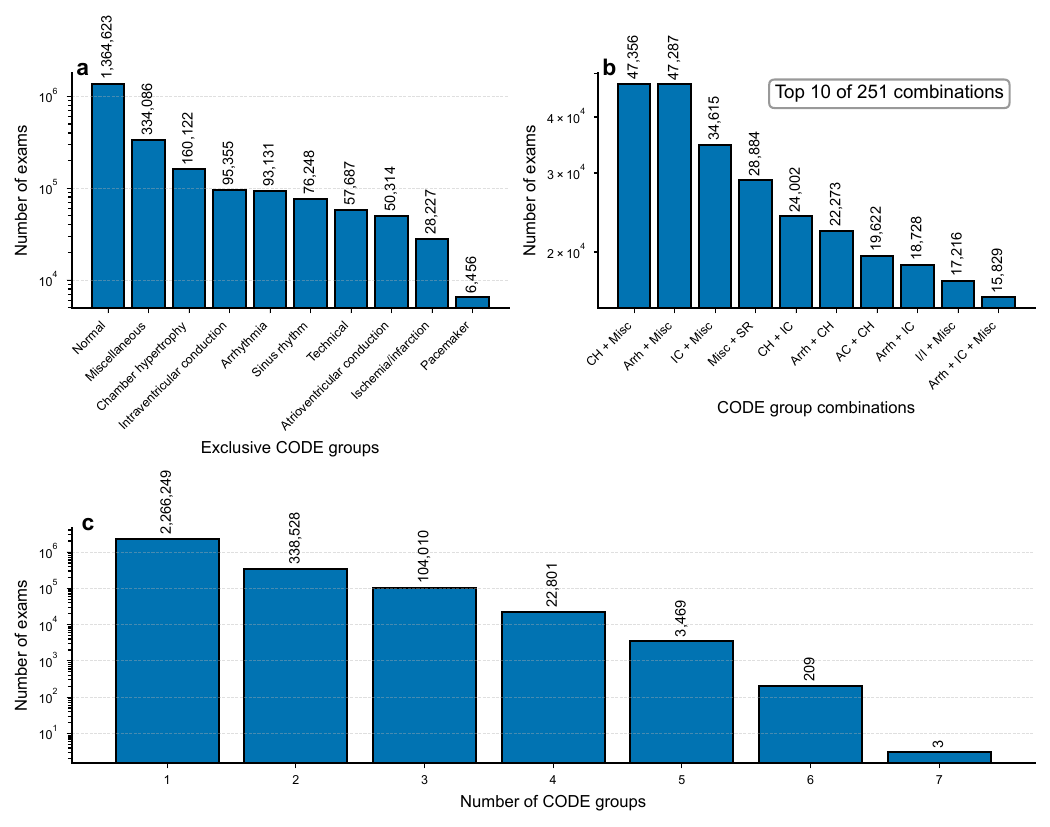} \vspace{-8mm}
		\caption{Analysis of group-level diagnostic label assignments in the CODE-II dataset. (a) Number of exams exclusively assigned to each of the 10 CODE groups. (b) Ten most frequent non-exclusive group combinations, among 251 distinct combinations observed. Group names were abbreviated (using the first four letters or a combination of initials) and joined by “+” to denote co-occurrence. (c) Distribution of the number of CODE groups assigned per exam, indicating how frequently exams belong to multiple groups.}
		\label{Figure:SupplementaryFigure6}
	}
\end{figure}

\begin{figure}[H]
	\centering{
		\includegraphics[scale=0.91]{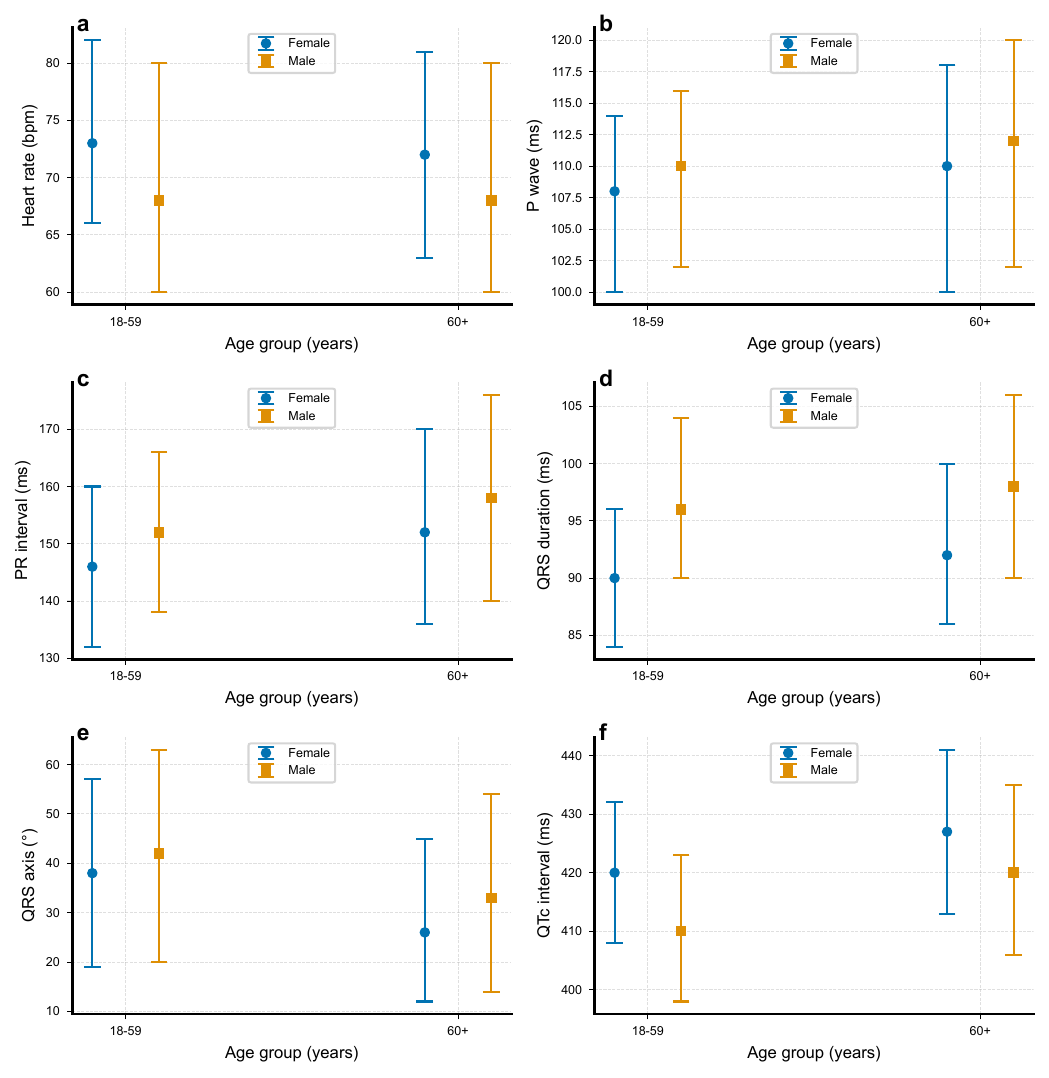} \vspace{-8mm}
		\caption{Median and interquartile range (IQR) of cardiac measurements stratified by age group and sex in the CODE-II dataset. (a) Heart rate (beats per minute); (b) P wave duration (milliseconds), reflecting atrial depolarization; (c) PR interval (ms), reflecting atrioventricular conduction time; (d) QRS duration (ms), representing ventricular depolarization; (e) QRS axis (degrees), showing the mean frontal plane electrical axis; and (f) Corrected QT interval (QTc, ms), adjusted for heart rate.}
		\label{Figure:SupplementaryFigure7}
	}
\end{figure}

\begin{figure}[H]
	\centering{
		\includegraphics[scale=1]{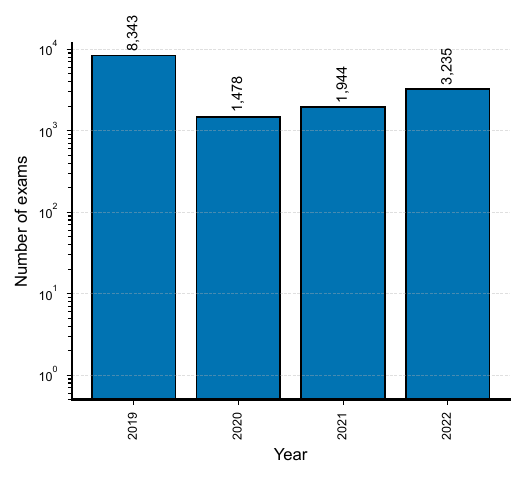} \vspace{-4mm}
		\caption{Number of ECG exams per year in the CODE-II-open dataset.}
		\label{Figure:SupplementaryFigure8}
	}
\end{figure}

\begin{figure}[H]
	\centering{
		\includegraphics[scale=1]{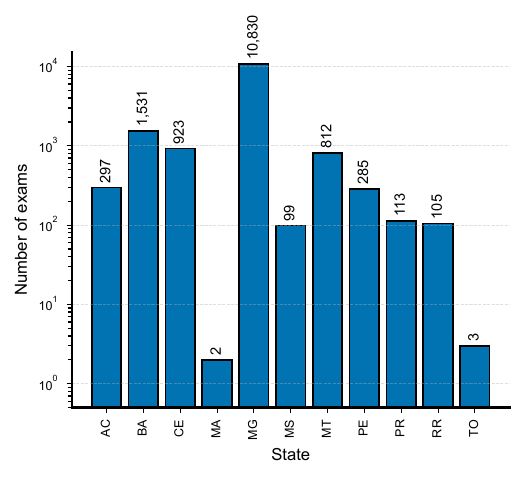} \vspace{-4mm}
		\caption{Geographic distribution of ECG exams in the CODE-II-open dataset across Brazilian states.}
		\label{Figure:SupplementaryFigure9}
	}
\end{figure}

\begin{figure}[H]
	\centering{
		\includegraphics[scale=1]{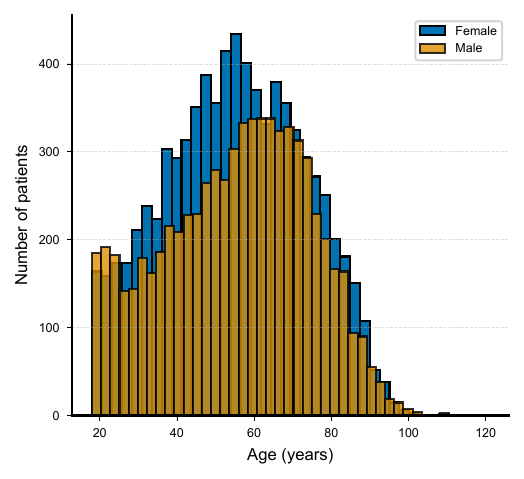} \vspace{-4mm}
		\caption{Age distribution of patients in the CODE-II-open dataset, stratified by sex.}
		\label{Figure:SupplementaryFigure10}
	}
\end{figure}

\begin{figure}[H]
	\centering{
		\includegraphics[scale=0.92]{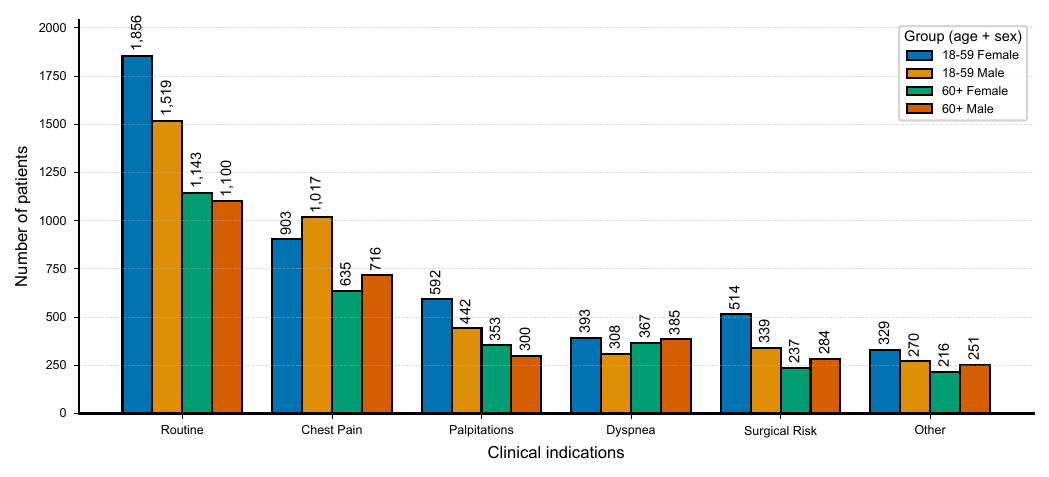} \vspace{-8mm}
		\caption{Distribution of ECG exam indications in the CODE-II-open dataset by age group and sex.}
		\label{Figure:SupplementaryFigure11}
	}
\end{figure}

\begin{figure}[H]
	\centering{
		\includegraphics[scale=1]{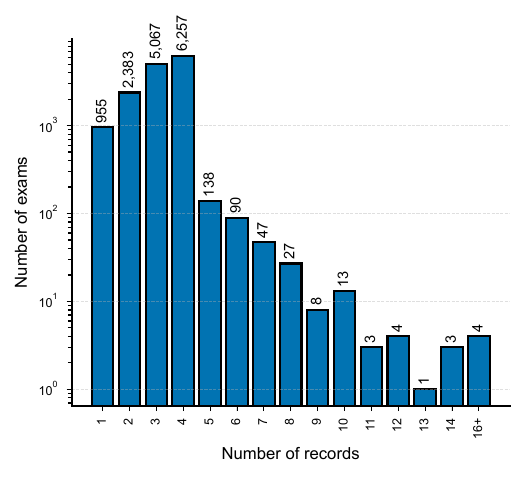} \vspace{-4mm}
		\caption{Distribution of the number of signal records per exam in the CODE-II-open dataset. Each exam may include multiple consecutive tracings recorded during the same clinical session.}
		\label{Figure:SupplementaryFigure12}
	}
\end{figure}

\begin{figure}[H]
	\centering{
		\includegraphics[scale=1]{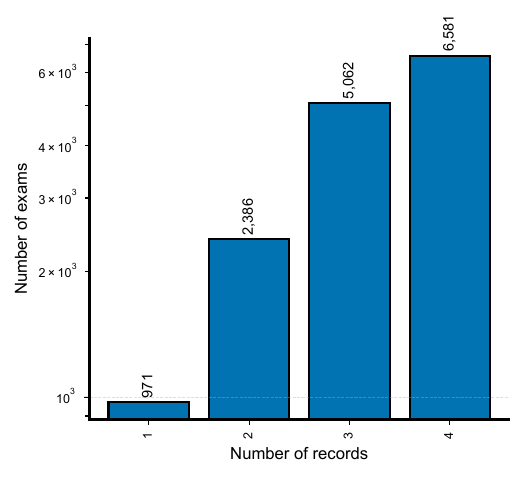} \vspace{-4mm}
		\caption{Final distribution of the number of tracings per exam in the CODE-II-open dataset after applying the pipeline that retains up to four tracings per exam.}
		\label{Figure:SupplementaryFigure13}
	}
\end{figure}

\begin{figure}[H]
	\centering{
		\includegraphics[scale=0.91]{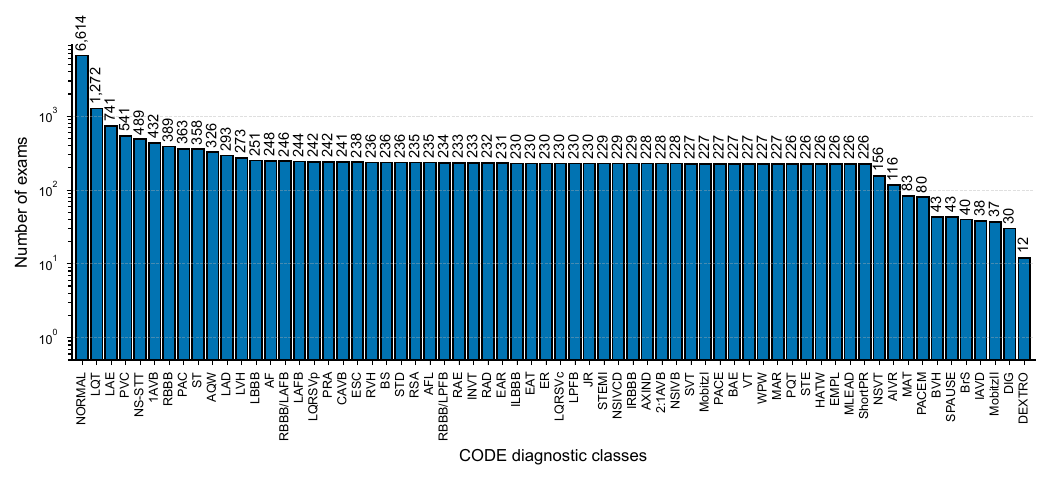} \vspace{-8mm}
		\caption{Number of exams per diagnostic class in the CODE-II-open dataset. Class assignments are not mutually exclusive: a single exam may be associated with multiple diagnoses, except for Normal ECGs, which are assigned exclusively.}
		\label{Figure:SupplementaryFigure14}
	}
\end{figure}

\begin{figure}[H]
	\centering{
		\includegraphics[scale=0.9]{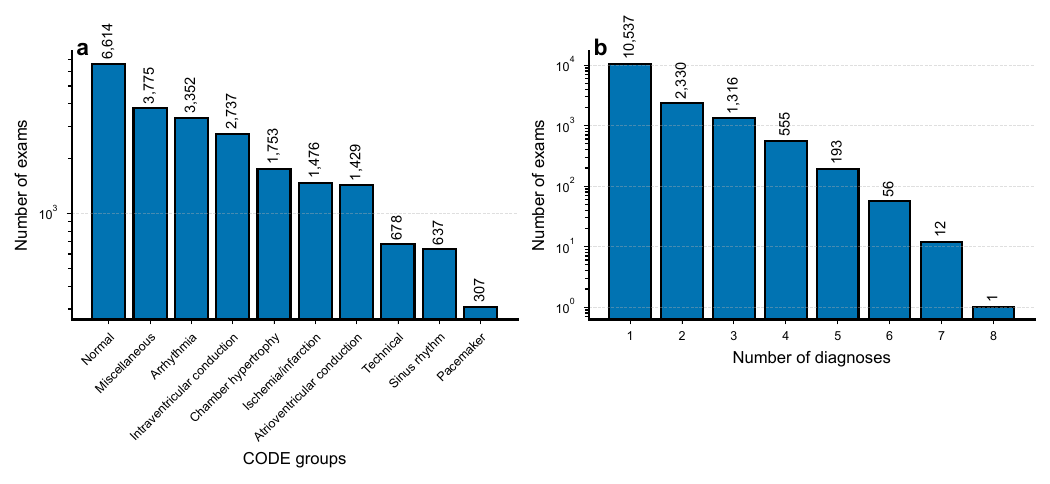} \vspace{-8mm}
		\caption{Distribution of CODE-II-open ECG exams by diagnostic group and number of assigned diagnoses. (a) Number of exams per CODE diagnostic group (note that exams may appear in multiple groups), and (b) the number of exams with a given number of assigned CODE diagnoses.}
		\label{Figure:SupplementaryFigure15}
	}
\end{figure}

\begin{figure}[H]
	\centering{
		\includegraphics[scale=0.9]{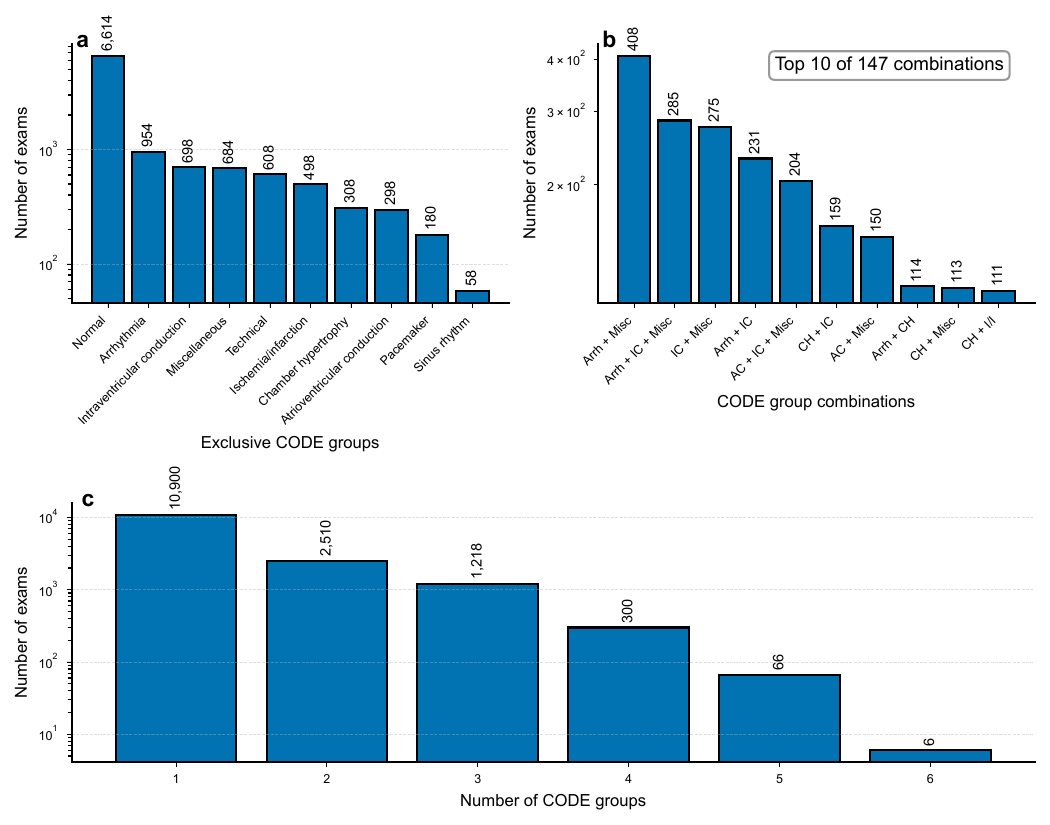} \vspace{-8mm}
		\caption{Analysis of group-level diagnostic label assignments in the CODE-II-open dataset. (a) Number of exams exclusively assigned to each of the 10 CODE groups. (b) Ten most frequent non-exclusive group combinations, among 147 distinct combinations observed. Group names were abbreviated (using the first four letters or a combination of initials) and joined by “+” to denote co-occurrence. (c) Distribution of the number of CODE groups assigned per exam, indicating how frequently exams belong to multiple groups.}
		\label{Figure:SupplementaryFigure16}
	}
\end{figure}

\begin{figure}[H]
	\centering{
		\includegraphics[scale=0.9]{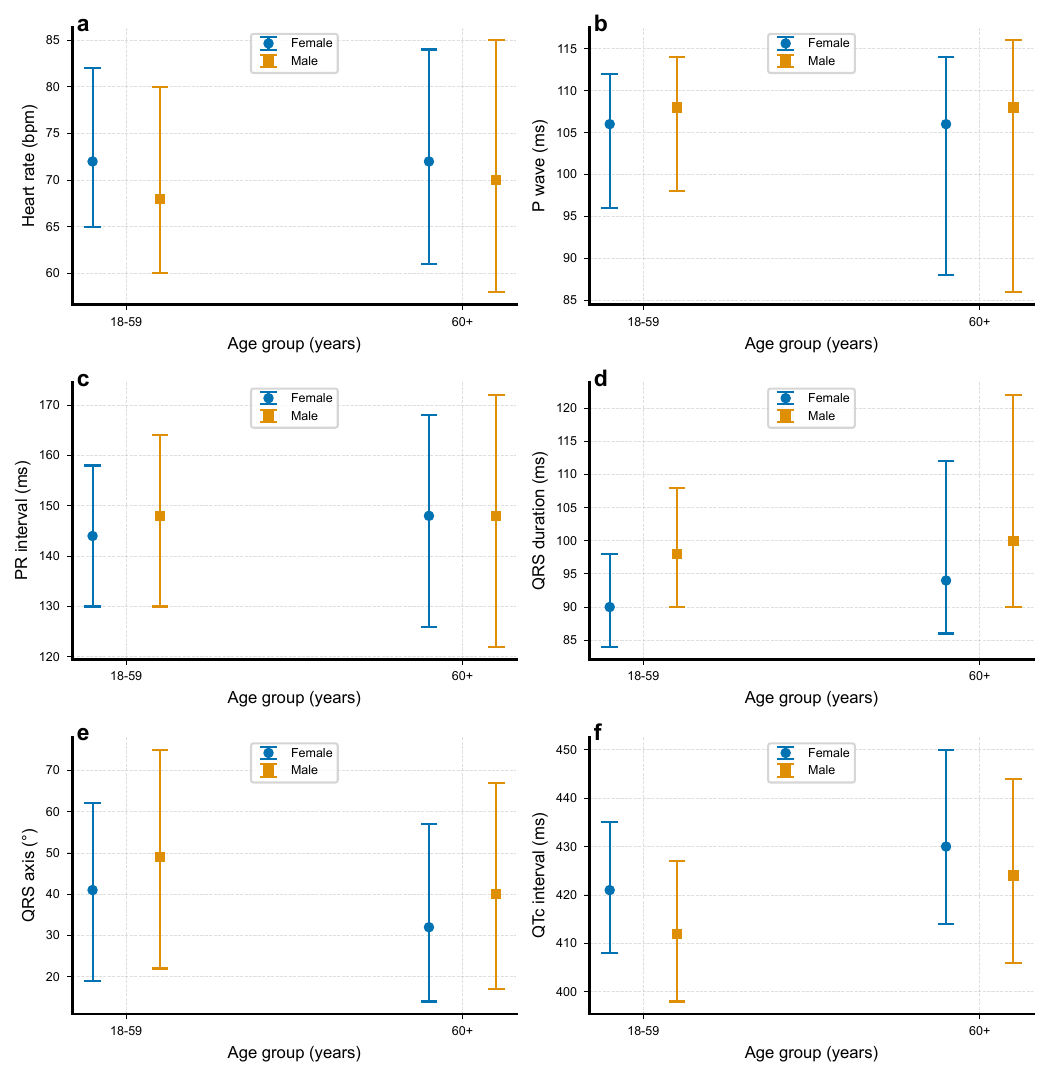} \vspace{-8mm}
		\caption{Median and interquartile range (IQR) of electrocardiographic measurements in the CODE-II-open dataset, stratified by age group and sex. (a) Heart rate (beats per minute); (b) P wave duration (milliseconds), reflecting atrial depolarization; (c) PR interval (ms), reflecting atrioventricular conduction time; (d) QRS duration (ms), representing ventricular depolarization; (e) QRS axis (degrees), showing the mean frontal plane electrical axis; and (f) Corrected QT interval (QTc, ms), adjusted for heart rate.}
		\label{Figure:SupplementaryFigure17}
	}
\end{figure}

\begin{figure}[H]
	\centering{
		\includegraphics[scale=1]{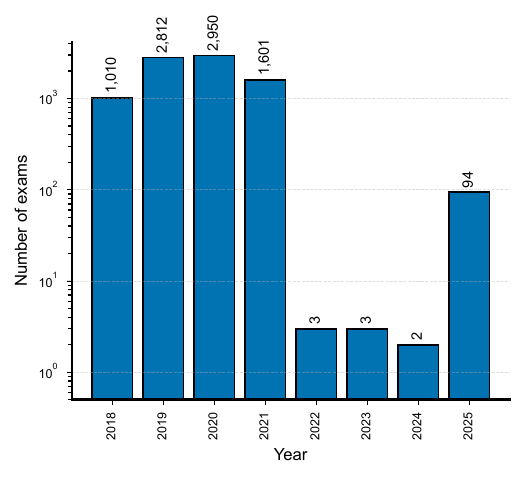} \vspace{-4mm}
		\caption{Annual distribution of ECG exams in the CODE-II-test dataset (2018–2025).}
		\label{Figure:SupplementaryFigure18}
	}
\end{figure}

\begin{figure}[H]
	\centering{
		\includegraphics[scale=1]{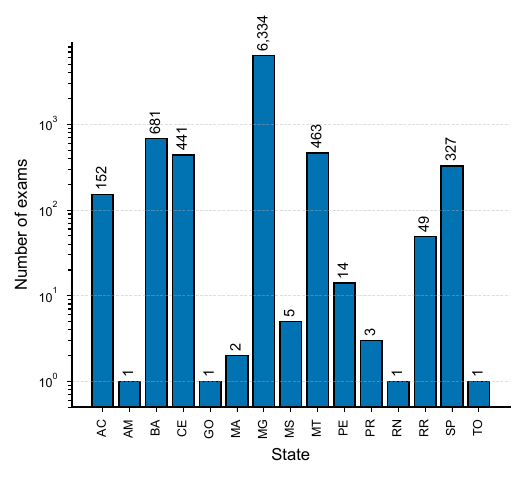} \vspace{-4mm}
		\caption{Geographic distribution of ECG exams in the CODE-II-test dataset across Brazilian states.}
		\label{Figure:SupplementaryFigure19}
	}
\end{figure}

\begin{figure}[H]
	\centering{
		\includegraphics[scale=1]{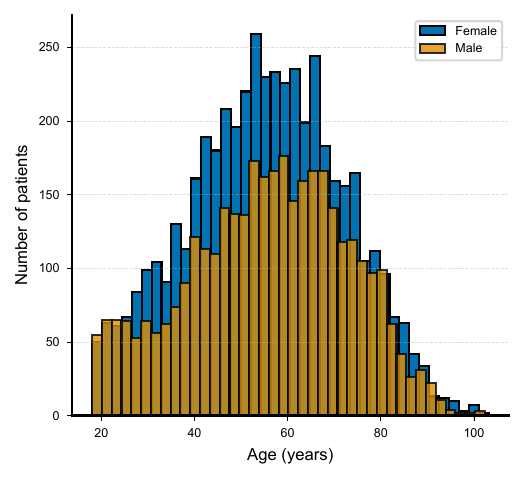} \vspace{-4mm}
		\caption{Age distribution of patients in the CODE-II-test dataset, stratified by sex.}
		\label{Figure:SupplementaryFigure20}
	}
\end{figure}

\begin{figure}[H]
	\centering{
		\includegraphics[scale=0.92]{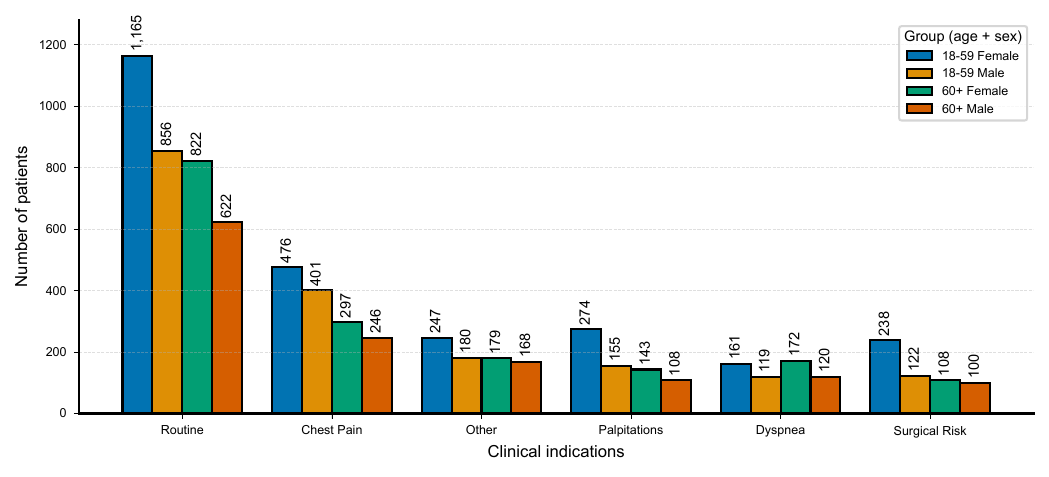} \vspace{-8mm}
		\caption{Distribution of ECG exam indications in the CODE-II-test dataset by age group and sex.}
		\label{Figure:SupplementaryFigure21}
	}
\end{figure}

\begin{figure}[H]
	\centering{
		\includegraphics[scale=1]{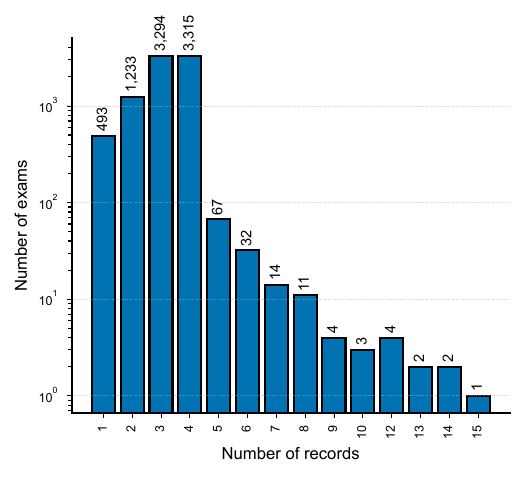} \vspace{-4mm}
		\caption{Number of signal records per exam in the CODE-II-test dataset.}
		\label{Figure:SupplementaryFigure22}
	}
\end{figure}

\begin{figure}[H]
	\centering{
		\includegraphics[scale=1]{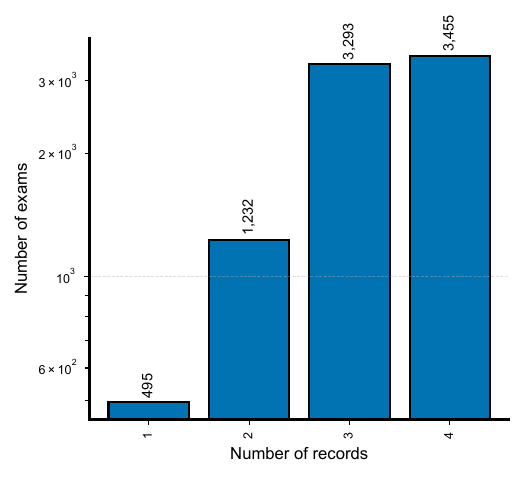} \vspace{-4mm}
		\caption{Final distribution of the number of tracings per exam in the CODE-II-test dataset after applying the pipeline that retains up to four tracings per exam.}
		\label{Figure:SupplementaryFigure23}
	}
\end{figure}

\begin{figure}[H]
	\centering{
		\includegraphics[scale=0.91]{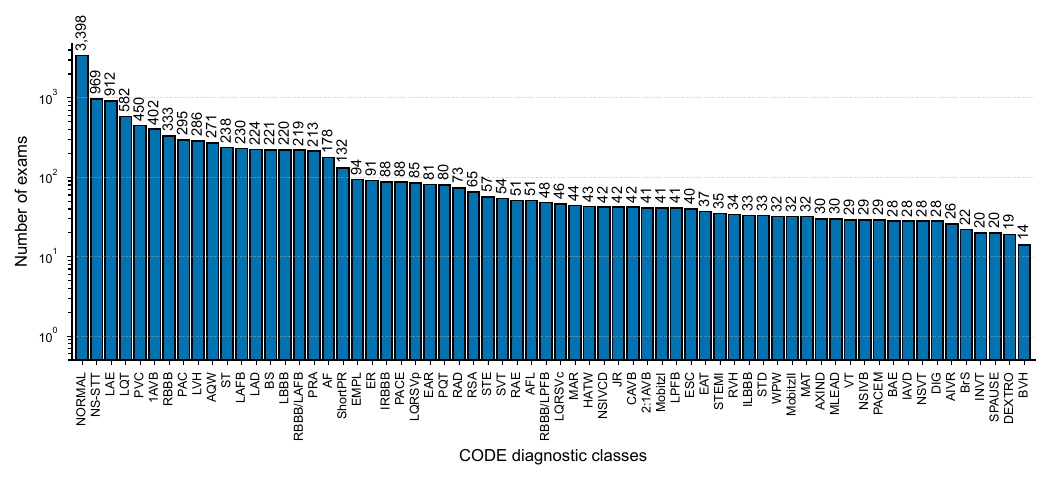} \vspace{-8mm}
		\caption{Distribution of ECG exams per diagnostic class in the CODE-II-test dataset. Class assignments are not mutually exclusive: a single exam may be associated with multiple diagnoses, except for Normal ECGs, which are assigned exclusively.}
		\label{Figure:SupplementaryFigure24}
	}
\end{figure}

\begin{figure}[H]
	\centering{
		\includegraphics[scale=0.9]{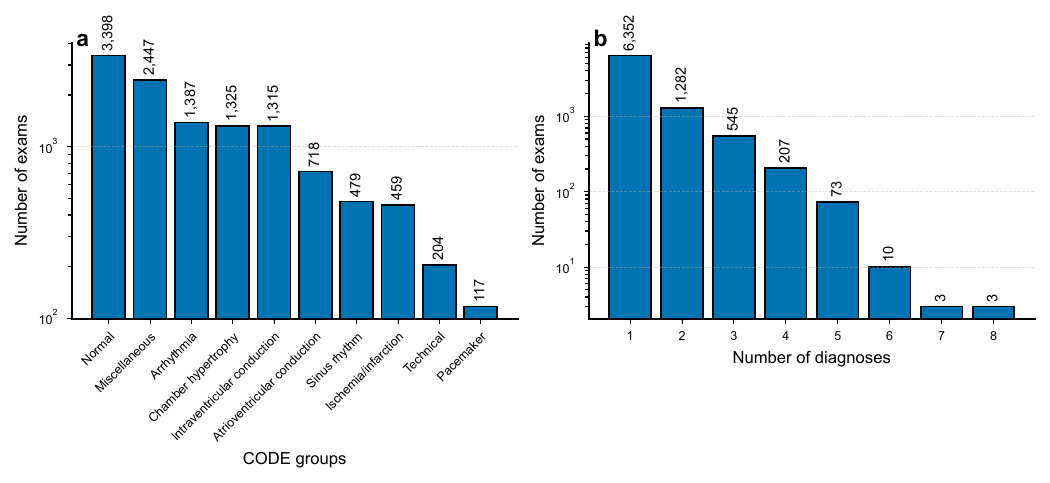} \vspace{-8mm}
		\caption{Distribution of ECG exams in the CODE-II-test dataset by diagnostic group and by number of assigned diagnoses. (a) Number of exams per CODE diagnostic group (exams may appear in multiple groups). (b) Distribution of exams by number of assigned CODE diagnoses.}
		\label{Figure:SupplementaryFigure25}
	}
\end{figure}

\begin{figure}[H]
	\centering{
		\includegraphics[scale=0.9]{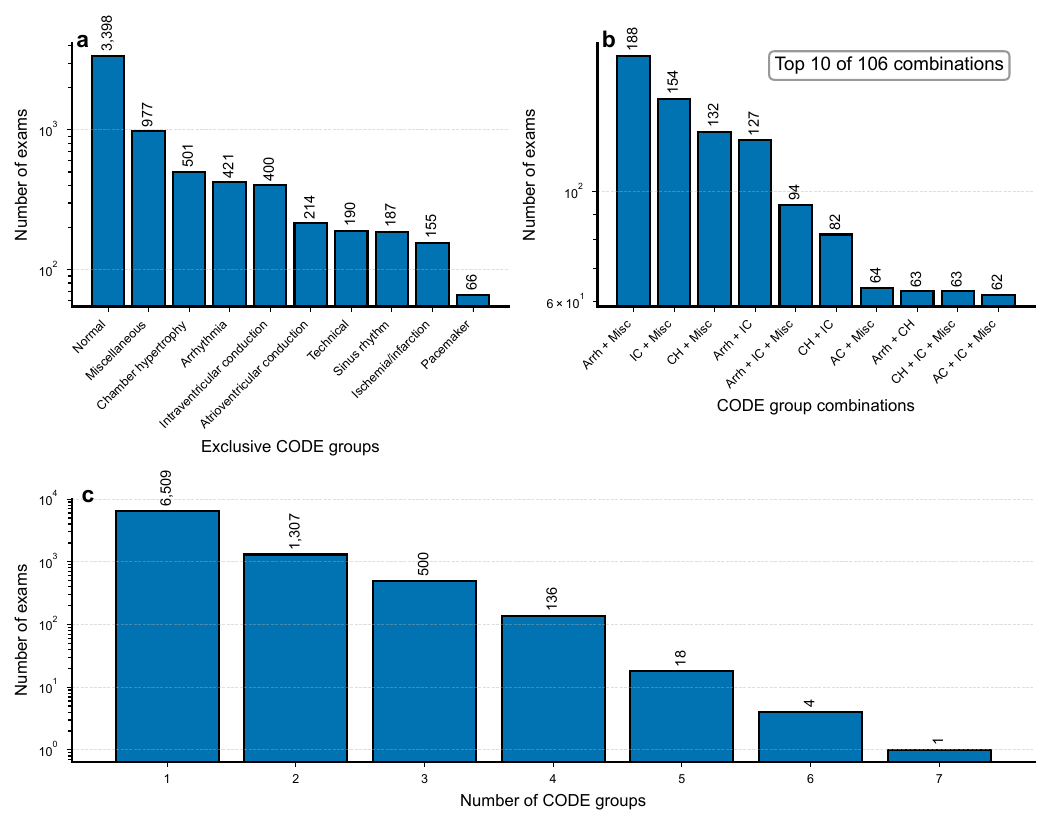} \vspace{-8mm}
		\caption{Group-level diagnostic label assignments in the CODE-II-test dataset. (a) Number of exams exclusively assigned to each of the 10 CODE groups. (b) Ten most frequent non-exclusive group combinations, among 106 distinct combinations observed. Group names were abbreviated (using the first four letters or a combination of initials) and joined by “+” to denote co-occurrence. (c) Distribution of the number of CODE groups assigned per exam, indicating how frequently exams belong to multiple groups.}
		\label{Figure:SupplementaryFigure26}
	}
\end{figure}

\begin{figure}[H]
	\centering{
		\includegraphics[scale=0.91]{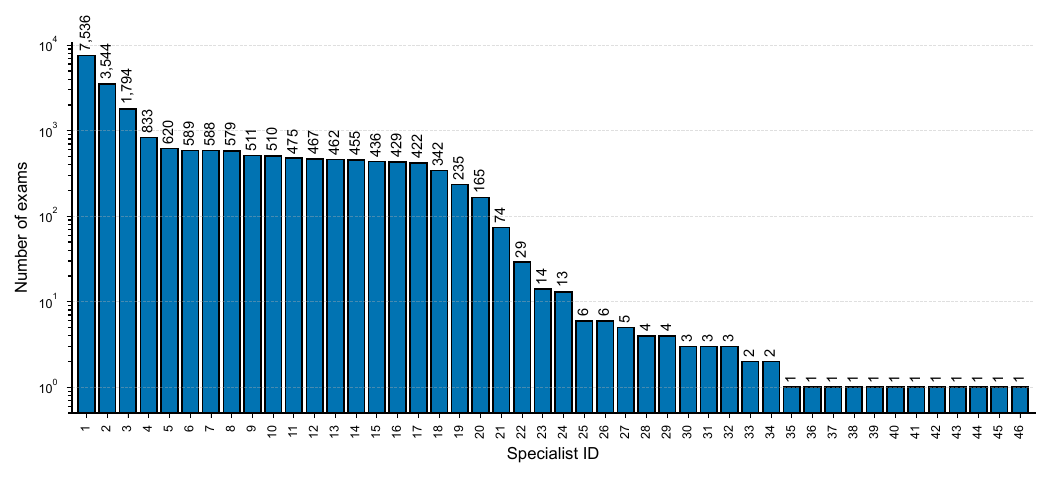} \vspace{-8mm}
		\caption{Distribution of the number of exams annotated per cardiologist in the CODE-II-test dataset (n = 46 cardiologists).}
		\label{Figure:SupplementaryFigure27}
	}
\end{figure}

\begin{figure}[H]
	\centering{
		\includegraphics[scale=1]{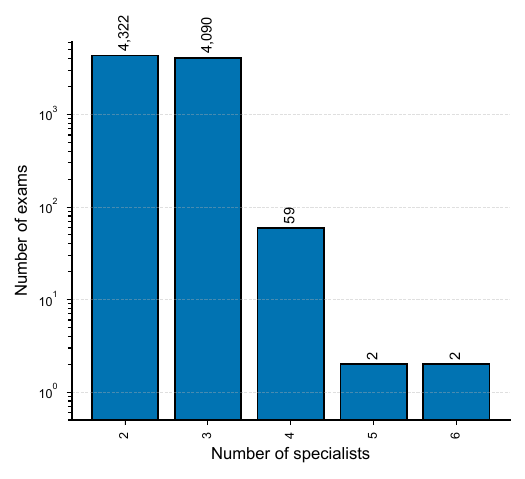} \vspace{-4mm}
		\caption{Distribution of the number of cardiologists per exam in the CODE-II-test dataset.}
		\label{Figure:SupplementaryFigure28}
	}
\end{figure}

% \begin{table}[ht]
% \centering
% \caption{Overview of pretraining characteristics for all pretrained models evaluated in this study. For each model, we report the total number of trainable parameters, the size of the pretraining dataset, and the corresponding data sources.}
% \label{Table:SupplementaryTable13}
% \resizebox{\textwidth}{!}{%
% \csvreader[
%   separator=tab,
%   tabular={lccc},
%   table head={%
%     \toprule
%     Model & \# Parameters & Pretraining Size & Pretraining Sources \\
%     \midrule
%   },
%   late after line=\\,
%   table foot=\bottomrule
% ]{tables/SupplementaryTable13.tsv}{}%
% {%
%   \csvcoli & \csvcolii & \csvcoliii & \csvcoliv
% }
% }
% \end{table}

% ================================
% Full content sections
% ================================
\newpage
\section*{Supplementary Tables}

\begin{table}[H]
\centering
\caption{Self-reported clinical comorbidities of patients in the CODE-II dataset, stratified by age group (18–59 years and $\geq\!60$ years) and sex, based on the first ECG exam. For hypertension, dyslipidemia, and diabetes, comorbidity status also inferred from reported medication usage.}
\label{Table:SupplementaryTable1}

\resizebox{\textwidth}{!}{%
  \begin{filecontents*}{tables/SupplementaryTable1.tsv}
Condition	Female_18_59	Male_18_59	Female_60_plus	Male_60_plus	Female_Total	Male_Total	All_Total	Pct_Female_Cond	Pct_Male_Cond	Pct_All_Patients
Hypertension	298,696	159,570	319,837	220,367	618,533	379,937	998,470	61.95	38.05	47.69
Diabetes	79,382	39,982	111,306	62,594	190,688	102,576	293,264	65.02	34.98	14.01
Smoking	50,485	56,017	26,775	37,793	77,260	93,810	171,070	45.16	54.84	8.17
Dyslipidemia	25,576	13,090	39,733	21,604	65,309	34,694	100,003	65.31	34.69	4.78
Myocardial Infarction	9,558	7,625	9,620	11,481	19,178	19,106	38,284	50.09	49.91	1.83
Chagas Disease	6,582	3,710	7,235	4,491	13,817	8,201	22,018	62.75	37.25	1.05
COPD	3,985	1,820	4,403	3,732	8,388	5,552	13,940	60.17	39.83	0.67
\end{filecontents*}
\csvreader[
    separator=tab,
    tabular={lrrrrrrrrrr},
    table head={%
      \toprule
      & \multicolumn{2}{c}{18--59} & \multicolumn{2}{c}{60+} & \multicolumn{3}{c}{Total} & \multicolumn{3}{c}{\% of Patients with Condition} \\
      \cmidrule(lr){2-3}\cmidrule(lr){4-5}\cmidrule(lr){6-8}\cmidrule(lr){9-11}
      Condition & Female & Male & Female & Male & Female & Male & All & Female & Male & All \\
      \midrule
    },
    late after line=\\,
    table foot=\bottomrule
  ]{tables/SupplementaryTable1.tsv}{}%
  {%
    \csvcoli & \csvcolii & \csvcoliii & \csvcoliv & \csvcolv &
    \csvcolvi & \csvcolvii & \csvcolviii & \csvcolix & \csvcolx & \csvcolxi
  }
}
\end{table}

\begin{table}[H]
\centering
\caption{Summary of selected CODE diagnostic classes in the CODE-II dataset, including class descriptions, labels, and the number and percentage of exams in the full dataset ($n = 2{,}735{,}269$), as well as class-wise counts of unique patients and ECG exams in the training and validation sets obtained using the proposed patient-exclusive stratified split. The table reports the 10 most prevalent classes plus 5 representative rare classes for illustration. The complete per-class distribution for all 66 CODE classes is provided in Supplementary Data 1 (external Excel file, Supplementary\_Data\_1.xlsx).}
\label{Table:SupplementaryTable2}

\resizebox{\textwidth}{!}{%
  \begin{filecontents*}{tables/SupplementaryTable2.tsv}
Description	CODElabel	TotalExams	UniquePatients	PctAllExams	ValidationExams	ValidationPatients	TrainingExams	TrainingPatients
ECG within normal limits for age and sex	NORMAL	1,364,623	1,174,017	49.89	272,925	267,612	1,091,698	906,405
Nonspecific ST-T abnormality	NS-STT	364,537	318,576	13.327	78,366	66,179	286,171	252,397
Left atrial enlargement	LAE	289,219	252,908	10.574	76,890	64,925	212,329	187,983
Left ventricular hypertrophy	LVH	110,180	94,222	4.028	25,492	21,598	84,688	72,624
Sinus tachycardia	ST	107,212	97,654	3.92	26,476	23,776	80,736	73,878
Premature ventricular complex(es)	PVC	105,608	92,877	3.861	31,248	26,398	74,360	66,479
First-degree atrioventricular block	1AVB	103,581	84,334	3.787	31,838	24,925	71,743	59,409
Prolonged QT interval	LQT	95,911	77,887	3.506	37,738	29,667	58,173	48,220
Premature atrial complex(es)	PAC	85,858	77,713	3.139	24,053	21,187	61,805	56,526
Pathological Q waves	AQW	72,558	60,443	2.653	22,007	17,679	50,551	42,764
Biventricular hypertrophy	BVH	57	56	0.002	18	17	39	39
Sinus pause	SPAUSE	56	55	0.002	14	13	42	42
Isorhythmic atrioventricular dissociation	IAVD	48	47	0.002	13	13	35	34
Digitalis effect	DIG	44	44	0.002	14	14	30	30
Dextrocardia	DEXTRO	29	28	0.001	6	6	23	22
\end{filecontents*}
\csvreader[
    separator=tab,
    tabular={llrrrrrrrr}, % 2 colunas de texto + 7 numéricas (à direita)
    table head={%
      \toprule
      \multicolumn{2}{c}{} & \multicolumn{3}{c}{CODE-II dataset} & \multicolumn{2}{c}{Validation set} & \multicolumn{2}{c}{Training set} \\
      \cmidrule(lr){3-5}\cmidrule(lr){6-7}\cmidrule(lr){8-9}
      CODE statement description & CODE label & Total Exams & Unique Patients & \% of All Exams & Validation Exams & Validation Patients & Training Exams & Training Patients \\
      \midrule
    },
    late after line=\\,
    table foot=\bottomrule
  ]{tables/SupplementaryTable2.tsv}{}%
  {%
    \csvcoli   & % CODE statement description
    \csvcolii  & % CODE label
    \csvcoliii & % Total Exams
    \csvcoliv  & % Unique Patients
    \csvcolv   & % % of All Exams
    \csvcolvi  & % Validation Exams
    \csvcolvii & % Validation Patients
    \csvcolviii& % Training Exams
    \csvcolix    % Training Patients
  }
}
\end{table}

\begin{table}[H]
\centering
\caption{Group-level distribution of CODE diagnostic assignments at the patient level, stratified by age (18–59 and $\geq 60$ years) and sex. Columns report counts by stratum, totals, the percentage of patients within each diagnosis group (by sex and overall), and the percentage relative to all patients in CODE-II. Diagnostic groups are not mutually exclusive; patients may belong to multiple groups, except for the Normal group, which is exclusive by definition.}
\label{Table:SupplementaryTable3}

\resizebox{\textwidth}{!}{%
  \begin{filecontents*}{tables/SupplementaryTable3.tsv}
DiagnosisGroup	Female_18_59	Male_18_59	Female_60_plus	Male_60_plus	Female_Total	Male_Total	All_Total	Pct_Female_Group	Pct_Male_Group	Pct_All_Patients
Normal	524,401	290,750	173,357	108,427	697,758	399,177	1,096,935	63.61	36.39	52.39
Miscellaneous	136,276	81,092	131,424	104,268	267,700	185,360	453,060	59.09	40.91	21.64
Chamber hypertrophy	48,350	65,685	72,809	83,572	121,159	149,257	270,416	44.8	55.2	12.92
Intraventricular conduction	24,776	38,086	56,863	66,707	81,639	104,793	186,432	43.79	56.21	8.9
Arrhythmia	28,529	24,156	57,770	66,572	86,299	90,728	177,027	48.75	51.25	8.45
Sinus rhythm	43,169	35,547	23,010	22,221	66,179	57,768	123,947	53.39	46.61	5.92
Atrioventricular conduction	27,767	18,439	27,641	31,981	55,408	50,420	105,828	52.36	47.64	5.05
Ischemia/infarction	9,054	12,555	16,810	20,846	25,864	33,401	59,265	43.64	56.36	2.83
Technical	14,306	10,173	10,023	8,652	24,329	18,825	43,154	56.38	43.62	2.06
Pacemaker	594	603	2,352	2,504	2,946	3,107	6,053	48.67	51.33	0.29
\end{filecontents*}
\csvreader[
    separator=tab,
    tabular={lrrrrrrrrrr}, % 1 texto + 10 numéricas
    table head={%
      \toprule
      & \multicolumn{2}{c}{18--59} & \multicolumn{2}{c}{60+} & \multicolumn{3}{c}{Total} & \multicolumn{3}{c}{\% of Patients in Diagnosis Group} \\
      \cmidrule(lr){2-3}\cmidrule(lr){4-5}\cmidrule(lr){6-8}\cmidrule(lr){9-11}
      Diagnosis Group & Female & Male & Female & Male & Female & Male & All & Female & Male & All \\
      \midrule
    },
    late after line=\\,
    table foot=\bottomrule
  ]{tables/SupplementaryTable3.tsv}{}%
  {%
    \csvcoli   & % Diagnosis Group
    \csvcolii  & % Female_18_59
    \csvcoliii & % Male_18_59
    \csvcoliv  & % Female_60_plus
    \csvcolv   & % Male_60_plus
    \csvcolvi  & % Female_Total
    \csvcolvii & % Male_Total
    \csvcolviii& % All_Total
    \csvcolix  & % Pct_Female_Group
    \csvcolx   & % Pct_Male_Group
    \csvcolxi    % Pct_All_Patients
  }
}
\end{table}

\begin{table}[H]
\centering
\caption{Prevalence of self-reported or inferred comorbidities across age and sex groups in the CODE-II-open dataset.}
\label{Table:SupplementaryTable4}

\resizebox{\textwidth}{!}{%
  \begin{filecontents*}{tables/SupplementaryTable4.tsv}
Condition	Female_18_59	Male_18_59	Female_60_plus	Male_60_plus	Female_Total	Male_Total	All_Total	Pct_Female_Cond	Pct_Male_Cond	Pct_All_Patients
Hypertension	1,676	1,151	2,326	1,974	4,002	3,125	7,127	56.15	43.85	47.51
Diabetes	429	260	771	535	1,200	795	1,995	60.15	39.85	13.3
Smoking	318	450	208	355	526	805	1,331	39.52	60.48	8.87
Dyslipidemia	157	97	266	201	423	298	721	58.67	41.33	4.81
Myocardial Infarction	54	77	77	141	131	218	349	37.54	62.46	2.33
Chagas Disease	67	58	116	75	183	133	316	57.91	42.09	2.11
COPD	41	13	58	40	99	53	152	65.13	34.87	1.01
\end{filecontents*}
\csvreader[
    separator=tab,
    tabular={lrrrrrrrrrr},
    table head={%
      \toprule
      & \multicolumn{2}{c}{18--59} & \multicolumn{2}{c}{60+} & \multicolumn{3}{c}{Total} & \multicolumn{3}{c}{\% of Patients with Condition} \\
      \cmidrule(lr){2-3}\cmidrule(lr){4-5}\cmidrule(lr){6-8}\cmidrule(lr){9-11}
      Condition & Female & Male & Female & Male & Female & Male & All & Female & Male & All \\
      \midrule
    },
    late after line=\\,
    table foot=\bottomrule
  ]{tables/SupplementaryTable4.tsv}{}%
  {%
    \csvcoli & \csvcolii & \csvcoliii & \csvcoliv & \csvcolv &
    \csvcolvi & \csvcolvii & \csvcolviii & \csvcolix & \csvcolx & \csvcolxi
  }
}
\end{table}

\begin{table}[ht]
\centering
\caption{Summary of selected CODE diagnostic classes in the CODE-II-open dataset, including class descriptions, labels, and the number and percentage of exams in the dataset ($n = 15{,}000$), as well as class-wise counts of unique patients and ECG exams in the training and validation sets obtained using the proposed patient-exclusive stratified split. The table reports the 10 most prevalent classes plus 5 representative rare classes for illustration. The complete per-class distribution for all 66 CODE classes is provided in Supplementary Data 2 (external Excel file, Supplementary\_Data\_2.xlsx).}
\label{Table:SupplementaryTable5}

\resizebox{\textwidth}{!}{%
  \begin{filecontents*}{tables/SupplementaryTable5.tsv}
Description	CODElabel	TotalExams	PctAllExams	ValidationExams	TrainingExams
ECG within normal limits for age and sex	NORMAL	6,614	44.09	1,488	5,126
Prolonged QT interval	LQT	1,272	8.48	398	874
Left atrial enlargement	LAE	741	4.94	213	528
Premature ventricular complex(es)	PVC	541	3.61	159	382
Nonspecific ST-T abnormality	NS-STT	489	3.26	179	310
First-degree atrioventricular block	1AVB	432	2.88	134	298
Right bundle branch block	RBBB	389	2.59	105	284
Premature atrial complex(es)	PAC	363	2.42	125	238
Sinus tachycardia	ST	358	2.39	130	228
Pathological Q waves	AQW	326	2.17	97	229
Brugada syndrome	BrS	40	0.27	7	33
Isorhythmic atrioventricular dissociation	IAVD	38	0.25	10	28
Second-degree atrioventricular block, Mobitz type II	MobitzII	37	0.25	12	25
Digitalis effect	DIG	30	0.2	12	18
Dextrocardia	DEXTRO	12	0.08	3	9
\end{filecontents*}
\csvreader[
    separator=tab,
    tabular={llrrrr}, % 2 colunas de texto + 4 numéricas (à direita)
    table head={%
      \toprule
      \multicolumn{2}{c}{} & \multicolumn{2}{c}{CODE-II-open dataset} & \multicolumn{1}{c}{Validation set} & \multicolumn{1}{c}{Training set} \\
      \cmidrule(lr){3-4}\cmidrule(lr){5-5}\cmidrule(lr){6-6}
      CODE statement description & CODE label & Total Exams & \% of All Exams & Validation Exams & Training Exams \\
      \midrule
    },
    late after line=\\,
    table foot=\bottomrule
  ]{tables/SupplementaryTable5.tsv}{}%
  {%
    \csvcoli   & % CODE statement description
    \csvcolii  & % CODE label
    \csvcoliii & % Total Exams
    \csvcoliv  & % % of All Exams
    \csvcolv   & % Validation Exams
    \csvcolvi   % Training Exams
  }
}
\end{table}

\begin{table}[ht]
\centering
\caption{Prevalence of self-reported or inferred comorbidities across age and sex groups in the CODE-II-test dataset.}
\label{Table:SupplementaryTable6}

\resizebox{\textwidth}{!}{%
  \begin{filecontents*}{tables/SupplementaryTable6.tsv}
Condition	Female_18_59	Male_18_59	Female_60_plus	Male_60_plus	Female_Total	Male_Total	All_Total	Pct_Female_Cond	Pct_Male_Cond	Pct_All_Patients
Hypertension	1,006	594	1,244	880	2,250	1,474	3,724	60.42	39.58	43.94
Chagas Disease	376	302	333	213	709	515	1,224	57.92	42.08	14.44
Diabetes	225	127	403	230	628	357	985	63.76	36.24	11.62
Smoking	160	212	103	144	263	356	619	42.49	57.51	7.3
Dyslipidemia	121	74	206	107	327	181	508	64.37	35.63	5.99
Myocardial Infarction	37	34	50	62	87	96	183	47.54	52.46	2.16
COPD	12	8	35	20	47	28	75	62.67	37.33	0.88
\end{filecontents*}
\csvreader[
    separator=tab,
    tabular={lrrrrrrrrrr},
    table head={%
      \toprule
      & \multicolumn{2}{c}{18--59} & \multicolumn{2}{c}{60+} & \multicolumn{3}{c}{Total} & \multicolumn{3}{c}{\% of Patients with Condition} \\
      \cmidrule(lr){2-3}\cmidrule(lr){4-5}\cmidrule(lr){6-8}\cmidrule(lr){9-11}
      Condition & Female & Male & Female & Male & Female & Male & All & Female & Male & All \\
      \midrule
    },
    late after line=\\,
    table foot=\bottomrule
  ]{tables/SupplementaryTable6.tsv}{}%
  {%
    \csvcoli & \csvcolii & \csvcoliii & \csvcoliv & \csvcolv &
    \csvcolvi & \csvcolvii & \csvcolviii & \csvcolix & \csvcolx & \csvcolxi
  }
}
\end{table}

\begin{table}[H]
\centering
\caption{Summary of selected CODE diagnostic classes in the CODE-II-test dataset, including class descriptions, labels, and the number and percentage of exams in the dataset ($n = 8{,}475$). The table reports the 10 most prevalent classes plus 5 representative rare classes for illustration. The complete per-class distribution for all 66 CODE classes is provided in Supplementary Data 3 (external Excel file, Supplementary\_Data\_3.xlsx).}
\label{Table:SupplementaryTable7}

\resizebox{0.9\textwidth}{!}{%
  \begin{filecontents*}{tables/SupplementaryTable7.tsv}
Description	CODElabel	TotalExams	PctAllExams
ECG within normal limits for age and sex	NORMAL	3,398	40.09
Nonspecific ST-T abnormality	NS-STT	969	11.43
Left atrial enlargement	LAE	912	10.76
Prolonged QT interval	LQT	582	6.87
Premature ventricular complex(es)	PVC	450	5.31
First-degree atrioventricular block	1AVB	402	4.74
Right bundle branch block	RBBB	333	3.93
Premature atrial complex(es)	PAC	295	3.48
Left ventricular hypertrophy	LVH	286	3.37
Pathological Q waves	AQW	271	3.2
Brugada syndrome	BrS	22	0.26
Sinus pause	SPAUSE	20	0.24
T-wave inversions	INVT	20	0.24
Dextrocardia	DEXTRO	19	0.22
Biventricular hypertrophy	BVH	14	0.17
\end{filecontents*}
\csvreader[
    separator=tab,
    tabular={llrr}, % 2 colunas de texto + 2 numéricas (à direita)
    table head={%
      \toprule
      \multicolumn{2}{c}{} & \multicolumn{2}{c}{CODE-II-test dataset}  \\
      \cmidrule(lr){3-4}
      CODE statement description & CODE label & Total Exams & \% of All Exams \\
      \midrule
    },
    late after line=\\,
    table foot=\bottomrule
  ]{tables/SupplementaryTable7.tsv}{}%
  {%
    \csvcoli   & % CODE statement description
    \csvcolii  & % CODE label
    \csvcoliii & % Total Exams
    \csvcoliv   % % of All Exams
  }
}
\end{table}

\begin{table}[H]
\centering
\caption{Cardiologist-level performance on the CODE-II-test under the fair-scope definition. For each cardiologist, we report the number of exams reviewed, the number of diagnostic classes encountered, and micro- and macro-averaged precision, recall, specificity, F1 score, and NPV computed only over classes actually seen by that cardiologist. To avoid small-sample inflation, the table lists cardiologists with $\geq 100$ reviewed exams. The complete list, including all cardiologists, is provided in Supplementary Data 4 (external Excel file, Supplementary\_Data\_4.xlsx).}
\label{Table:SupplementaryTable8}

\resizebox{\textwidth}{!}{%
  \begin{filecontents*}{tables/SupplementaryTable8.tsv}
CardiologistID	ExamsReviewed	ClassesSeen	Micro_Precision	Micro_Recall	Micro_Specificity	Micro_F1	Micro_NPV	Macro_Precision	Macro_Recall	Macro_Specificity	Macro_F1	Macro_NPV
1	8,368	65	0.821	0.924	0.996	0.869	0.998	0.805	0.915	0.995	0.844	0.998
2	3,583	52	0.986	0.990	1.000	0.988	1.000	0.961	0.973	0.999	0.966	1.000
3	1,831	65	0.962	0.985	0.999	0.973	1.000	0.955	0.980	0.999	0.964	1.000
4	847	63	0.966	0.991	0.999	0.978	1.000	0.955	0.993	0.999	0.972	1.000
5	627	56	0.981	0.991	1.000	0.986	1.000	0.967	0.974	0.999	0.970	1.000
6	602	38	0.975	0.985	0.999	0.980	0.999	0.974	0.981	0.999	0.975	0.999
7	596	62	0.928	0.993	0.998	0.959	1.000	0.935	0.996	0.998	0.961	1.000
8	587	51	0.974	0.990	0.999	0.982	1.000	0.946	0.974	0.999	0.957	1.000
9	529	52	0.962	0.974	0.999	0.968	0.999	0.970	0.949	0.999	0.954	0.999
10	522	48	0.980	0.981	0.999	0.981	0.999	0.991	0.970	0.999	0.974	0.999
11	483	50	0.982	0.988	0.999	0.985	1.000	0.982	0.990	0.999	0.984	1.000
12	470	57	0.987	0.993	1.000	0.990	1.000	0.966	0.968	1.000	0.964	1.000
13	469	35	0.983	0.987	0.999	0.985	0.999	0.964	0.984	0.999	0.969	0.999
14	464	53	0.980	0.985	0.999	0.982	1.000	0.961	0.974	0.999	0.965	1.000
15	453	43	0.963	0.961	0.999	0.962	0.999	0.934	0.936	0.998	0.932	0.999
16	447	49	0.959	0.962	0.999	0.960	0.999	0.927	0.915	0.998	0.911	0.999
17	431	35	0.973	0.981	0.999	0.977	0.999	0.942	0.958	0.999	0.946	0.999
18	347	57	0.882	0.990	0.995	0.933	1.000	0.867	0.975	0.995	0.912	1.000
19	245	29	0.954	0.964	0.998	0.959	0.999	0.932	0.953	0.997	0.940	0.998
20	178	29	0.921	0.935	0.997	0.928	0.997	0.861	0.801	0.994	0.820	0.997
\end{filecontents*}
\csvreader[
    separator=tab,
    tabular={rrrrrrrrrrrrr}, % 13 colunas: 3 contagens + 10 métricas
    table head={%
      \toprule
      & & & \multicolumn{5}{c}{Micro-averaged} & \multicolumn{5}{c}{Macro-averaged} \\
      \cmidrule(lr){4-8}\cmidrule(lr){9-13}
      Cardiologist ID & Exams reviewed & Classes seen & Precision & Recall & Specificity & F1 score & NPV & Precision & Recall & Specificity & F1 score & NPV \\
      \midrule
    },
    late after line=\\,
    table foot=\bottomrule
  ]{tables/SupplementaryTable8.tsv}{}%
  {%
    \csvcoli   & % Cardiologist ID
    \csvcolii  & % Exams reviewed
    \csvcoliii & % Classes seen
    \csvcoliv  & % Micro Precision
    \csvcolv   & % Micro Recall
    \csvcolvi  & % Micro Specificity
    \csvcolvii & % Micro F1
    \csvcolviii& % Micro NPV
    \csvcolix  & % Macro Precision
    \csvcolx   & % Macro Recall
    \csvcolxi  & % Macro Specificity
    \csvcolxii & % Macro F1
    \csvcolxiii  % Macro NPV
  }
}
\end{table}

\begin{table}[H]
\centering
\caption{Per-class model performance on the CODE-II-test under the F1-max thresholding rule. For each selected diagnostic class, we report Test prevalence (number and percentage), threshold-independent metrics (AUROC, AUPRC), and threshold-dependent metrics (F1-score, Recall/Sensitivity, Specificity, Precision, and NPV) computed using class-specific thresholds selected on the CODE-II validation split and applied unchanged to the test set. All metrics are reported with 95\% confidence intervals estimated from 1,000 bootstrap resamples (shown in parentheses). The table lists the 10 most prevalent classes and 5 representative rare classes for illustration. The complete results for all 66 classes are provided in Supplementary Data 5 (external Excel file, Supplementary\_Data\_5.xlsx).}
\label{Table:SupplementaryTable9}

\resizebox{\textwidth}{!}{%
  \begin{filecontents*}{tables/SupplementaryTable9.tsv}
Description	Label	TotalExams	PercentAllExams	Threshold	AUROC	AUPRC	F1Score	Recall	Specificity	Precision	NPV
ECG within normal limits for age and sex	NORMAL	3,398	40.09	0.2572	0.959 (0.955-0.963)	0.931 (0.923-0.938)	0.850 (0.841-0.858)	0.967 (0.960-0.972)	0.793 (0.782-0.804)	0.758 (0.745-0.770)	0.973 (0.967-0.977)
Nonspecific ST-T abnormality	NS-STT	969	11.43	0.2462	0.938 (0.932-0.945)	0.700 (0.669-0.732)	0.629 (0.608-0.650)	0.801 (0.775-0.826)	0.904 (0.897-0.911)	0.518 (0.494-0.542)	0.972 (0.968-0.976)
Left atrial enlargement	LAE	912	10.76	0.2886	0.898 (0.889-0.907)	0.531 (0.497-0.567)	0.541 (0.514-0.566)	0.638 (0.608-0.670)	0.913 (0.907-0.919)	0.469 (0.441-0.498)	0.954 (0.950-0.960)
Prolonged QT interval	LQT	582	6.87	0.2247	0.975 (0.971-0.978)	0.755 (0.717-0.788)	0.688 (0.660-0.714)	0.802 (0.773-0.831)	0.961 (0.957-0.965)	0.603 (0.568-0.636)	0.985 (0.982-0.988)
Premature ventricular complex(es)	PVC	450	5.31	0.1900	0.986 (0.979-0.992)	0.864 (0.833-0.892)	0.834 (0.808-0.859)	0.908 (0.881-0.933)	0.985 (0.982-0.988)	0.771 (0.736-0.807)	0.995 (0.993-0.996)
First-degree atrioventricular block	1AVB	402	4.74	0.3812	0.988 (0.982-0.993)	0.885 (0.858-0.910)	0.794 (0.763-0.824)	0.850 (0.816-0.887)	0.986 (0.983-0.988)	0.745 (0.705-0.786)	0.992 (0.990-0.994)
Right bundle branch block	RBBB	333	3.93	0.3898	0.996 (0.995-0.997)	0.921 (0.898-0.941)	0.829 (0.798-0.855)	0.898 (0.864-0.931)	0.989 (0.987-0.991)	0.770 (0.728-0.810)	0.996 (0.994-0.997)
Premature atrial complex(es)	PAC	295	3.48	0.1549	0.968 (0.957-0.978)	0.619 (0.559-0.671)	0.640 (0.599-0.678)	0.759 (0.712-0.809)	0.978 (0.975-0.981)	0.554 (0.507-0.602)	0.991 (0.989-0.993)
Left ventricular hypertrophy	LVH	286	3.37	0.2426	0.981 (0.975-0.986)	0.776 (0.733-0.819)	0.701 (0.659-0.742)	0.748 (0.697-0.797)	0.987 (0.984-0.989)	0.660 (0.606-0.712)	0.991 (0.989-0.993)
Pathological Q waves	AQW	271	3.20	0.2523	0.978 (0.971-0.984)	0.681 (0.619-0.738)	0.643 (0.593-0.690)	0.659 (0.603-0.715)	0.987 (0.985-0.990)	0.629 (0.569-0.686)	0.989 (0.986-0.991)
Brugada syndrome	BrS	22	0.26	0.0080	0.992 (0.976-0.999)	0.672 (0.461-0.854)	0.162 (0.000-0.381)	0.092 (0.000-0.235)	1.000 (1.000-1.000)	0.864 (0.000-1.000)	0.998 (0.997-0.999)
T-wave inversions	INVT	20	0.24	0.1086	0.993 (0.982-0.999)	0.505 (0.278-0.734)	0.365 (0.143-0.593)	0.300 (0.105-0.500)	0.999 (0.999-1.000)	0.493 (0.200-0.800)	0.998 (0.998-0.999)
Sinus pause	SPAUSE	20	0.24	0.0001	0.833 (0.743-0.909)	0.024 (0.008-0.054)	0.041 (0.000-0.130)	0.051 (0.000-0.167)	0.997 (0.995-0.998)	0.036 (0.000-0.118)	0.998 (0.997-0.999)
Dextrocardia	DEXTRO	19	0.22	0.0002	0.998 (0.997-0.999)	0.454 (0.281-0.666)	0.161 (0.000-0.378)	0.105 (0.000-0.273)	1.000 (0.999-1.000)	0.400 (0.000-1.000)	0.998 (0.997-0.999)
Biventricular hypertrophy	BVH	14	0.17	0.0002	0.961 (0.936-0.982)	0.077 (0.018-0.200)	0.160 (0.000-0.353)	0.143 (0.000-0.350)	0.999 (0.998-1.000)	0.198 (0.000-0.500)	0.999 (0.998-0.999)
\end{filecontents*}
\csvreader[
    separator=tab,
    tabular={llrrrrrrrrrr},
    table head={%
      \toprule
      CODE statement description & CODE label & Total exams & Prevalence & Threshold & AUROC & AUPRC & F1 score & Recall & Specificity & Precision & NPV \\
      \midrule
    },
    late after line=\\,
    table foot=\bottomrule
  ]{tables/SupplementaryTable9.tsv}{}%
  {%
    \csvcoli   & % Description
    \csvcolii  & % Label
    \csvcoliii & % Total Exams
    \csvcoliv  & % % of All Exams
    \csvcolv   & % AUROC
    \csvcolvi  & % AUPRC
    \csvcolvii & % Threshold
    \csvcolviii& % Precision
    \csvcolix  & % Recall
    \csvcolx   & % Specificity
    \csvcolxi  & % F1 Score
    \csvcolxii   % NPV
  }
}
\end{table}

\begin{table}[H]
\centering
\caption{Aggregate model performance on the CODE-II-test under two threshold rules: the primary F1-max thresholds and an alternative based on Youden’s J (both selected on the CODE-II validation split and applied unchanged to the Test set). Micro- and macro-averaged metrics are reported---Precision, Recall, F1-score, AUROC, and AUPRC---for each rule.}
\label{Table:SupplementaryTable10}

\resizebox{0.8\textwidth}{!}{%
  \begin{filecontents*}{tables/SupplementaryTable10.tsv}
Metric	F1max_micro	F1max_macro	Youden_micro	Youden_macro
Precision	0.632	0.517	0.228	0.182
Recall	0.801	0.563	0.946	0.945
F1-score	0.706	0.512	0.368	0.274
AUROC	0.983	0.978	0.983	0.978
AUPRC	0.776	0.553	0.776	0.553
\end{filecontents*}
\csvreader[
    separator=tab,
    tabular={lrrrr},
    table head={%
      \toprule
      & \multicolumn{2}{c}{F1-max thresholds} & \multicolumn{2}{c}{Youden's J thresholds} \\
      \cmidrule(lr){2-3}\cmidrule(lr){4-5}
      Metric & micro-averaged & macro-averaged & micro-averaged & macro-averaged \\
      \midrule
    },
    late after line=\\,
    table foot=\bottomrule
  ]{tables/SupplementaryTable10.tsv}{}%
  {%
    \csvcoli   & % Metric
    \csvcolii  & % F1-max micro
    \csvcoliii & % F1-max macro
    \csvcoliv  & % Youden micro
    \csvcolv     % Youden macro
  }
}
\end{table}

\begin{table}[H]
\centering
\caption{Per-class model performance on the CODE-II-test for the three diagnostic classes that produced degenerate metrics under the F1-max rule (precision, recall, and F1 equal to zero). For each class, we report Test-set prevalence (number and \%), AUROC and AUPRC, and---under two thresholding rules, F1-max and Youden's J, both selected on the CODE-II validation split and applied unchanged to the Test set---the selected threshold and the resulting Precision, Recall, Specificity, F1-score, and NPV.}
\label{Table:SupplementaryTable11}

\resizebox{\textwidth}{!}{%
\begin{filecontents*}{tables/SupplementaryTable11.tsv}
Description	Label	TotalExams	PercentAllExams	AUROC	AUPRC	ThresholdRule	Threshold	Precision	Recall	Specificity	F1Score	NPV
Multifocal atrial tachycardia	MAT	32	0.38	0.994	0.452	F1-max	0.0312	0.000	0.000	0.999	0.000	0.996
Multifocal atrial tachycardia	MAT	32	0.38	0.994	0.452	Youden's J	3.9976e-08	0.068	1.000	0.948	0.128	1.000
Isorhythmic atrioventricular dissociation	IAVD	28	0.33	0.952	0.074	F1-max	0.0079	0.000	0.000	0.999	0.000	0.997
Isorhythmic atrioventricular dissociation	IAVD	28	0.33	0.952	0.074	Youden's J	7.0004e-08	0.037	0.786	0.933	0.071	0.999
Digitalis effect	DIG	28	0.33	0.926	0.119	F1-max	0.0016	0.000	0.000	0.999	0.000	0.997
Digitalis effect	DIG	28	0.33	0.926	0.119	Youden's J	2.9463e-09	0.012	0.964	0.733	0.023	0.999
\end{filecontents*}
\csvreader[
  separator=tab,
  head to column names,
  tabular={llrrrrlrrrrrr},  % 13 colunas: 2l + 4r + 1l + 6r
  table head={%
    \toprule
    CODE statement description & CODE label & Total exams & \% of all exams & AUROC & AUPRC &
    Thresholding rule & Threshold & Precision & Recall & Specificity & F1 score & NPV \\
    \midrule
  },
  table foot=\bottomrule
]{tables/SupplementaryTable11.tsv}%
{Description=\Desc,Label=\Lab,TotalExams=\Tot,PercentAllExams=\Pct,
 AUROC=\AUROC,AUPRC=\AUPRC,ThresholdRule=\Rule,Threshold=\Thr,
 Precision=\Prec,Recall=\Rec,Specificity=\Spec,F1Score=\Fone,NPV=\NPV}%
{%
  % 1ª linha do par se a regra for F1-max; 2ª linha caso contrário
  \IfStrEq{\Rule}{F1-max}{%
    % imprime todas as 13 colunas
    \Desc & \Lab & \Tot & \Pct & \AUROC & \AUPRC &
    \Rule & \Thr & \Prec & \Rec & \Spec & \Fone & \NPV \\
  }{%
    % imprime só da 7ª à 13ª; nas 6 primeiras deixa vazio
    & & & & & &
    \Rule & \Thr & \Prec & \Rec & \Spec & \Fone & \NPV \\
  }%
}%
}
\end{table}

\begin{table}[H]
\centering
\caption{Macro-AUROC results for all supervised and pre-trained models on the five PTB-XL tasks---diagnosis, subclass, superclass, form, and rhythm---and on the full CPSC 2018 dataset. All models were trained using 100\% of the available training data. For each model, we report the mean macro-AUROC over three runs, with the run-to-run variation shown in parentheses (maximum minus minimum). The table also includes the average macro-AUROC across tasks and the total number of model parameters. The complete set of results, including each model under few-shot training (5\% and 10\%) on PTB-XL and the improvement relative to its randomly initialized counterpart (when applicable), is provided in Supplementary Data~6 (external Excel file, \textit{Supplementary\_Data\_6.xlsx}).}

\label{Table:SupplementaryTable12}
\resizebox{\textwidth}{!}{%
\begin{filecontents*}{tables/SupplementaryTable12.tsv}
Model	PTBXL_diag	PTBXL_sub	PTBXL_sup	PTBXL_form	PTBXL_rhythm	CPSC2018	AVG_AUC_macro	Params
BiLSTM	0.9153 (0.016)	0.9202 (0.006)	0.9224 (0.001)	0.8493 (0.017)	0.9386 (0.033)	0.9581 (0.008)	0.9173	2,335,660
LSTM	0.9168 (0.004)	0.9161 (0.020)	0.9241 (0.003)	0.8266 (0.065)	0.9602 (0.009)	0.9617 (0.010)	0.9176	908,716
XResNet1d101	0.9142 (0.008)	0.8990 (0.008)	0.9148 (0.002)	0.8248 (0.019)	0.9640 (0.009)	0.9607 (0.005)	0.9129	1,877,292
ResNet1d-Wang	0.9291 (0.007)	0.9312 (0.006)	0.9253 (0.003)	0.8609 (0.011)	0.9460 (0.004)	0.9592 (0.006)	0.9253	450,860
InceptionTime	0.9216 (0.007)	0.9134 (0.014)	0.9251 (0.003)	0.8608 (0.031)	0.9589 (0.006)	0.9615 (0.006)	0.9236	512,684
ResNet1d-Lima	0.9195 (0.010)	0.9171 (0.006)	0.9209 (0.002)	0.8041 (0.016)	0.9009 (0.009)	0.9347 (0.003)	0.8995	7,144,908
ResNet1d-Ribeiro	0.9030 (0.008)	0.9027 (0.008)	0.9166 (0.008)	0.8533 (0.012)	0.9543 (0.004)	0.9481 (0.022)	0.913	6,616,508
ResNet1d-Merl	0.9312 (0.008)	0.9332 (0.003)	0.9247 (0.005)	0.8752 (0.021)	0.9144 (0.017)	0.9446 (0.004)	0.9206	3,871,404
ViT1d-Merl	0.9078 (0.005)	0.9128 (0.005)	0.9094 (0.006)	0.7933 (0.009)	0.8925 (0.018)	0.9337 (0.006)	0.8916	5,481,644
ECG-FM	0.8962 (0.005)	0.9148 (0.019)	0.9163 (0.003)	0.8692 (0.024)	0.9666 (0.001)	0.9694 (0.006)	0.9221	90,916,908
Heartlang	0.9188 (0.005)	0.9181 (0.006)	0.9047 (0.006)	0.8582 (0.003)	0.9323 (0.016)	0.9465 (0.012)	0.9131	39,083,820
HuBERT-Small	0.9113 (0.007)	0.9070 (0.003)	0.9046 (0.001)	0.7948 (0.028)	0.9322 (0.005)	0.9310 (0.006)	0.8968	30,231,724
HuBERT-Base	0.8927 (0.032)	0.9054 (0.008)	0.9040 (0.007)	0.8139 (0.013)	0.9448 (0.011)	0.9362 (0.011)	0.8995	92,831,916
Ours	\textbf{0.9399 (0.001)}	\textbf{0.9336 (0.001)}	\textbf{0.9343 (0.003)}	\textbf{0.8932 (0.010)}	\textbf{0.9691 (0.001)}	\textbf{0.9731 (0.002)}	\textbf{0.9405}	7,140,556
\end{filecontents*}
\csvreader[
  separator=tab,
  tabular={lcccccccr},
  table head={%
    \toprule
    & \multicolumn{5}{c}{PTB-XL classification tasks} & \multicolumn{1}{c}{CPSC 2018} & \multicolumn{1}{c}{Average Macro-AUROC} & \multicolumn{1}{c}{\# Parameters} \\
    \cmidrule(lr){2-6} \cmidrule(lr){7-7} \cmidrule(lr){8-8} \cmidrule(lr){9-9}
    Model & Diagnosis & Subclass & Superclass & Form & Rhythm & Diagnosis & Avg. AUROC & Params \\
    \midrule
  },
  late after line=\\,
  table foot=\bottomrule
]{tables/SupplementaryTable12.tsv}{}%
{%
  \csvcoli & \csvcolii & \csvcoliii & \csvcoliv & \csvcolv & \csvcolvi & \csvcolvii & \csvcolviii & \csvcolix 
}
}
\end{table}

\begin{table}[H]
\centering
\caption{Model sizes and pre-training datasets for baseline ECG models. The table reports the number of parameters (estimated from full-data training on PTB-XL diagnostic classes), the size of the pre-training set (in ECG records or ECG–report pairs), and the corresponding data sources used for model pre-training.}
\label{Table:SupplementaryTable13}
\resizebox{\textwidth}{!}{%
\begin{filecontents*}{tables/SupplementaryTable13.tsv}
Model	Parameters	Pretraining Size	Pretraining Sources
ResNet1d-Lima	7,144,908	2,470,424 ECG records	CODE-I
ResNet1d-Ribeiro	6,616,508	2,322,513 ECG records	CODE-I
ResNet1d-Merl	3,871,404	771,693 pairs of ECG records and reports	MIMIC-IV ECG
ViT1d-Merl	5,481,644	771,693 pairs of ECG records and reports	MIMIC-IV ECG
ECG-FM	90,916,908	2,503,136 ECG records	UHN-ECG, MIMIC-IV ECG, CPSC, CPSC-Extra, PTB-XL, Georgia, Ningbo, and Chapman
Heartlang	39,083,820	800,035 ECG records	MIMIC-IV ECG
HuBERT-Small	30,231,724	9.1 million ECG records	CODE-I, CPSC and CPSC-Extra, PTB and PTB-XL, Georgia, Chapman, Ningbo, the partition from the Tianchi Arrhythmia Competition, Shandong Provincial Hospital, MIMIC-IV ECG, and SaMi-Trop
HuBERT-Base	92,831,916	9.1 million ECG records	CODE-I, CPSC and CPSC-Extra, PTB and PTB-XL, Georgia, Chapman, Ningbo, the partition from the Tianchi Arrhythmia Competition, Shandong Provincial Hospital, MIMIC-IV ECG, and SaMi-Trop
Ours	7,140,556	2,735,269 ECG records	CODE-II
\end{filecontents*}
\csvreader[
  separator=tab,
  tabular={lccp{8.0cm}}, % ← Ajuste aqui: 4ª coluna usa largura fixa com quebra de linha
  table head={%
    \toprule
    Model & \# Parameters & Pre-training Size & Pre-training Sources \\
    \midrule
  },
  late after line=\\,
  table foot=\bottomrule
]{tables/SupplementaryTable13.tsv}{}%
{%
  \csvcoli & \csvcolii & \csvcoliii & \csvcoliv
}
}
\end{table}

\begin{table}[H]
\centering
\caption{Training and fine-tuning configurations for pre-trained ECG models. For each model, we report the learning rates used under full-data and few-shot scenarios, the input frequency and segment length, and the adopted fine-tuning strategy. When applicable, model-specific exceptions are noted. The same configurations were used for both PTB-XL and CPSC 2018 experiments unless otherwise specified. Input signals shorter than the target input length were symmetrically zero-padded, while longer signals were symmetrically truncated.}
\label{Table:SupplementaryTable14}
\resizebox{\textwidth}{!}{%
\begin{filecontents*}{tables/SupplementaryTable14.tsv}
Model	LR_FullData	LR_FewShot	Freq_Hz	InputLength	FinetuningStrategy
ResNet1d-Lima	1e-3	1e-3	400	4,096	Full-tuning
ResNet1d-Ribeiro	1e-3	1e-3	400	4,096	Full-tuning
ResNet1d-Merl	1e-3	1e-3	500	N/A	Full-tuning
ViT1d-Merl	1e-3	1e-3	500	N/A	Full-tuning
ECG-FM	1e-6 for pre-trained models,  1e-5 for models with random initialization	1e-3	500	N/A	Full-tuning
Heartlang	1e-3	1e-3	100	N/A	Full-tuning
HuBERT-Small	1e-5	1e-3, 1e-4	500	2,500	Full-tuning was performed except for the convolutional embedder, as specified in the original paper
HuBERT-Base	1e-5	1e-3, 1e-4	500	2,500	Full-tuning was performed except for the convolutional embedder, as specified in the original paper
Ours	1e-3	1e-3	400	4,096	Full-tuning
\end{filecontents*}
\csvreader[
  separator=tab,
  tabular={lp{4.5cm}cccp{5.0cm}},
  table head={%
    \toprule
    & \multicolumn{2}{c}{Learning Rate} & \multicolumn{2}{c}{Input Configuration} & \multicolumn{1}{c}{Fine-tuning Strategy} \\
    \cmidrule(lr){2-3} \cmidrule(lr){4-5} \cmidrule(lr){6-6}
    Model & Full-data & Few-shot & Frequency (Hz) & Input length & Fine-tuning strategy \\
    \midrule
  },
  late after line=\\,
  table foot=\bottomrule
]{tables/SupplementaryTable14.tsv}{}%
{\csvcoli & \csvcolii & \csvcoliii & \csvcoliv & \csvcolv & \csvcolvi}
}
\end{table}

\begin{table}[H]
\centering
\caption{Performance of the baseline and reference ECG models on the CODE-II-open dataset under their best-performing configuration, either trained from scratch or fine-tuned from pre-trained weights. The same evaluation pipeline used for external benchmarking was applied to ensure comparability across experiments. Metrics are reported as the mean macro-AUROC and macro-AUPRC over three independent runs, with the range width (maximum–-minimum) shown in parentheses. For reference, the performance of the baseline model trained on the full CODE-II cohort is also reported, denoted as Ours (CODE-II) and highlighted in bold. The complete results, including all individual runs, are provided in Supplementary Data 7 (external Excel file, \textit{Supplementary\_Data\_7.xlsx}).}
\label{Table:SupplementaryTable15}

\resizebox{0.8\textwidth}{!}{% 
\begin{filecontents*}{tables/SupplementaryTable15.tsv}
Model	training_type	AUROC_macro	AUPRC_macro
BiLSTM	from scratch	0.9372 (0.011)	0.3709 (0.069)
LSTM	from scratch	0.9382 (0.017)	0.3659 (0.083)
XResNet1d101	from scratch	0.9467 (0.005)	0.4184 (0.012)
ResNet1d-Wang	from scratch	0.9405 (0.001)	0.3909 (0.013)
InceptionTime	from scratch	0.9489 (0.003)	0.4168 (0.022)
ResNet1d-Lima	fine-tuned	0.9496 (0.008)	0.4092 (0.030)
ResNet1d-Ribeiro	fine-tuned	0.9531 (0.001)	0.4281 (0.005)
ResNet1d-Merl	fine-tuned	0.9425 (0.002)	0.3917 (0.014)
ViT1d-Merl	fine-tuned	0.9278 (0.002)	0.3451 (0.019)
ECG-FM	fine-tuned	0.9248 (0.003)	0.3292 (0.011)
Heartlang	fine-tuned	0.9125 (0.014)	0.2993 (0.038)
HuBERT-Small	fine-tuned	0.9410 (0.003)	0.3767 (0.027)
HuBERT-Base	fine-tuned	0.9364 (0.004)	0.3562 (0.021)
Ours (CODE-II-open)	from scratch	0.9505 (0.003)	0.4097 (0.012)
Ours (CODE-II)	from scratch	\textbf{0.9779}	\textbf{0.5608}
\end{filecontents*}
\csvreader[
    separator=tab,
    tabular={lccc},
    table head={
        \toprule
        Model & Training type & macro-AUROC & macro-AUPRC \\
        \midrule
    },
    late after line=\\,
    table foot=\bottomrule
]{tables/SupplementaryTable15.tsv}{}%
{\csvcoli & \csvcolii & \csvcoliii & \csvcoliv}
}
\end{table}

\newpage
%====================================================================================
% REFERENCES - SECTION
%====================================================================================
% Bibliografia do SI (numeração reinicia em [1] e título próprio)
% \bibliographystyleSI{naturemag}
% \bibliographySI{references}

% --- Bibliografia do SI: prefixo [S…] e numeração correta ---
\begingroup
\small
% \makeatletter
%   \@ifundefined{NAT@ctr}{}{\setcounter{NAT@ctr}{0}}
%   \makeatother
 % Prefixa o rótulo do item na lista do SI: [S1], [S2], ...
  %\renewcommand{\bibnumfmt}[1]{[S#1]}
  \bibliographystyleSI{naturemag}
  \bibliographySI{references} % mesmo .bib do principal
\endgroup

%\FinishSupplement
%-------------------------------------

%====================================================================================
% Final point
%====================================================================================
\end{document}